\documentclass[useAMS,usenatbib]{mn2e}
\usepackage{graphicx}
\usepackage{multirow, bigstrut}%
\usepackage[para,online,flushleft]{threeparttable}
\usepackage{amssymb}
\usepackage[usenames,dvipsnames]{xcolor}%
\usepackage[figuresright]{rotating}
\usepackage{pdflscape}
\usepackage{hyperref}

\begin{document}

\title[3D MHD shock-filament simulations]{The interaction of a magnetohydrodynamical shock with a filament}
\author[K. J. A. Goldsmith and J. M. Pittard] 
  {K. J. A. ~Goldsmith \thanks{pykjag@leeds.ac.uk} and J. M. ~Pittard\\
   School of Physics and Astronomy, University of Leeds, 
   Woodhouse Lane, Leeds LS2 9JT, UK}
   
\date{Accepted ... Received ...; in original form ...}

\pagerange{\pageref{firstpage}--\pageref{lastpage}} \pubyear{2016}
\newtheorem{theorem}{Theorem}[section]
\label{firstpage}

\maketitle

\begin{abstract}
We present 3D magnetohydrodynamic numerical simulations of the adiabatic interaction of a shock with a dense, filamentary cloud. We investigate the effects of various filament lengths and orientations on the interaction using different orientations of the magnetic field, and vary the Mach number of the shock, the density contrast of the filament $\chi$, and the plasma beta, in order to determine their effect on the evolution and lifetime of the filament. We find that in a parallel magnetic field filaments have longer lifetimes if they are orientated more ``broadside'' to the shock front, and that an increase in $\chi$ hastens the destruction of the cloud, in terms of the modified cloud-crushing timescale, $t_{cs}$. The combination of a mild shock and a perpendicular or oblique field provides the best condition for extending the life of the filament, with some filaments able to survive almost indefinitely since they are cocooned by the magnetic field. A high value for $\chi$ does not initiate large turbulent instabilities in either the perpendicular or oblique field cases but rather draws the filament out into long tendrils which may eventually fragment. In addition, flux ropes are only formed in parallel magnetic fields. The length of the filament is, however, not as important for the evolution and destruction of a filament.

\end{abstract}

\section{Introduction}
The interstellar medium (ISM) is known to be a highly dynamic and non-uniform entity containing regions of varying temperature and density (see the review paper by \citet{Ferriere01}). Studies of the interaction of hot, high-velocity gas with cooler, dense material (often referred to as ``clouds") are of great interest for a complete understanding of the gas dynamics of the ISM since it is evident that the evolution and morphology of large-scale flows can be determined by the far smaller clouds \citep{Elmegreen04, MacLow04, Scalo04, McKee07, Hennebelle12, Padoan14}. Clouds may either accrete material from, or lose material to, the ambient medium: clouds which are hit by shocks or winds are likely to be destroyed, with such destruction affecting the flow by ``mass-loading" it via processes such as hydrodynamic ablation, whereas clouds may also collapse after being struck by a shock and therefore trigger star formation, thus removing material from the ISM \citep{Elmegreen77, Federrath10, Federrath12}. 

Shock-cloud interactions have been previously inferred from observations (e.g. \citealt{Baade54}; \citealt{vandenBergh71}) while more recent observations have provided direct evidence, e.g. bow shocks, for shock waves interacting with clouds (e.g. \citealt{Levenson02}). Recently, \textit{Herschel} images have revealed the ubiquitous presence of filamentary structures throughout the ISM in both star-forming and non-star-forming regions (e.g. \citealt{Andre10, Andre14}).

There is now a large amount of literature, beginning in the 1970s, concerning the idealised case of a planar adiabatic shock striking an isolated spherical cloud. Numerical studies where the shock Mach number $M$ and cloud density contrast $\chi$ were varied include \citet{Stone92} and \citet{Klein94}. Other studies have reported on the effects of additional processes on the interaction, such as magnetic fields (e.g. \citealt{MacLow94}; \citealt{Shin08}), radiative cooling (e.g. \citealt{Mellema02}; \citealt{Fragile04}; \citealt{Yirak10}) and thermal conduction (e.g. \citealt{Orlando05, Orlando08}). \citet{Pittard09, Pittard10} also explored the turbulent nature of cloud destruction, whilst \citet{Poludnenko02} and \citet{Aluzas12, Aluzas14} investigated the interaction of shocks with multiple clouds, and \citet{vanLoo10} explored the interaction of a weak, radiative shock with a magnetised cloud.

The purely hydrodynamic shock-cloud interactions lead to the cloud becoming initially compressed, as the shock strikes it, and over-pressured before the cloud re-expands. The cloud is then destroyed via the growth of dynamical instabilities such as Kelvin-Helmholtz (KH) and Rayleigh-Taylor (RT) instabilities which deposit vorticity at the cloud surface, leading to the mixing of the cloud material with the ambient medium. The interaction is milder at lower shock Mach numbers (e.g. \citealt{Nakamura06}; \citealt{Pittard10}) and more marked differences are observed when the post-shock gas is subsonic with respect to the cloud.

The presence of magnetic fields can strongly change the nature of the interaction. 2D axisymmetric simulations have shown that if there is a magnetic field present then the formation of the KH and Richtmyer-Meshkov (RM) instabilities are impeded and the mixing of the cloud with the flow is reduced \citep{MacLow94}. Thus the presence of a magnetic field can prevent the complete destruction of the cloud, allowing it to survive as a coherent structure, as opposed to mixing completely with the ambient flow (as in the field-free case). Furthermore, if the field is parallel to the shock normal a ``flux rope" is formed behind the cloud since the field is preferentially amplified at that point due to shock-focussing. 3D simulations show that when the magnetic field is strong and aligned either perpendicularly or obliquely to the shock normal the cloud takes on a sheet-like appearance at late times and becomes orientated parallel to the post-shock field \citep{Shin08}. A perpendicular field can better deflect the flow around the cloud and reduce mixing, whereas a parallel field allows the cloud to be permeated by the flow and this enhances mixing \citep{Li13}. This effect was also noted in the paper on wind-cloud interactions by \citet{Banda16}, who found that cloud models where the magnetic field component was transverse to the wind direction had higher mixing fractions and velocity component dispersions than models where the field component was aligned with the flow. More recent work has considered the optimum field strength needed to produce cloud fragments which can survive the destructive processes and has found that intermediate-strength fields are most effective, since strong fields prevent compression and weak fields do not insulate the cloud from cooling \citep{Johansson13}. 

There are very few numerical studies in the current literature which consider interactions involving non-spherical clouds, and (to our knowledge) none which describe the effects of a magnetic field on these interactions. One of the first such studies concerned a shock interacting with a cylindrical cloud of aspect ratio 3:1 \citep{Klein94}. The cloud was orientated along the axis of propagation. \citet{Klein94} used a modified equation for the cloud-crushing time and found their results comparable to those of a spherical cloud; thus they concluded that small changes to the initial shape of the cloud did not alter their main conclusions.

Another study \citep{Xu95} focussed on 3D simulations of shock-cloud interactions for clouds with varying morphologies and orientations. Unlike \citet{Klein94}, who assumed a cylindrical cloud aligned in the direction of shock propagation, \citet{Xu95} were able to orientate their cloud of aspect ratio 2:1 in all directions. They found that by modifying the cross-section of the cloud its evolution could be significantly altered depending on the cloud geometry. They also found that, whilst the formation of a vortex ring is a feature of interactions with spherical clouds, a prolate cloud aligned perpendicularly to the shock normal does not form a vortex ring since the interaction of the shock is more complex. Additionally, an aligned cloud was also accelerated to the post-shock flow velocity at a much faster rate than a spherical cloud. In contrast, the evolution of an inclined prolate cloud was substantially different from the aligned cloud: in this case the cloud's inclination caused it to be spun around, drastically altering the development of instabilities. 

The most recent study, \citet{Pittard16}, investigated shock-filament interactions and studied the formation of turbulent vortices behind the filaments as a result of the shock-filament interaction. They found that varying the filament length and angle of orientation to the shock front significantly changed the nature of the interaction. Filaments orientated at $\theta \lesssim 60^{\circ}$ formed three parallel rolls, whilst filaments orientated sideways-on expanded preferentially along their minor axis and in the direction of shock propagation. Slightly oblique filaments tended to spill the high vorticity flow around the upstream end of the filament. These filaments had longer wakes and were less symmetrical. Highly oblique filaments, in contrast, had a dominant vortex ring at the upstream end of the filament which aided their subsequent fragmentation.

The current study extends the purely hydrodynamic work conducted by \citet{Pittard16}. By nature, it represents an idealised scenario before more realistic simulations of filaments are conducted. We investigate the effects that magnetic fields have on shock-filament interactions by varying the Mach number, density contrast, and plasma beta, in addition to varying the orientation and length of the filament, for parallel, perpendicular, and oblique magnetic fields.

The outline of this paper is as follows: in Section 2 we introduce our numerical method, initial conditions and the results of a convergence study. In Section 3 we present the results of our simulations. A discussion of the relevance of our work to shock-filament and wind-filament studies is given in Section 4. Section 5 summarises and concludes, and addresses the motivation for further work.

\section{The numerical setup}
The computations were performed using the MG magnetohydrodynamic code which utilises adaptive mesh refinement (AMR). The code solves a Riemann problem at each cell interface in order to determine the conserved fluxes for the time update, using piecewise linear cell interpolation. The scheme is second-order accurate in space and time. A linear solver is used in most instances, with an exact solver where there is a large difference between the two states \citep{Falle91, Falle98}. The code solves numerically the ideal magnetohydrodynamic (MHD) equations of inviscid flow. In this study we limit ourselves to a purely MHD case, ignoring the effects of thermal conduction, radiative cooling, and self-gravity. Computations were performed for an adiabatic, ideal gas, with a ratio of specific heats $\gamma = 5/3$.

A hierarchy of $n$ grid levels, $G^{0} \cdots G^{n-1}$, is used and the two coarsest grids ($G^{0}$ and $G^{1}$) cover the entire domain, with finer grids being added where needed and removed where they are not. The amount of refinement is increased at points in the mesh where shocks or discontinuities exist, i.e. where the variables associated with the fluid show steep gradients. At these points, the number of computational grid cells produced by the previous level is increased by a factor of 2 in each spatial direction. Thus, fine grids are only utilised in regions where the mesh is highly variable, with much coarser grids used where the flow is relatively uniform. Refinement and derefinement are performed on a cell-by-cell basis and are controlled by the differences in the solutions on the two coarsest grids. Refinement occurs when there is a difference of more than 1\% between a conserved variable in the finest grid and its projection from a grid one level down. If the difference in the two preceding levels falls to below 1\%, the cell is derefined. In order to maintain accuracy and ensure a smooth transition between multiple levels, the refinement criteria are, to an extent, diffused, flux corrections are applied at the boundaries between coarse and fine cells, and the solution in the coarser cells is over-written by that in the finer cells. The time step on grid $G^{n}$ is $\Delta t_{0}/2n$ where $\Delta t_{0}$ is the time step on grid $G^{0}$. The effective resolution is taken to be the resolution of the finest grid and is given as $R_{cr}$, where `cr' is half the number of cells per filament semi-minor axis in the finest grid, equivalent to the number of cells per cloud radius for a spherical cloud. In the following sections we refer to this cloud radius as the ``filament radius''. All length scales are, therefore, measured in units of the filament radius, $r_{c}$, where $r_{c}=1$, and the unit of density is taken to be the density of the surrounding unshocked gas, $\rho_{amb}$. We impose no inherent scale on our simulations, thus our results are applicable to a broad range of scenarios.

\subsection{Initial conditions}

A three-dimensional XYZ cartesian grid is used with constant inflow from the negative $x$ direction and free inflow/outflow conditions at other boundaries. The numerical domain is set to be large enough so that the main features of the interaction occur before the shock reaches the edge of the grid. Since the grid extent is $\chi$-dependent (because, for example, a larger value of $\chi$ means that a hydrodynamical cloud takes longer to be destroyed, and therefore a larger grid is needed - see \citet{Pittard10} \S4.1.2. for a discussion on how the nature of the interaction changes with $\chi$ for hydrodynamic cases) and $M$-dependent the grid extent for each simulation is given in Table~\ref{Table1}. 

\begin{table}
 \caption{The grid extent for each of the simulations. $M$ is the sonic Mach number and $\chi$ is the cloud density contrast. The unit of length is the initial filament radius, $r_{c}$.}
 \label{Table1}
 \begin{threeparttable} 
 \begin{tabular}{@{}llccc}
  \hline
  $M$ & $\chi$
        & $X$
        & $Y$ & $Z$   \\
  \hline
10	& 10 &  $-20 < X < 560 $& $-10 < Y < 10$ & $-12 < Z < 10$ \\
10	& $10^{2, \dag}$ & $-20 < X < 500$ & $-14 < Y < 14$ & $-23 < Z < 15$ \\
10	& $10^{2, \ddag}$ & $-20 < X < 1000$ & $-14 < Y < 14$ & $-30 < Z < 14$ \\
10	& $10^{3, \dag}$ & $-20 < X < 300$ & $-14 < Y < 14$ & $-41 < Z < 15$ \\
10	& $10^{3, \ddag}$ & $-20 < X < 800$ & $-14 < Y < 14$ & $-40 < Z < 20$ \\
3	& 10 & $-20 < X < 500$	& $-14 < Y < 14$ & $-15 < Z < 13$ \\
1.5	& 10 & $-20 < X < 800$	& $-12 < Y < 20$ & $-20 < Z < 20$ \\
  \hline
 \end{tabular}
 \begin{tablenotes}{Notes: \dag parallel magnetic field; \ddag perpendicular/oblique magnetic field} 
 \end{tablenotes}
\end{threeparttable} 
 \end{table}
 
\begin{figure} 
\centering
\includegraphics[width=89mm]{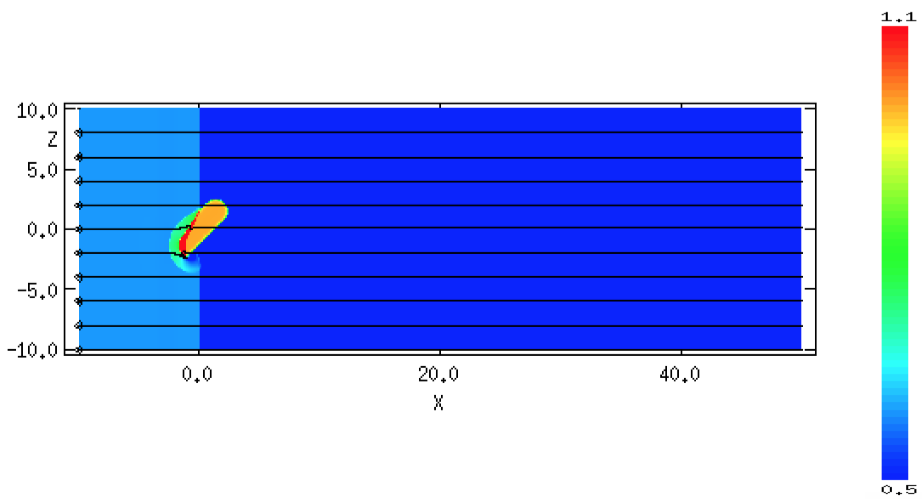}
\vspace{-5mm}
\caption{The interaction at $t = 0\,t_{cs}$ for model $m10c1b1l4o45pa$ (see \S3 for the model naming convention). The scale shows logarithmic density, from red (highest density) to blue (lowest density). The density has been scaled with respect to the ambient density, so that a value of $0$ represents the value of $\rho_{amb}$ and $1$ represents $10\times \rho_{amb}$. The filament is initially positioned at the origin, with the spatial scale in units of the initial filament radius $r_{c}$. The shock front moves from $-x$ to $+x$ and the magnetic field lines are parallel to the shock front.}
\label{Fig1}
\end{figure}

The simulated cloud is a cylinder of length $l$ with hemispherical caps, representing an idealised filament, and the total length of the filament is given by $(l+2)r_{c}$. We are therefore able to vary the aspect ratio and orientation of the filament in order to investigate how such a change might alter the interaction. The filament has been given smooth edges over about 10\% of its radius, using the density profile from \citet{Pittard09}, with $p_{1}=10$ giving a reasonably sharp-edged cloud. The filament and surrounding ambient medium are in pressure equilibrium. The filament is centred on the grid origin $x,\,y,\,z=(0,0,0)$ with the planar shock front (propagating through a magnetised ambient medium) imposed on the grid at $x = -10$. Figure~\ref{Fig1} shows the interaction at $t = 0\,t_{cs}$ (see Eq.~\ref{eq3} for the definition of this timescale). The simulations are described by the sonic Mach number of the shock $M$, the cloud density contrast $\chi$, the filament length $l$, and the ratio of thermal to magnetic pressure (also known as the ``plasma beta'') $\beta_{0}=8\pi P_{0}/B_{0}^{2}$, where $P_{0}$ is the ambient thermal pressure and $B_{0}$ is the ambient magnetic field strength. The filament orientation with respect to the $z$ axis (or shock front), $\theta$, and the magnetic field orientation with respect to the shock normal, are also considered. The simulations are scale-free and expressed in dimensionless units.

Various diagnostic quantities are used to follow the evolution of the interaction (see \citealt{Klein94}; \citealt{Nakamura06}; \citealt{Pittard09}). These quantities include the filament mass ($m$), mean density ($\langle \rho \rangle$), filament volume ($V$), mean velocity along each axis (e.g. $\langle v_{x}\rangle$), and velocity dispersions along each orthogonal axis (e.g. $\delta v_{x}$). An advected scalar is used to trace the filament material in the flow, allowing the whole filament along with its denser core to be distinguished from the ambient medium. Therefore each of the global quantities is able to be computed for the cells associated with either the filament core (using the subscript ``core'', e.g. $m_{core}$) or the entire filament (using the subscript ``cloud'', e.g. $m_{cloud}$).

\citet{Klein94} defined a characteristic timescale for a spherical cloud to be crushed by the shock being driven into it (the ``cloud-crushing time"):
\begin{equation}
t_{cc} = \frac{\chi^{1/2} r_{c}}{v_{b}}\, ,
\end{equation}
where $v_{b}$ is the shock velocity in the ambient medium. A second timescale was defined by \citet{Klein94}, namely a modified cloud-crushing time for cylindrically-shaped clouds:
\begin{equation}
t_{cc}^{'} = \frac{(\chi \, a_{0} c_{0})^{1/2}}{v_{b}}\, ,
\end{equation}
where $a_{0}$ and $c_{0}$ are the initial radii of the cloud in the radial and axial directions respectively. \citet{Xu95} instead provided a modified cloud-crushing time for prolate clouds:
\begin{equation}
t_{cs} = \frac{r_{s} \chi ^{1/2}}{v_{b}}\, ,
\label{eq3}
\end{equation}
where $r_{s}$ is the radius of a sphere of equivalent mass. \citet{Pittard16} compared all three timescales and found that the one defined by \citet{Xu95} for prolate clouds gave a slightly better reduction in variance between the simulations. Therefore, this timescale, $t_{cs}$, has been adopted for this paper, with the assumption that the smooth edges to the filament can be approximated as reasonably sharp edges (\citealt{Pittard09}).

Several other timescales are available. For example, the ``drag time'', $t_{drag}$, is the time taken for the average cloud velocity relative to the post-shock flow to decrease by a factor of $e$ (i.e. the time when the average cloud velocity $\langle v \rangle_{cloud}\,=(1-1/e)\,v_{ps}$, where $v_{ps}$ is the velocity of the post-shock flow as measured in the frame of the pre-shock ambient medium); the ``mixing time'', $t_{mix}$, is the time when the filament core mass is half that of its initial value, and the cloud ``lifetime'', $t_{life}$, is the time taken for the filament core mass to reach 1\% of its initial value.

Time zero in our calculations is taken to be the time when the inter-cloud shock is level with the centre of the filament.

\subsection{Convergence studies}
In numerical studies it is important to show that the quantities from the simulation under consideration are converged and do not change as the resolution increases, and that therefore the calculations are being performed at a resolution great enough to resolve clearly the main features of the interaction, e.g. the growth of magnetohydrodynamic instabilities. The growth of such instabilities at the cloud surface generates turbulence and any increase in resolution could lead to increasingly small scales with respect to the turbulence. Diagnostic quantities such as the mixing rate between cloud and ambient medium are sensitive to small-scale instabilities and are therefore less likely to show convergence. Resolution tests of numerical shock-cloud interactions for 2D adiabatic, hydrodynamic, spherical clouds have revealed that such simulations require a resolution of at least 100 cells per cloud radius ($R_{100}$) for converged results (e.g. \citealt{Klein94}; \citealt{Nakamura06}), with more complex cases requiring even higher resolutions (e.g. \citealt{Yirak10}). However, it is very computationally expensive to run 3D simulations to such high resolutions.

3D studies of spherical clouds have shown that convergence at resolutions as low as $R_{32}$ is achievable, though to properly capture the behaviour of the interaction a resolution of $R_{64}$ is necessary \citep{Pittard15}. Even more encouragingly, these authors found very little difference between inviscid and $k - \epsilon$ turbulence model\footnote{The subgrid $\kappa-\epsilon$ turbulence model is used to model the mean flow in fully-developed, high Reynolds number turbulence. It has been calibrated by comparing the growth of shear layers determined experimentally with computed values \citep{Dash83}. Details of its implementation in \textit{MG} can be found in \citet{Falle94}} simulations (it had previously been established that 2D studies which include the $k - \epsilon$ model are convergent at lower resolutions, in contrast with inviscid studies \citep{Pittard09}). The non-turbulent, hydrodynamic 3D \citet{Xu95} study found that the evolution of the effective size of a prolate cloud was resolution-dependent and that a resolution of at least $R_{27}$ was needed for convergence of all the diagnostic quantities. However, because a large grid was required for their cloud they were unable to run a ``high'' resolution simulation to test this. One of the few 3D MHD resolution tests in the literature was performed by \citet{Shin08} for a spherical cloud using a non-AMR code at resolutions of $R_{120}$ and $R_{60}$ and concluded that most aspects of the MHD shock-cloud interaction were well converged at both resolutions. To our knowledge, the only resolution tests for a 3D purely hydrodynamic shock-filament interaction were performed by \citet{Pittard16}, who demonstrated that convergence was possible at a resolution of $R_{32}$.  

We extend these resolution tests to a 3D MHD shock-filament interaction. We focus on two measures, the mean cloud velocity, $\langle v_{x}\rangle$, and the core mass of the cloud, $m_{core}$, which are affected by the cloud material becoming mixed with the flow and which are therefore suitable indicators of convergence. 

\begin{figure}
\centering
     \includegraphics[width=68mm]{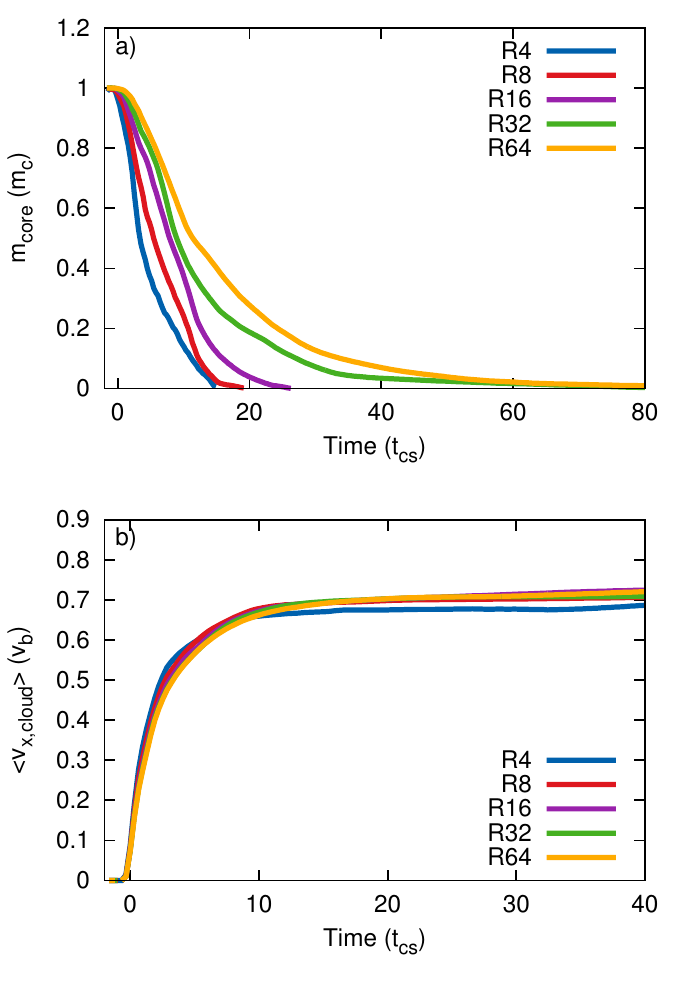}
  \caption{Convergence tests for 3D MHD simulations of a Mach 10 shock hitting a filament with density contrast $\chi = 10$ in a parallel field. The time evolution of the core mass (a) (normalised to the value of the initial filament mass, $m_{core, 0}$), and mean cloud velocity (b) are shown.}
  \label{Fig2}
  \end{figure}

\begin{figure} 
\centering
\includegraphics[width=85mm]{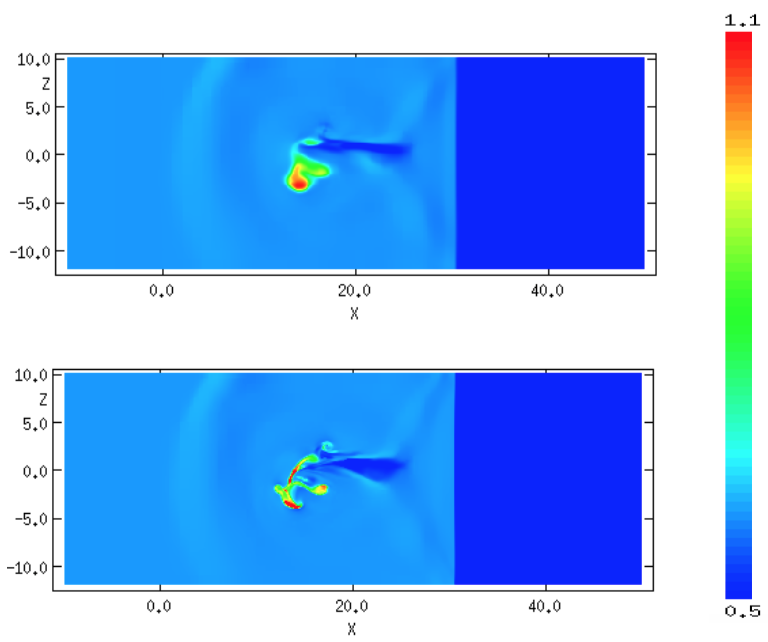}
\caption{Resolution test for a Mach 10 shock overrunning a filament, using the initial setup shown in Fig.~\ref{Fig1}. A logarithmic density plot, scaled in terms of the ambient density, is shown at $t=6.11 \,t_{cs}$ for resolutions $R_{8}$ (top) and $R_{32}$ (bottom).}
\label{Fig3}
\end{figure}

It is known that simulations run with lower density contrasts are much more resolution-dependent. When $\chi=10$ (which is the case for the majority of our simulations) the filament is destroyed faster at lower resolutions. Figure~\ref{Fig2} shows the time evolution of the core mass (a) and mean cloud velocity (b) as a function of the spatial resolution for simulations with $M=10$, $\beta_{0} = 1$, $\chi = 10$, $l=4$, a parallel field orientation, and a filament orientation of $45^{\circ}$ to the $z$ axis. Figure~\ref{Fig3} illustrates the difference in resolution, in terms of the main features of the evolution of the filament, between resolutions $R_{8}$ and $R_{32}$. It can be seen from Fig.~\ref{Fig2}(b) that, with the exception of $R_{4}$, all resolutions are reasonably convergent until approximately $30 \,t_{cs}$, after which there is some slight divergence. However, from Fig.~\ref{Fig2}(a), it is clear that there are much larger differences between each of the simulations. There appears to be some convergence between $R_{32}$ and $R_{64}$, at least until approximately $15 \,t_{cs}$ when a fifth of the core mass has been lost, and the filaments in these simulations initially lose their core mass much more slowly than the filaments in the lower resolution simulations. However, we were restricted from comparing even higher resolution runs because of the large computational requirements. 

\begin{figure}
\centering
     \includegraphics[width=131mm]{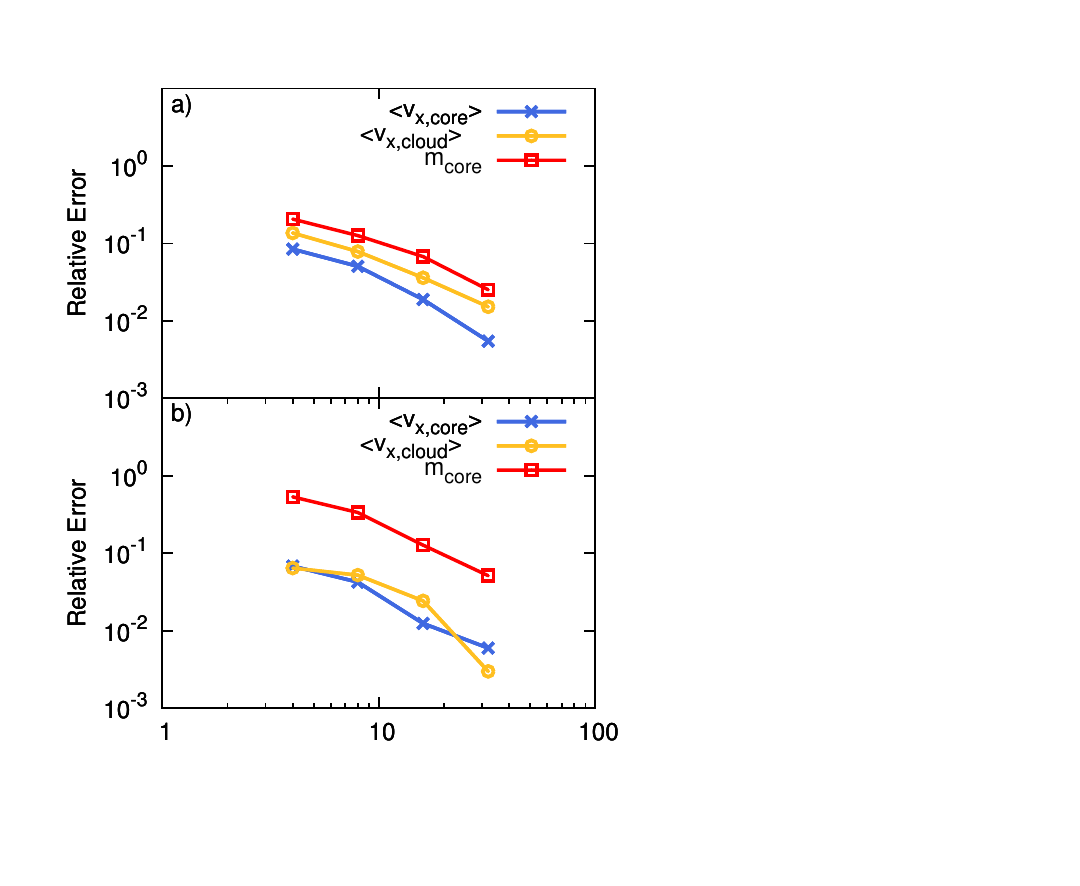}
     \vspace{-20mm}     
  \caption{Relative error (compared to the highest resolution simulation) versus spatial resolution (the number of cells per filament radius on the finest grid) for a number of global quantities measured from a shock-filament interaction with $\chi=10$, $M=10$, and $\beta_{0}=1$ at $t=2\,t_{cs}$ (top) and $t=5\,t_{cs}$ (bottom).}
  \label{Fig4}
  \end{figure}

Figure~\ref{Fig4} shows the relative error, which is defined as the fractional difference between the value of a global parameter measured at a resolution $N$ and that measured at the finest resolution $f$:

\begin{equation}
\Delta Q_{N}= \frac{|Q_{N}-Q_{f}|}{|Q_{f}|} \, ,
\end{equation}
where, for simulations with $M=10$, $\chi=10$, and $\beta_{0}=1$, $f=64$. It can be seen that, in general, the relative error decreases with increasing resolution, and thus manifests convergence. This is in line with the results from \citet{Pittard15} and \citet{Pittard16}. Figure~\ref{Fig4}(a) shows that for a resolution of $R_{32}$ all quantities have a relative error of below $5\%$ at $t=2\,t_{cs}$. As the simulations progress, the relative error in the core mass increases overall. However, for $R_{32}$, the relative error in the mass is still $\sim5\%$ (and is even lower for the other quantities), indicating that a resolution of $R_{32}$ provides reasonably-converged results, and adding support for the adoption of this resolution in all subsequent simulations.

\section{Results}
In this section we present the results of various simulations where we have varied $M$, $\chi$, $\beta_{0}$, $l$, and $\theta$. Table~\ref{Table2} summarises the calculations performed. We adopt a naming convention for each simulation such that $m10c1b1l2o45pa$ refers to a simulation with $M=10$, $\chi=10$, $\beta_{0} =1$, $l=2$, a filament orientation of $\theta=45^{\circ}$ and a parallel magnetic field. The majority of the simulations performed are for $M=10$, $\chi=10$, and $\beta_{0} =1$, whilst the length and orientation of the filament are varied. Towards the end of each section we will also discuss the results from simulations with different Mach numbers, density contrasts and plasma betas. A simulation of a spherical cloud of radius $r_{c}=1$ is also included for comparison with filaments of varying length (note that these simulations were run with a resolution of $R_{16}$).

\begin{table*}
\setlength{\tabcolsep}{5pt}
\centering
 \caption{A summary of the shock-filament simulations performed for a parallel magnetic field. $M$ is the sonic Mach number, $\chi$ is the density contrast of the filament to the surrounding ambient medium, $\beta_{0}$ is the ratio of thermal to magnetic pressure, $l$ defines the length of the filament, and $\theta$ defines the angle of orientation of the filament between its major-axis and the shock surface. $v_{b}$ is the shock speed through the inter-cloud medium (in code units).$v_{ps}$ is the post-shock flow velocity, and is given in units of $v_{b}$. $M_{A}$ is the Alfv\'{e}nic Mach number, $M_{slow/fast}$ are the slow/fast magnetosonic Mach numbers. $t_{cc}$ is the cloud-crushing timescale of \citet{Klein94}, while $t_{cs}$ is the cloud-crushing timescale for a spherical cloud of equivalent mass introduced by \citet{Xu95}. Key filament timescales are additionally noted. Values appended by $\dag$ denote that the true value was greater than that given but that the simulation had ended before this point was reached.}
 \label{Table2}
 \begin{tabular}{@{}lccccccccccccll}
  \hline
  Simulation & $M$ & $\chi$ & $\beta_{0}$
        & $l$ ($r_{c}$) & $\theta$ ($^{\circ}$) & $v_{b}$ & $v_{ps} (v_{b})$ & $M_{A}$ & $M_{slow}$ & $M_{fast}$ & $t_{cs}/t_{cc}$ & $t_{drag}/t_{cs}$ & $t_{mix}/t_{cs}$ & $t_{life}/t_{cs}$ \\
  \hline
m10c1b1l2o45 & 	10 & 	$10$ & 	1 & 	2 & 	$45^{\circ}$  & 13.6 & 0.74 & 9.13  & 10.0 & 9.13 & 1.36 & 2.98 & 8.32 & 25.4 \\
m10c1b1l4o45 & 	10 & 	$10$ & 	1 & 	4 & 	$45^{\circ}$ & 13.6 & 0.74 & 9.13 & 10.0 & 9.13 & 1.59 & 2.55 & 9.06 & 69.5 \\
m10c1b1l8o45 & 	10 & 	$10$ & 	1 & 	8 & 	$45^{\circ}$ & 13.6 & 0.74 & 9.13 & 10.0 & 9.13 & 1.91 & 2.36 & 8.86 & 37.4 \\
m10c1b1l4o0 & 	10 & 	$10$ & 	1 & 	4 & 	$0^{\circ}$ & 13.6 & 0.74 & 9.13 & 10.0 & 9.13 & 1.59 &1.27 & 7.10 & 91.1 \\
m10c1b1l4o30 & 	10 & 	$10$ & 	1 & 	4 & 	$30^{\circ}$ & 13.6 & 0.74 & 9.13 & 10.0 & 9.13 & 1.59 & 1.90 & 10.4 & 104 \\
m10c1b1l4o70 & 	10 & 	$10$ & 	1 & 	4 & 	$70^{\circ}$ & 13.6 & 0.74 & 9.13 & 10.0 & 9.13 & 1.59 & 3.19 & 7.10 & 20.7 \\
m10c1b1l4o85 & 	10 & 	$10$ & 	1 & 	4 & 	$85^{\circ}$ & 13.6 & 0.74 & 9.13 & 10.0 & 9.13 & 1.59 & 2.56 & 6.46 & 19.1 \\
m10c1b1l4o90 & 	10 & 	$10$ & 	1 & 	4 & 	$90^{\circ}$ & 13.6 & 0.74 & 9.13 & 10.0 & 9.13 & 1.59 & 2.56 & 6.11 & 19.1 \\
m10c2b1l4o45 & 	10 & 	$10^{2}$ & 	1 & 	4 & 	$45^{\circ}$ & 13.6 & 0.74 & 9.13 & 10.0 & 9.13 & 1.59 & 4.35 & 5.17 & 11.7 \\
m10c3b1l4o45 & 	10 & 	$10^{3}$ & 	1 & 	4 & 	$45^{\circ}$ & 13.6 & 0.74 & 9.13 & 10.0 & 9.13 & 1.59 & 4.72 & 4.49 & 7.30 \\
m10c1b0.5l4o45 & 	10 & 	$10$        & 	0.5 & 4 & 	$45^{\circ}$ & 13.6 & 0.74 & 6.45 & 10.0 & 6.46 & 1.59 & 2.55 & 35.7 & 79.1 \\
m10c1b10l4o45 & 	10 & $10$        & 	10 & 	4 & 	$45^{\circ}$ & 13.6 & 0.74 & 28.9 & 28.9 & 10.0 & 1.59 & 2.55 & 7.42 & 19.1 \\
m1.5c1b1l4o45 & 	1.5 & $10$ & 	1 & 	4 & 	$45^{\circ}$ & 2.04 & 0.42 & 1.37 & 1.50 & 1.37 & 1.59 & 2.26 & 127$^{\dag}$ & 127$^{\dag}$ \\
m3c1b1l4o45 & 	3    & 10          &        1 & 	4 & 	$45^{\circ}$ & 4.07 & 0.67 & 2.74 & 3.00 & 2.74 & 1.59 & 2.70 & 212 & 213$^{\dag}$ \\
  \hline
 \end{tabular}
\end{table*}

\begin{table*}
\setlength{\tabcolsep}{2pt}
\centering
 \caption{As Table~\ref{Table2} but for perpendicular and oblique magnetic fields. All columns apply to both perpendicular and oblique fields, except columns which contain parentheses - in these columns, values without (with) parentheses indicate perpendicular (oblique) simulations. Values appended by $\dag$ denote that the true value was greater than that given but that the simulation had ended before this point was reached.}
 \label{Table3}
 \begin{tabular}{@{}lcccccccccccccc}
  \hline
  Simulation & $M$ & $\chi$ & $\beta_{0}$
        & $l$ ($r_{c}$) & $\theta$ ($^{\circ}$) & $v_{b}$ & $v_{ps} (v_{b})$ & $M_{A}$ & $M_{slow}$ & $M_{fast}$ & $t_{cs}/t_{cc}$ & $t_{drag}/t_{cs}$ & $t_{mix}/t_{cs}$ & $t_{life}/t_{cs}$ \\
  \hline
m10c1b1l2o45 & 	10 & 	$10$ & 	1 & 	2 & 	$45^{\circ}$  & 13.6 & 0.73 (0.74) & 9.13  & $\infty$ (17.7) & 6.74 (7.29) & 1.36 & 1.86 (2.23) & 181 (112) & 181.70$^{\dag}$ (149.48$^{\dag}$) \\
m10c1b1l4o45 & 	10 & 	$10$ & 	1 & 	4 & 	$45^{\circ}$ & 13.6 & 0.73 (0.74) & 9.13  & $\infty$ (17.7) & 6.74 (7.29) & 1.59 & 1.59 (1.91) & 128 (128$^{\dag}$) & 127.71$^{\dag}$ (127.81$^{\dag}$) \\
m10c1b1l8o45 & 	10 & 	$10$ & 	1 & 	8 & 	$45^{\circ}$ & 13.6 & 0.73 (0.74) & 9.13  & $\infty$ (17.7) & 6.74 (7.29) & 1.91 & 1.32 (1.58) & 104$^{\dag}$ (106$^{\dag}$) & 104.08$^{\dag}$ (106.06$^{\dag}$) \\
m10c1b1l4o0 & 	10 & 	$10$ & 	1 & 	4 & 	$0^{\circ}$ & 13.6 & 0.73 (0.74) & 9.13  & $\infty$ (17.7) & 6.74 (7.29) & 1.59 & 0.95 (1.27) & 71.7 (73.3) & 128$^{\dag}$ (128$^{\dag}$) \\
m10c1b1l4o30 & 	10 & 	$10$ & 	1 & 	4 & 	$30^{\circ}$ & 13.6 & 0.73 (0.74) & 9.13  & $\infty$ (17.7) & 6.74 (7.29) & 1.59 & 1.28 (1.59) & 119 (80.8) & 1190$^{\dag}$ (128$^{\dag}$) \\
m10c1b1l4o70 & 	10 & 	$10$ & 	1 & 	4 & 	$70^{\circ}$ & 13.6 & 0.73 (0.74) & 9.13  & $\infty$ (17.7) & 6.74 (7.29) & 1.59 & 2.55 (3.19) & 107 (87.9) & 111$^{\dag}$ (116$^{\dag}$) \\
m10c1b1l4o85 & 	10 & 	$10$ & 	1 & 	4 & 	$85^{\circ}$ & 13.6 & 0.73 (0.74) & 9.13  & $\infty$ (17.7) & 6.74 (7.29) & 1.59 & 2.55 (2.56) & 53.5 (90.8) & 112$^{\dag}$ (128$^{\dag}$) \\
m10c1b1l4o90 & 	10 & 	$10$ & 	1 & 	4 & 	$90^{\circ}$ & 13.6 & 0.73 (0.74) & 9.13  & $\infty$ (17.7) & 6.74 (7.29) & 1.59 & 2.56 (2.56) & 62.0 (47.4) & 95.7$^{\dag}$ (128$^{\dag}$) \\
m10c2b1l4o45 & 	10 & 	$10^{2}$ & 	1 & 	4 & 	$45^{\circ}$ & 13.6 & 0.73 (0.74) & 9.13  & $\infty$ (17.7) & 6.74 (7.29) & 1.59 & 4.15 (4.15) & 78.5 (30.0) & 92.7 (89.5$^{\dag}$) \\
m10c3b1l4o45 & 	10 & 	$10^{3}$ & 	1 & 	4 & 	$45^{\circ}$ & 13.6 & 0.73 (0.74) & 9.13  & $\infty$ (17.7) & 6.74 (7.29) & 1.59 & 4.89 (5.42) & 14.8 (1.58) & 24.4$^{\dag}$ (18.8$^{\dag}$) \\
m10c1b0.5l4o45 & 	10 & 	$10$        & 	0.5 & 4 & 	$45^{\circ}$ & 13.6 & 0.72 (0.73) & 6.45 & $\infty$ (1.58) & 5.42 (5.77) & 1.59 & 1.27 (1.59) & 128$^{\dag}$ (98.5) & 128$^{\dag}$ (128$^{\dag}$) \\
m10c1b10l4o45 & 	10 & $10$        & 	10 & 	4 & 	$45^{\circ}$ & 13.6 & 0.74 (0.74) & 28.9  & $\infty$ (4.21) & 9.45 (9.70) & 1.59 & 2.87 (2.87) & 13.3 (12.0) & 128$^{\dag}$ (128$^{\dag}$) \\
m1.5c1b1l4o45 & 	1.5 & $10$ & 	1 & 	4 & 	$45^{\circ}$ & 2.04 & 0.02 (0.12) & 1.37 & $\infty$ (2.66) & 1.01 (1.09) & 1.59 & 1.10 (1.49) & 90.4$^{\dag}$ (229$^{\dag}$) & 90.4$^{\dag}$ (229$^{\dag}$) \\
m3c1b1l4o45 & 	3    & 10          &        1 & 	4 & 	$45^{\circ}$ & 4.07 & 0.55 (0.59) & 2.74  & $\infty$ (5.31) & 2.02 (2.19) & 1.59 & 0.91 (1.24) & 183$^{\dag}$ (114$^{\dag}$) & 183$^{\dag}$ (114$^{\dag}$) \\
  \hline
 \end{tabular}
\end{table*}

\subsection{Parallel field}

\subsubsection{Filament morphology}
\begin{figure*}
\centering 
\includegraphics[width=160mm]{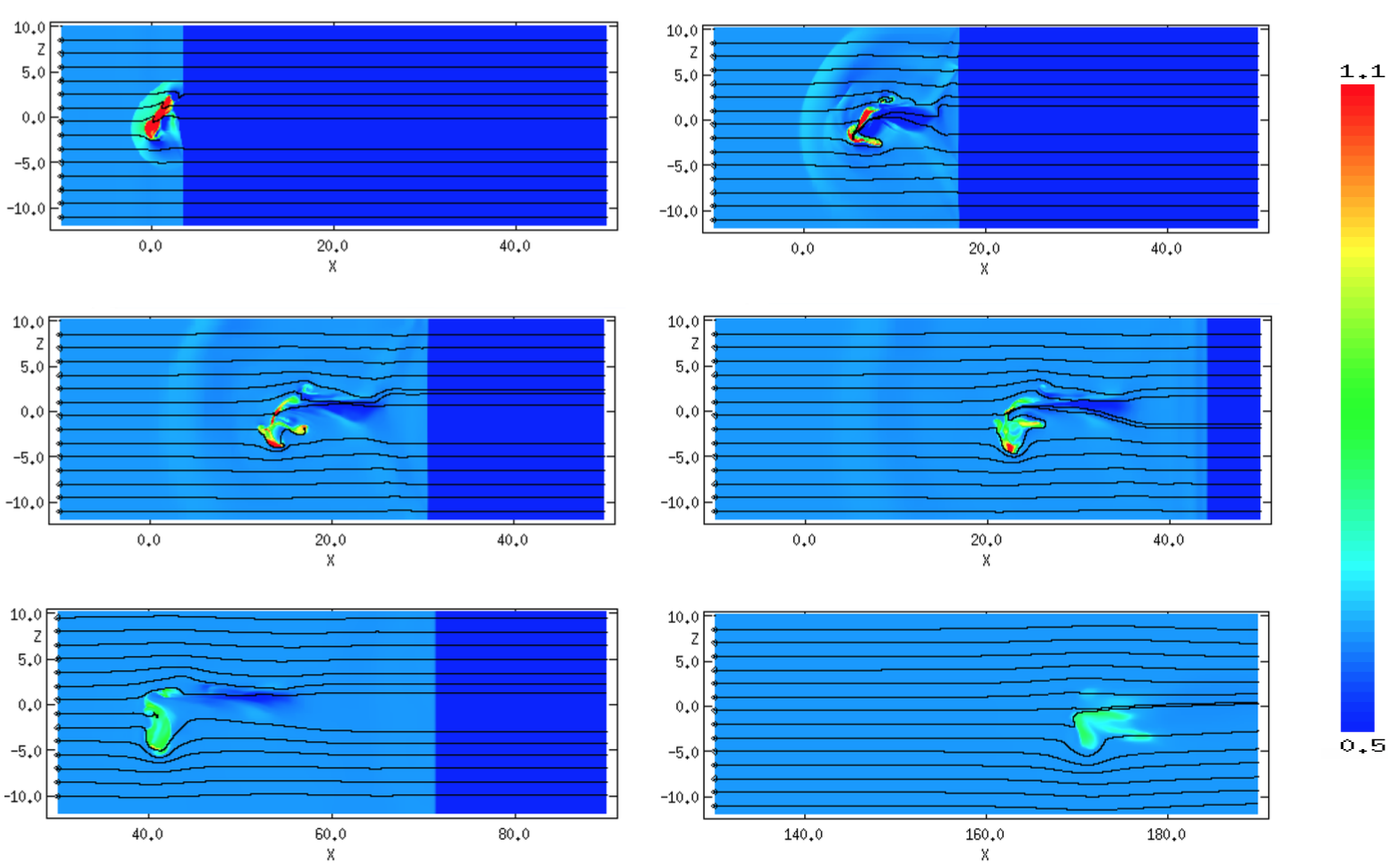}
\caption{The time evolution of the logarithmic density, scaled with respect to the ambient density, for model $m10c1b1l4o45pa$. The evolution proceeds left to right, top to bottom, with $t=0.95 \,t_{cs}$, $t=3.54\, t_{cs}$, $t=6.11\, t_{cs}$, $t=9.06 \,t_{cs}$, $t=14.2 \,t_{cs}$, and $t=52.5 \,t_{cs}$. Note the shift in the $x$ axis scale for the final two panels. The initial magnetic field is parallel to the shock normal.}
\label{Fig5}
\end{figure*}

We first review the morphology of filaments embedded in an initially parallel (i.e. at $0^{\circ}$ to the shock normal) magnetic field . Figure~\ref{Fig5} presents snapshots of the time evolution of the density distribution for simulation $m10c1b1l4o45pa$. The evolution of the filament broadly follows the stages outlined in \S4.1 of \citet{Pittard09}. Firstly, the filament is struck and compressed by the shock front, and a bow shock is formed. Then the filament expands until $t \approx 6.46\,t_{cs}$. However, unlike the hydrodynamical spherical cloud case where the cloud broadly maintains its shape, the filament is instead contorted out of shape and the expansion of the cloud is less evident. The filament is swept downstream in the ambient flow, showing very little fragmentation due to the parallel magnetic field but continually being stripped of material. The presence of parallel magnetic field lines means that, unlike the hydrodynamic case, the MHD filament exhibits little or no surface instabilities, ensuring that the filament core survives for a far longer timescale than would otherwise be possible. MHD filaments in a parallel field do not tend to form long tails of cloud material, but instead a linear ``void'' is created which comprises an area of low density and high magnetic pressure. In non-oblique filaments (henceforth known as ``axisymmetric'' filaments), and in particular filaments orientated at $\theta=90^{\circ}$, this region forms a very clear ``flux rope'', but where the filament is angled to the shock front (``oblique'' filaments) such a structure is less well defined because the contortion of the filament in the ambient flow is not symmetric.  

\begin{figure*}
\centering
\includegraphics[width=160mm]{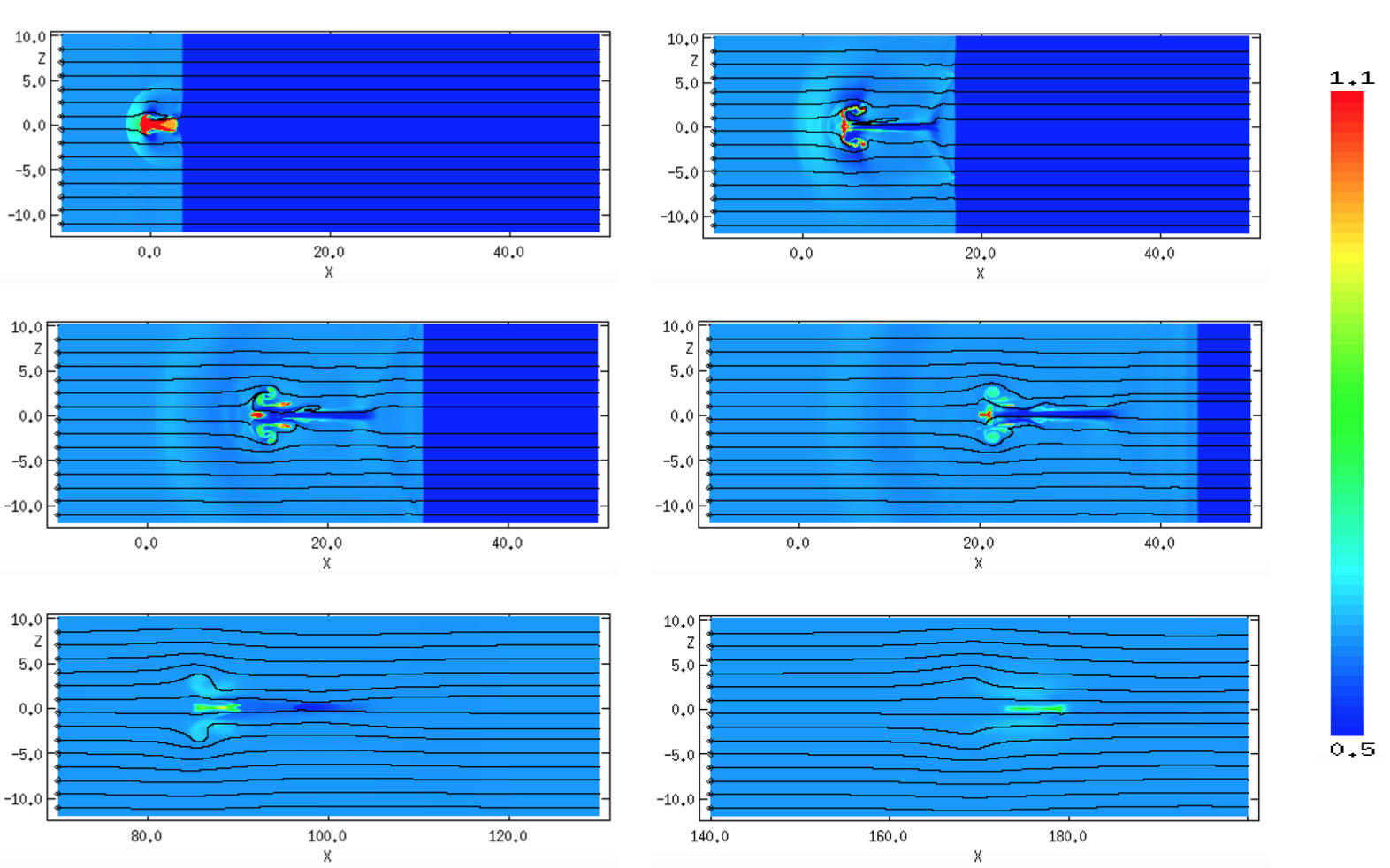} 
\caption{The time evolution of the logarithmic density, scaled with respect to the ambient density, for model $m10c1b1l4o90pa$. The evolution proceeds left to right, top to bottom, with $t=0.95 \,t_{cs}$, $t=3.54 \,t_{cs}$, $t=6.11\, t_{cs}$, $t=9.06 \,t_{cs}$, $t=27.9\, t_{cs}$, and $t=52.2 \,t_{cs}$. Note the shift in the $x$ axis scale for the bottom two panels. The initial magnetic field is parallel to the shock normal.}
\label{Fig6}
\end{figure*}

\begin{figure*}
\centering
\includegraphics[width=160mm]{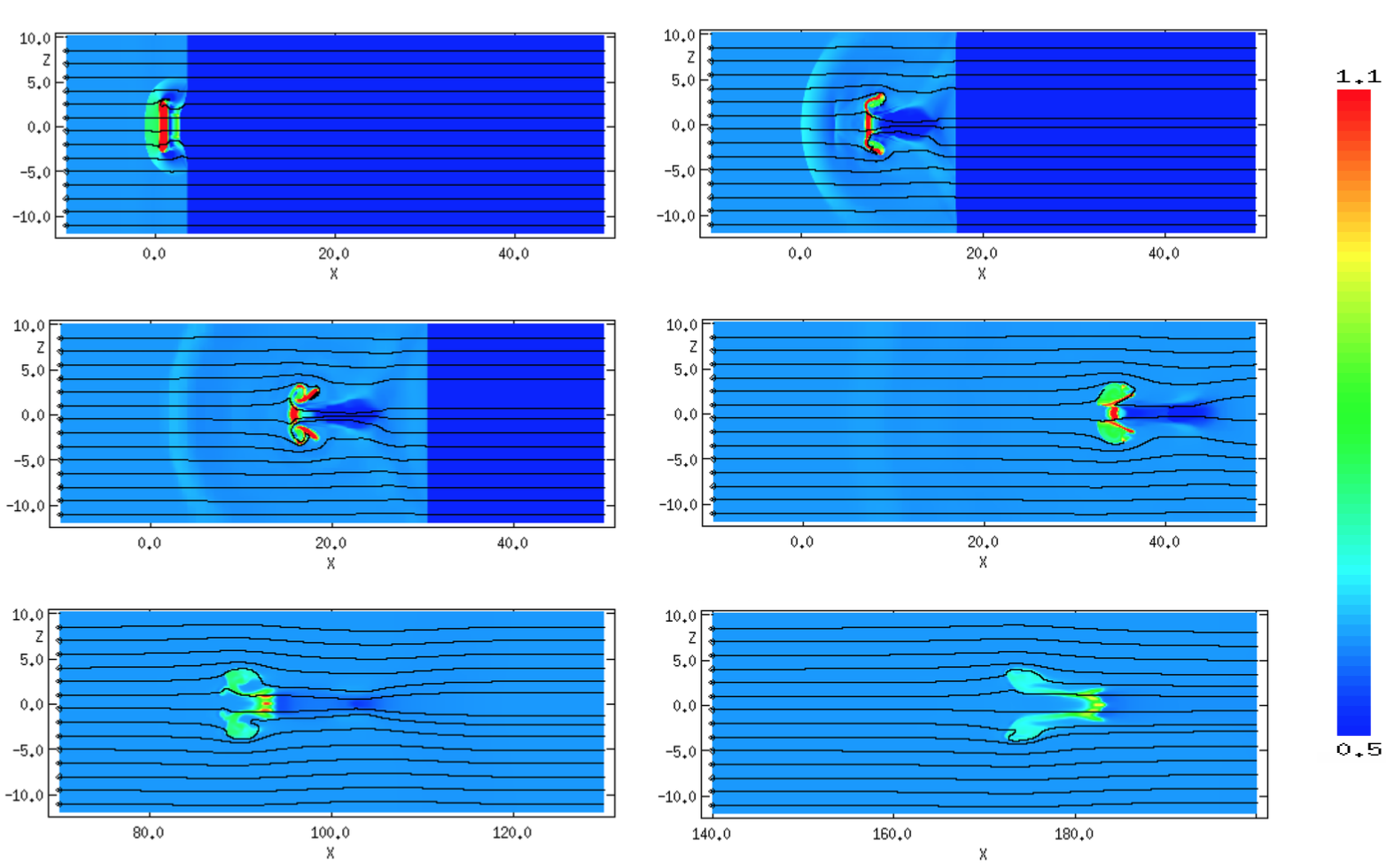} 
\caption{The time evolution of the logarithmic density, scaled with respect to the ambient density, for model $m10c1b1l4o0pa$. The evolution proceeds left to right, top to bottom, with $t=0.95 \,t_{cs}$, $t=3.54 \,t_{cs}$, $t=6.11 \,t_{cs}$, $t=11.7 \,t_{cs}$, $t=27.9 \,t_{cs}$, and $t=52.2 \,t_{cs}$. Note the shift in the $x$ axis scale for the bottom two panels. The initial magnetic field is parallel to the shock normal.}
\label{Fig7}
\end{figure*}

Figures~\ref{Fig6} and ~\ref{Fig7} show the density distribution at various times for simulations $m10c1b1l4o90pa$ and $m10c1b1l4o0pa$, respectively. The orientation of these two filaments leads to many more interesting features than those seen with the obliquely-orientated clouds. For the interaction in Fig.~\ref{Fig6} the filament is struck end-first, while in Fig.~\ref{Fig7} the filament is struck on its broadside. The initial filament structure in Fig.~\ref{Fig6}, after it has been struck by the shock, is very similar to that of the other runs, since the mechanical energy of the shock is driving the interaction rather than the magnetic energy of the filament. Compressed filament material is seen to form a column or ``flux rope'' behind the filament head but the level of compression is limited in comparison with the purely hydrodynamic case due to the magnetic field lines which surround the filament and resist compression by the converging flow. The post-shock flow is prevented from entering the flux rope by the build-up of magnetic pressure in that area. The surface of the filament, by contrast, shows shear instabilities (though damped because of the field) which serve to create ``wings'' - areas either side of the filament where the material is being ablated and bent by the surrounding flow (see \S 3.1.1. of \citet{Aluzas14}). Although the level of instability is greater than in the cases where the filament was orientated obliquely, the filament nonetheless remains relatively coherent and does not fragment. Instead it undergoes continual ablation to the surrounding flow until no substantial mass remains. The filament with $l=4$ and $\theta = 85^{\circ}$ begins to follow this evolution, and an initial well-defined flux rope is formed. However, since the filament is oriented at a slight angle to the shock front the structures forming on the axis behind the filament are quickly destabilised and the evolution proceeds as in the obliquely-orientated cases described above.

The filament in Fig.~\ref{Fig7} also forms ``wings''. However, since the shock front strikes the entire length of the filament, the wings are far more substantial and act to shield the far side of the filament from the flow. Therefore, the column of compressed material forming the flux rope in this instance is much broader than in the previous case. The filament is then dragged downstream by the post-shock flow, becoming elongated before finally being destroyed.  

\begin{figure*}
\centering
\includegraphics[width=160mm]{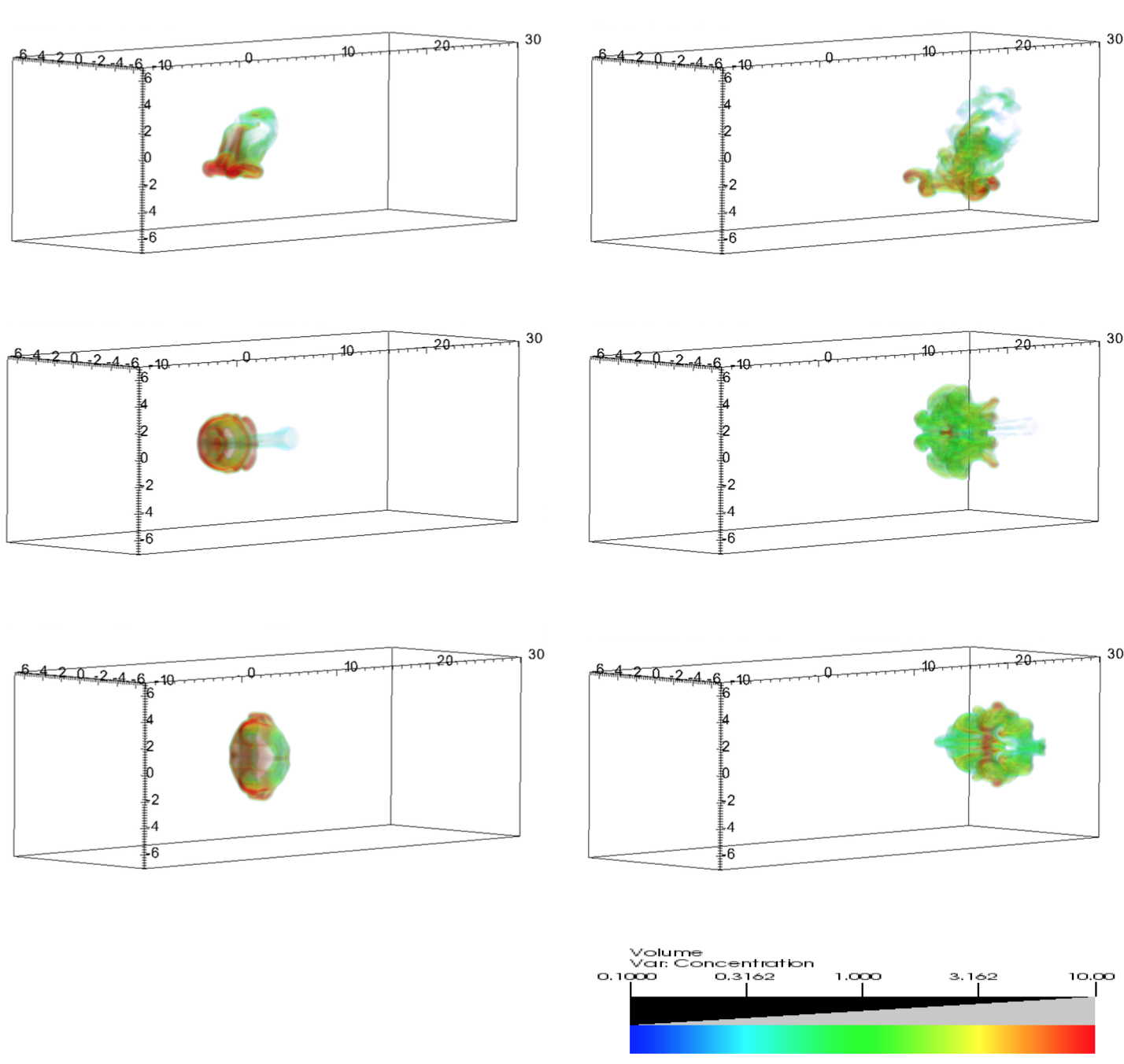} 
\caption{3D volumetric renderings of models $m10c1b1l4o45pa$ (top), $m10c1b1l4o90pa$ (middle), and $m10c1b1l4o0pa$ (bottom) at $t=3.54\, t_{cs}$ (left-hand column) and $t=9.06 \,t_{cs}$ (right-hand column). The initial magnetic field is parallel to the shock normal.}
\label{Fig8}
\end{figure*}

Figure~\ref{Fig8} shows a 3D volumetric rendering of the time evolution of the density of filament material in simulations $m10c1b1l4o45pa$, $m10c1b1l4o90pa,$ and $m10c1b1l4o0pa$, showing clearly the flux rope associated with the filament orientated at $\theta=90^{\circ}$, and also that material is forced out of the side of the filament in simulation $m10c1b1l4o45pa$. Because only the filament material is shown, other features such as the bow shock are not displayed in these plots.

\subsubsection{Effect of filament length and orientation on the evolution of the core mass}
\begin{figure*} 
\centering
     \includegraphics[width=110mm]{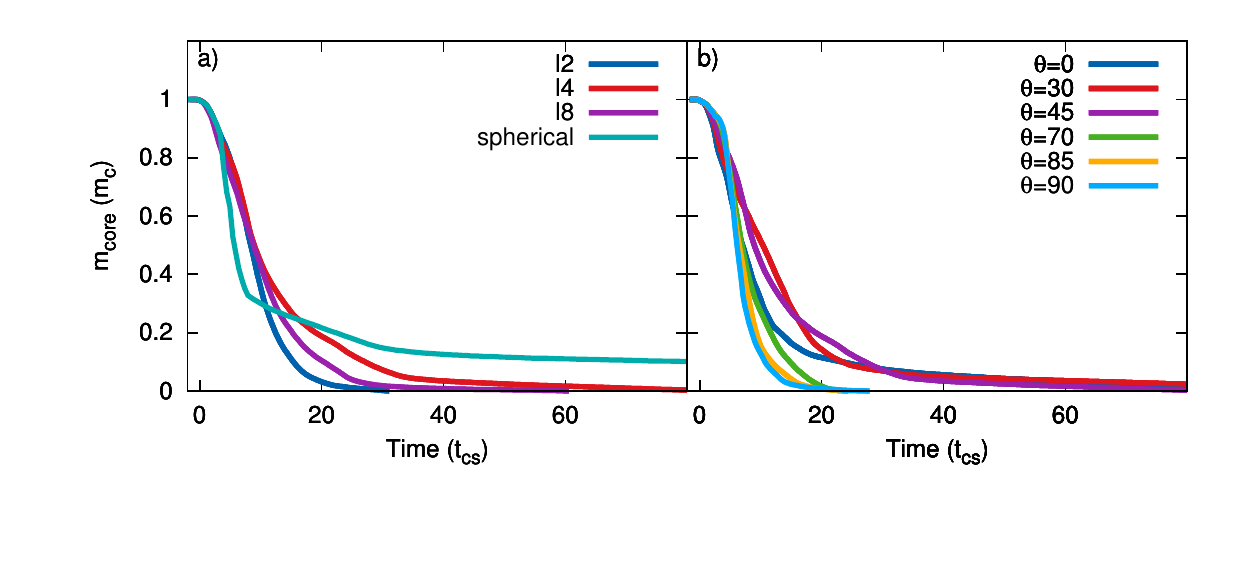} 
      \vspace{-8mm}
        \caption{Time evolution of the core mass, $m_{core}$, for (a) a filament with variable length and an orientation of $45^{\circ}$, and (b) $l=4$ with variable orientation, in an initial parallel magnetic field.}
  \label{Fig9}
  \end{figure*}

In a purely hydrodynamical case with a Mach 10 shock the filament is destroyed within a short timescale of $ t \sim 10\, t_{cs}$ (the filament survives for longer when hit by a weaker shock - see \citet{Pittard16}). This is because turbulent instabilities are able to build up at the surface of the filament and encourage the ablation of mass from it. However, when magnetic fields are present instabilities are damped, and filaments survive over far longer timescales. Figure~\ref{Fig9} shows the evolution of the filament core mass over time for filaments with different lengths and orientations. It can be seen that the timescale for destruction in these cases is far greater than in the hydrodynamical scenario presented in \citet{Pittard16}. 

We see that in terms of the core mass, the filament with $l=4$ and an orientation of $\theta=90^{\circ}$, and that with a length of $l=2$ and an orientation of $\theta = 45^{\circ}$, are destroyed at $t \approx 31 \,t_{cs}$ and $t \approx 28 \,t_{cs}$, respectively. However, the filament with $l=4$ and $\theta=0^{\circ}$, and that with $l=8$ and $\theta=45^{\circ}$, are not destroyed until $t \approx 104 \,t_{cs}$ (not visible in the figure) and $t \approx 61 \,t_{cs}$, respectively.

The orientation of the filament to the shock normal plays an important role in the core mass evolution and the lifetime of the filament (Fig.~\ref{Fig9}(b)). Whilst all filament orientations show a similar initial decrease in mass until $t \approx 5 \,t_{cs}$ the filament orientated at $\theta=90^{\circ}$ (i.e. end on), although initially the slowest to lose mass, thereafter shows the most rapid drop in mass until its destruction (cf. Fig. 28(i) in \citet{Pittard16}). It is noticeable that those filaments with orientations of $0^{\circ} <\theta \lesssim 45^{\circ}$ are much slower overall to lose the majority of their core mass (with the mass loss rate decreasing significantly once less than 5$\%$ of the initial filament mass remains), whilst those with orientations of $\theta > 45^{\circ}$ are destroyed much more quickly.

Unless the filament is very short (in which case it begins to approximate a spherical cloud), the length of the filament has less of an influence on the mass loss than the orientation. From Fig.~\ref{Fig9}(a) it can be seen that all three filaments initially show a similar decrease in their core mass. However, the filament with length $l=2$ subsequently loses mass at a much faster rate than the other two lengths. This differs from the hydrodynamic case in \citet{Pittard16}, where the filament of length $l=8$ loses mass faster than the other filaments. Interestingly, the spherical cloud, whilst incurring a faster mass-loss rate than the filament with $l=2$, begins to level off at $\sim 7 \,t_{cs}$ and retains approximately one tenth of its initial mass by the end of the simulation. In this case, although the ``length'' of the filament is short, it is axisymmetric to the shock front and behaves in a similar manner to the filament of length $l=4$ and $\theta = 0^{\circ}$.

\subsubsection{Effect of filament length and orientation on the mean velocity and the velocity dispersion}
\begin{figure*} 
\centering
     \includegraphics[width=110mm]{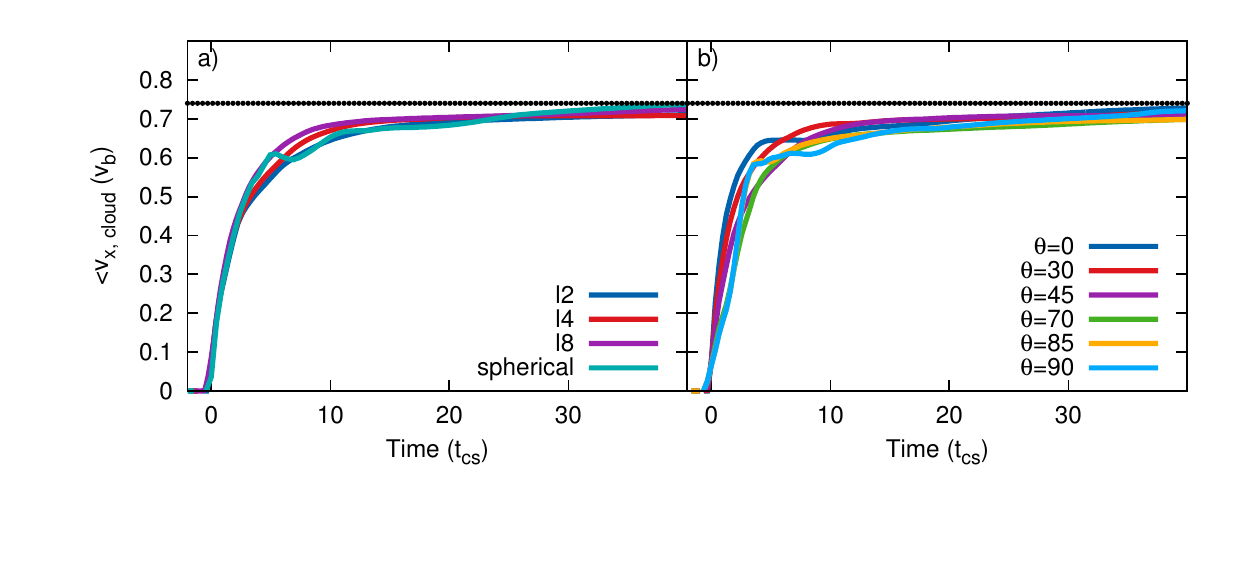}
      \vspace{-8mm}     
  \caption{Time evolution of the filament mean velocity, $\langle v_{x} \rangle$, for (a) a filament with variable length and an orientation of $45^{\circ}$, and (b) $l=4$ with variable orientation, in an initial parallel magnetic field. The dotted black line indicates the velocity of the post-shock flow.}
  \label{Fig10}
  \end{figure*}

There are two stages to the acceleration of the filament through the ambient flow. The filament is first accelerated to the velocity of the transmitted shock, $ \propto v_{b}/\sqrt{\chi}$, as the shock is driven through it, and then further accelerated by the flow of post-shock gas until it reaches the velocity of the flow, e.g. $0.743\, v_{b}$ for $M=10$, $\beta_{0}=1$ and a parallel field. Figure~\ref{Fig10} shows the time evolution of the mean cloud velocity in the $x$ direction, $\langle v_{x}\rangle$. It can be seen that filaments with orientations of $\theta\lesssim 45^{\circ}$ are initially accelerated faster than those with orientations of $\theta > 45^{\circ}$, and this is likely to be because there is a greater surface area presented to the shock front with these orientations, i.e. the filament is `broadside' to the shock front. It is interesting to note that the filament hit end on is initially accelerated the least rapidly, but that the rate of velocity gain does not level off as much as in some of the other models until the filament experiences a drastic reduction in acceleration at $v \simeq 0.6\, v_{b}$. It is clear that the filaments with $l=4$ and $\theta = 0^{\circ}\, , 90^{\circ}$ display more overtly the two-stepped nature of the acceleration. At $t > 40\, t_{cs}$, the filaments with $\theta=30^{\circ}$ and $\theta=90^{\circ}$ slightly overshoot and then decelerate to the velocity of the post-shock flow (not visible in Fig.~\ref{Fig10}), possibly due to the release of some built-up tension in the field lines. 

In comparison with the filament orientation, the length appears to have no significant effect on the mean velocity, with all filaments being accelerated at approximately the same rate. This is in contrast to the spherical cloud which displays a profile similar to the end on filament in Fig.~\ref{Fig10}(b). 

\begin{figure*} 
\centering    
     \includegraphics[width=110mm]{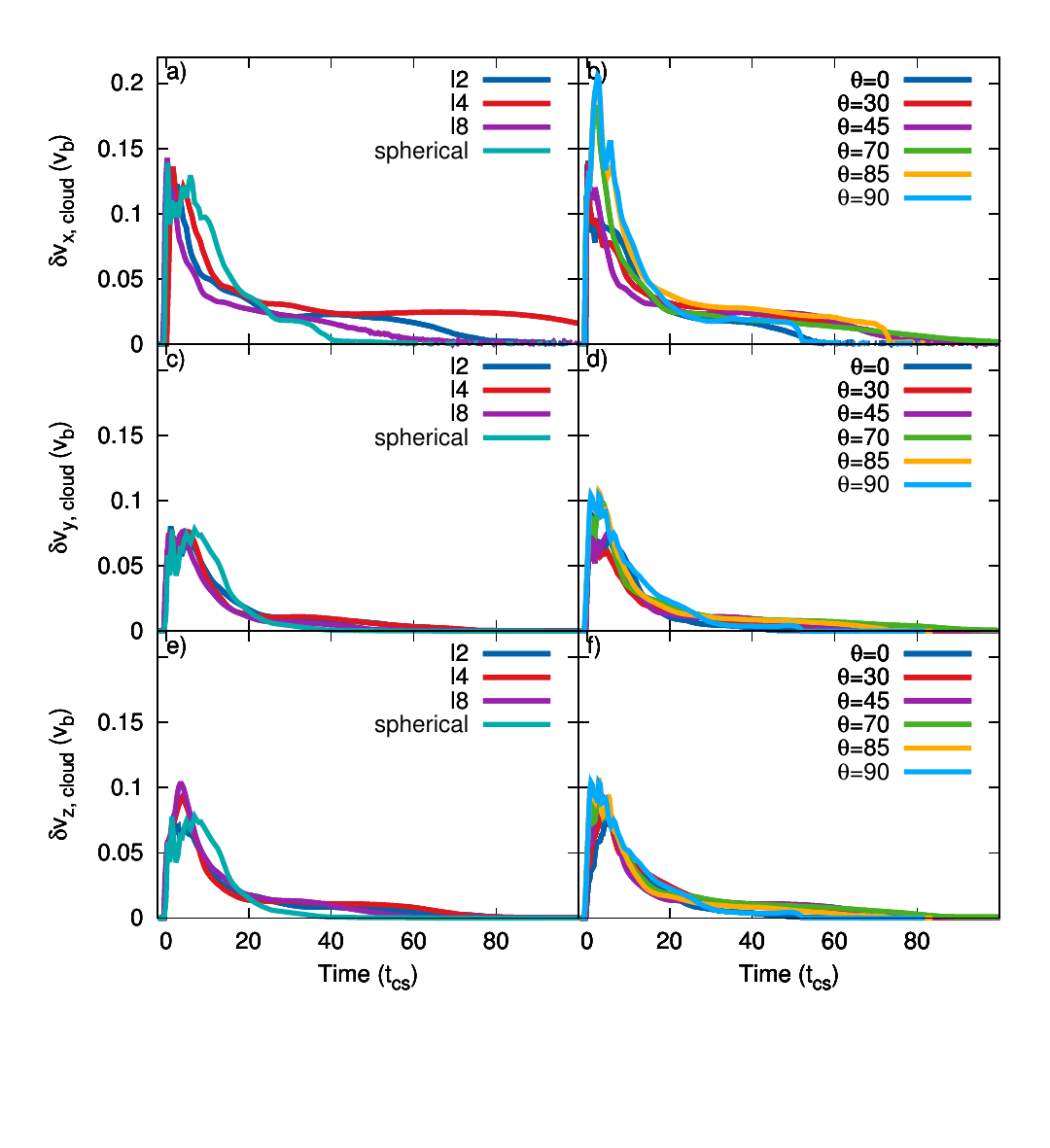}      
      \vspace{-12mm}     
  \caption{Time evolution of the filament velocity dispersion in the $x$, $y$, and $z$ directions, $\delta v_{x,y,z}$, for a filament with variable length and an orientation of $45^{\circ}$ (left-hand column), and $l=4$ with variable orientation (right-hand column), struck by a parallel shock.}
  \label{Fig11}
   \end{figure*}

The interaction of shocks with filaments creates substantial velocity dispersions and reveals the presence of instabilities. In the $x$ direction, the filaments with orientations $\theta \gtrsim 70^{\circ}$ have the highest peaks (Fig.~\ref{Fig11}(d)), with the $\theta=0^{\circ}$ and $\theta=30^{\circ}$ filaments showing the least dispersion in the $x$ direction. This is in agreement with \citet{Pittard16} where, for end on or nearly end on filaments, their $\delta v_{x}/v_{b}$ also reaches $\simeq 0.2$ (cf. their Fig. 28(e)). Figures~\ref{Fig11}(e,f), by contrast, indicate much less overall dispersion in the $y$ and $z$ directions. This is because, in the $x$ direction, the initial peak occurs as the transmitted shock travels through the filament. Thus, there is a large dispersion between the shocked and unshocked filament material at that time. A similar effect is produced in the $y$ and $z$ directions, although slightly later, when the filament is undergoing compression.

A comparison of the top and bottom panels of Fig.~\ref{Fig11} reveals that the velocity dispersion is more sensitive to filament orientation than length in the $x$ direction, and more sensitive to length rather than orientation in the $z$ direction.

\subsubsection{Effect of filament length and orientation on the mean density} 
\begin{figure*} 
\centering
    \includegraphics[width=110mm]{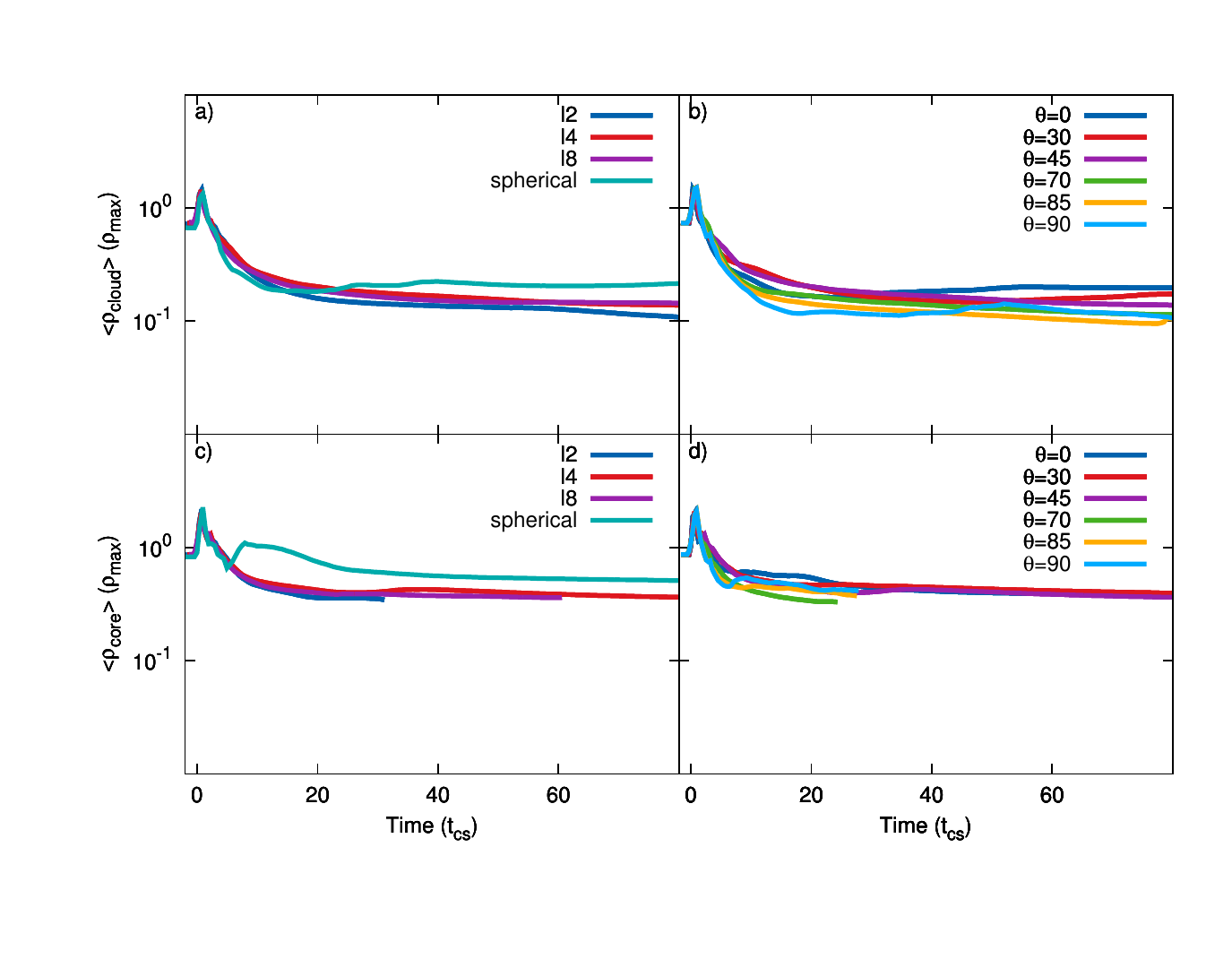}   
     \vspace{-10mm}    
     \caption{Time evolution of the mean density of the filament, $\langle \rho_{cloud}\rangle$ (top), and filament core, $\langle \rho_{core}\rangle$ (bottom), normalised to the initial maximum filament density, for filaments with (left-hand column) variable length and an orientation of $45^{\circ}$, and (right-hand column) $l=4$ and a variable orientation, in a parallel magnetic field.}
  \label{Fig12}
  \end{figure*}

Figure~\ref{Fig12} shows the time evolution of the mean density of the filament, $\langle \rho_{cloud}\rangle$, and filament core, $\langle \rho_{core}\rangle$. The peak mean densities, after the shock has hit and compressed the filament, for various lengths and orientations of the filament are similar. However, the mean densities of filaments with $l=4$ and $\theta = 90^{\circ}$, or $l=2$ and $\theta=45^{\circ}$, decline more rapidly, with a lower final value of $\langle \rho \rangle /\rho_{max}$ being reached in these cases (though in Fig.~\ref{Fig12}(d) the filament with $\theta = 70^{\circ}$ reaches a lower mean density level by the end of the simulation). It is noticeable in Fig.~\ref{Fig12}(b) that for filaments with orientations of $\theta = 0^{\circ}$, $\theta = 30^{\circ}$, or $\theta = 90^{\circ}$ there is a subsequent increase in the mean density after reaching their lowest value, and this is mirrored in the spherical cloud mean density in Fig.~\ref{Fig12}(a). The initial peak of the spherical cloud mean density in Fig.~\ref{Fig12}(c) is slightly higher than for the filaments, and a second, broader, peak is present also. The difference in the height of the peak mean densities may be due to the fact that the shocks driven into the filaments do not converge as well as those driven into the spherical cloud.

\subsubsection{$\chi$ dependence of the filament evolution}
\begin{figure*}
\centering
\includegraphics[width=160mm]{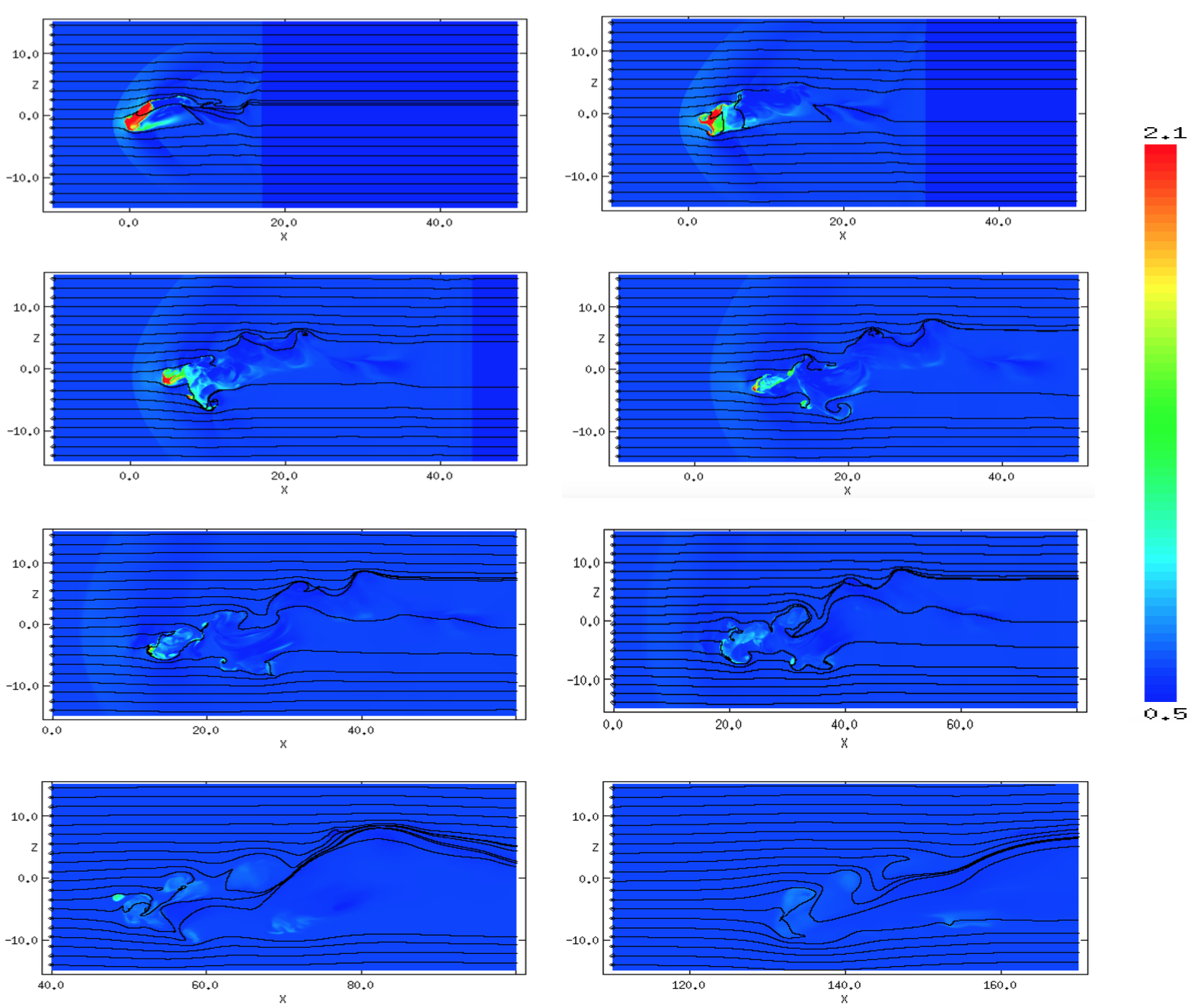}
\caption{The time evolution of the logarithmic density, scaled with respect to the ambient density, for model $m10c2b1l4o45pa$. The evolution proceeds left to right, top to bottom, with $t=1.09\, t_{cs}$, $t=1.97\, t_{cs}$, $t=2.86 \,t_{cs}$, $t=3.65\, t_{cs}$, $t=4.57 \,t_{cs}$, $t=5.36\, t_{cs}$, $t=8.85\, t_{cs}$, and $t=16.5 \,t_{cs}$. Note the shift in the $x$ axis scale for the final four panels, and the change in the logarithmic density scale compared to previous cases. The initial magnetic field is parallel to the shock normal.}
\label{Fig13}
\end{figure*}

\begin{figure*}
\centering
\includegraphics[width=160mm]{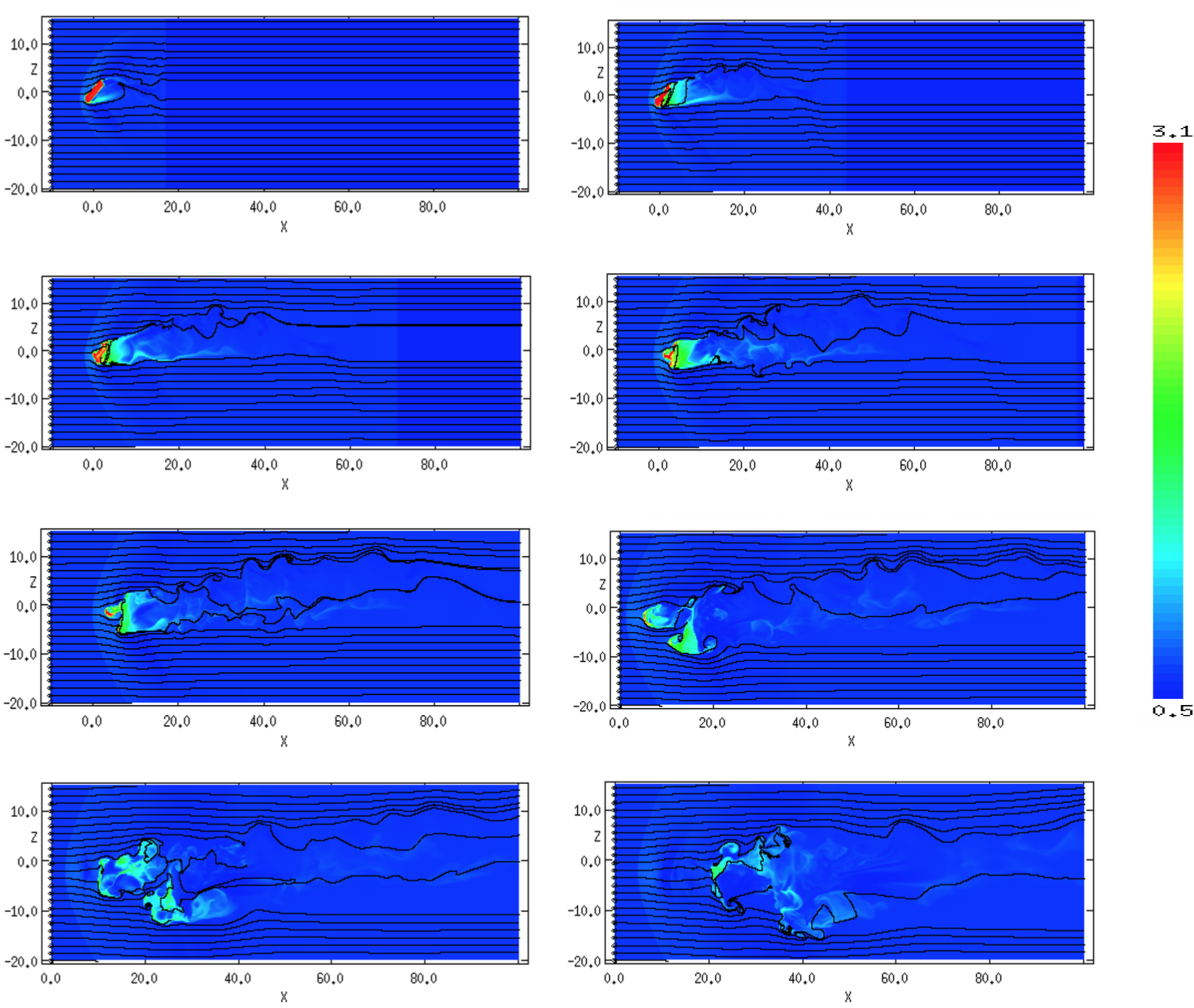}
\caption{The time evolution of the logarithmic density, scaled with respect to the ambient density, for model $m10c3b1l4o45pa$ using a resolution of $R_{16}$. The evolution proceeds left to right, top to bottom, with $t=0.33 \,t_{cs}$, $t=0.88\, t_{cs}$, $t=1.43 \,t_{cs}$, $t=1.95 \,t_{cs}$, $t=2.50 \,t_{cs}$, $t=3.03 \,t_{cs}$, $t=3.57\, t_{cs}$, and $t=4.11 \,t_{cs}$. Note the shift in the $x$ axis scale for the final three panels, and the change in the logarithmic density scale compared to previous cases. The initial magnetic field is parallel to the shock normal.}
\label{Fig14}
\end{figure*}

Varying the cloud density contrast radically alters the evolution of the filament. This is clearly seen in Figs.~\ref{Fig13} and \ref{Fig14}, where the filament downstream of the bow shock evolves in a highly turbulent manner, not dissimilar to previous hydrodynamical shock-cloud simulations (e.g. \citealt{Pittard16}). The tail of turbulent cloud material follows the pattern of the field lines at that point which are highly contorted and tangled. Since instabilities are able to form on the surface of the filament to a much greater degree than the other simulations run with a parallel magnetic field, the core mass of the filaments in these cases are destroyed in very short timescales of $t=17.2 \,t_{cs}$ and $t=8.4\, t_{cs}$ for $\chi=100$ and $\chi=1000$, respectively, though they are first drawn out into long strands, or tails, of cloud material before being broken up into clumps and eventually mixed with the post-shock flow. Indeed, the development of turbulent instabilities increases with increasing $\chi$. This is in complete contrast to the $\chi=10$ case shown in Fig.~\ref{Fig5}, where the evolving filament in that case forms a compact and smooth structure and does not display pronounced turbulent instabilities. The decreased destruction time of the filament (in units of $t_{cs}$) with increasing $\chi$ follows the trend in \citet{Pittard16}, where $t_{life}$ reduces as $\chi$ increases when $M=10$.\footnote{It should be noted that, owing to computational difficulties with running the $\chi=1000$ simulation at such a high resolution, we used a slightly lower resolution of $R_{16}$ for this case. Thus, it should be borne in mind that this filament may be destroyed more rapidly than would be the case with a resolution of $R_{32}$.} However, this is in direct contrast with \citet{Pittard15}, which revealed that spherical clouds do not show a clear trend with $\chi$ for $t_{life}$ at $M=10$. This shows that $t_{mix}$ and $t_{life}$ do not exhibit monotonic behaviour with varying $\chi$ when $M=10$.

\begin{figure*} 
\centering     
     \includegraphics[width=160mm]{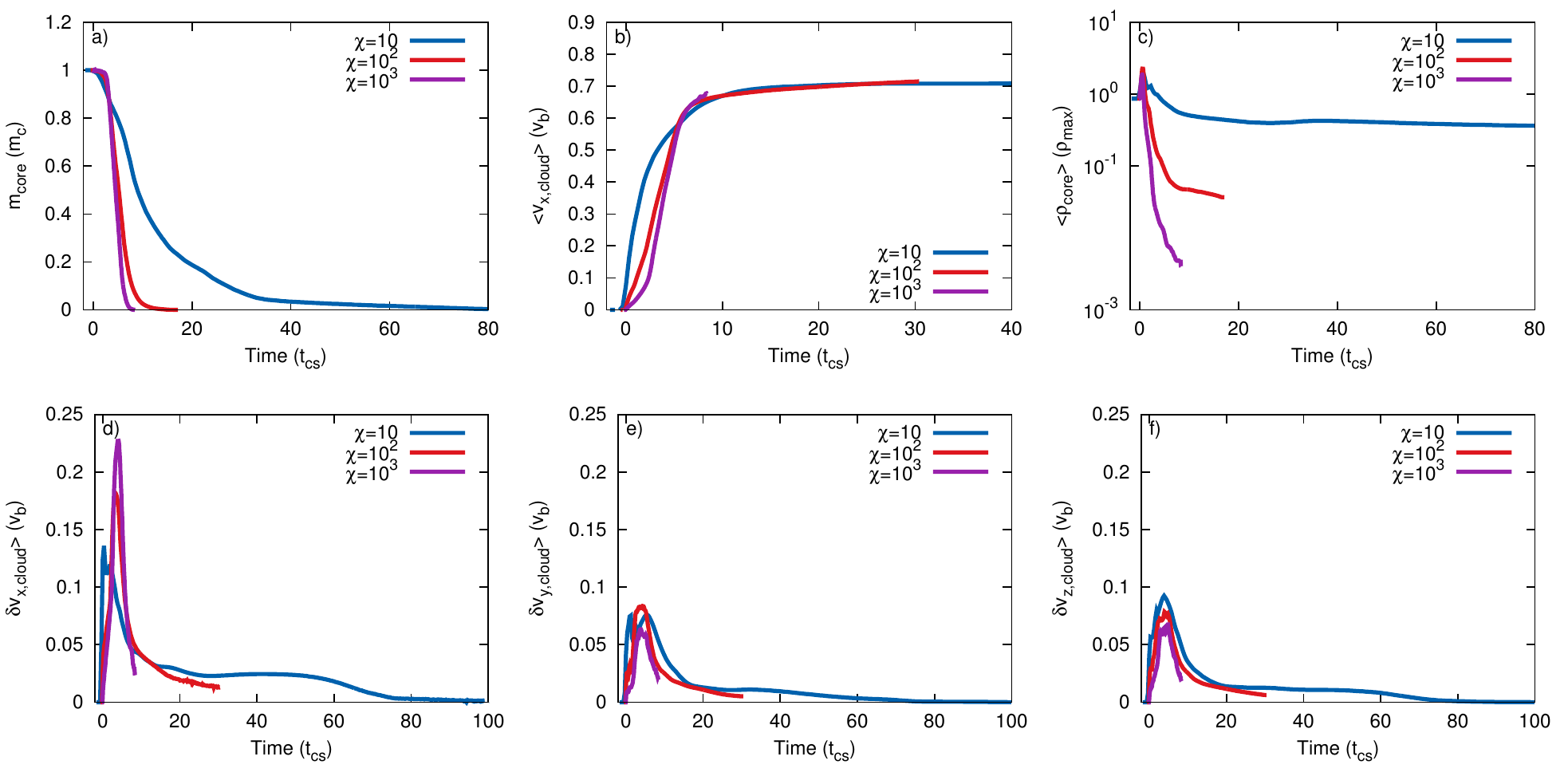} 
      \vspace{-3mm}            
   \caption{$\chi$ dependence of the evolution for filaments with $l=4$ and $\theta=45^{\circ}$. The initial magnetic field is parallel to the shock normal, $M=10$, and $\beta_{0}=1$. Note that model $m10c3b1l4o45pa$ was run at a resolution of $R_{16}$.}
  \label{Fig15}
  \end{figure*}

The demise of the $\chi=100$ and $\chi=1000$ filaments is seen in the mean density plot (Fig.~\ref{Fig15}(c)), which shows that although these two filaments initially have a much higher mean density in comparison with $\rho_{amb}$, their mean density thereafter quickly reduces, while the filament with $\chi=10$ maintains a much higher mean density after its initial compression by the shock front. In addition, Fig.~\ref{Fig15}(b) shows that the filament with $\chi=1000$ is destroyed before it has reached the velocity of the post-shock flow. The presence of instabilities is, however, present in the velocity dispersion plots (Fig.~\ref{Fig15}(d-f)) with both the higher $\chi$ filaments producing a higher dispersion peak in the $x$ direction than the $\chi=10$ filament. In addition, the peak dispersion for higher values of $\chi$ is shifted from the $\chi=10$ case in the $x$ and $y$ directions, indicating that turbulent instabilities take longer to form and are more important for the dispersal of the filament than its initial compression.

\subsubsection{Mach dependence of the filament evolution}
The Mach number of the shock can affect the growth rate of KH and RT instabilities, and can also affect the speed at which material is stripped from the filament and the time taken for the filament to become fully mixed with the surrounding flow. The post-shock conditions are dependent on the Mach number. In the purely hydrodynamic case, low Mach numbers (i.e. $M \leq 2.76$ \citep{Pittard10}) lead to a subsonic post-shock flow with respect to a stationary obstacle. Conversely, high Mach numbers provide a supersonic post-shock flow.

\begin{figure*} 
\centering    
      \includegraphics[width=160mm]{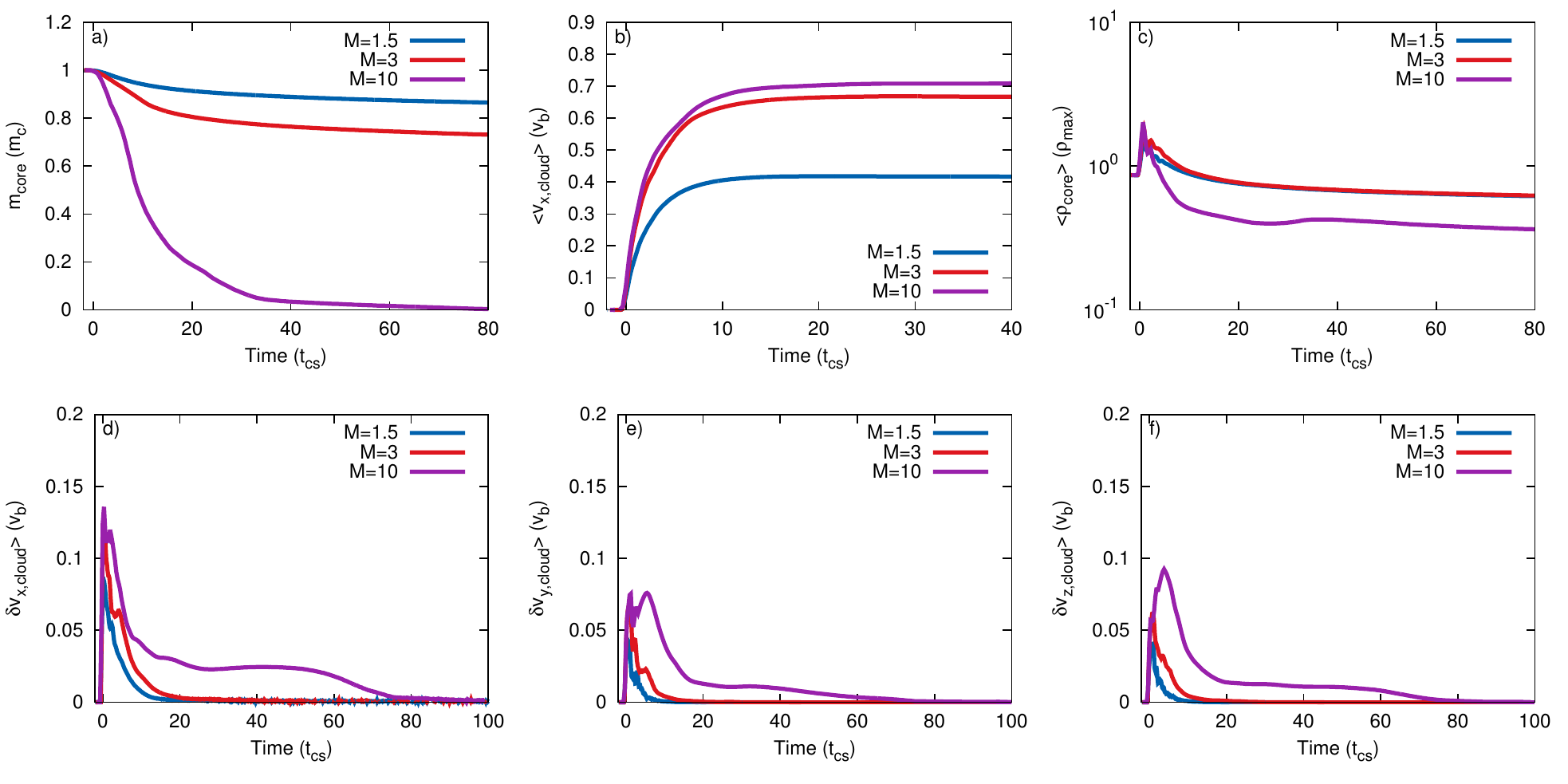}    
       \vspace{-3mm} 
   \caption{Mach number dependence of the evolution for filaments with $l=4$ and $\theta=45^{\circ}$. The initial magnetic field is parallel to the shock normal, t$\chi=10$, and $\beta_{0}=1$.}
  \label{Fig16}
  \end{figure*}

We investigated three values for the shock Mach number: $M=1.5, \, 3,$ and $10$. Figure~\ref{Fig16} shows the Mach number dependence of the evolution. It is evident from Fig.~\ref{Fig16}(a) that the core mass declines much more rapidly for $M=10$ than for $M=1.5$, indicating that core material exists for far longer with a low Mach number because of the milder interaction of the shock with the filament. The morphology of the filaments with $M = 1.5$ and $M = 3$ does not radically alter over time, with the filament merely being bent into a horseshoe shape and experiencing very little compression or ablation of cloud material until the end of the simulation at $t = 126.9\,t_{cs}$ (for $M = 1.5$) and $t = 212.7\,t_{cs}$ (for $M = 3$). It is clear, therefore, that the interaction of the shock with the cloud is much more gentle in these cases than for $M = 10$. Figure~\ref{Fig16}(b) illustrates the differing values for the velocity of the post-shock flow according to Mach number, with very low Mach numbers resulting in a much slower acceleration to the (smaller) normalised velocity of the post-shock flow. The more gentle interaction at the lower Mach numbers results in the acceleration of the filament up to the post-shock flow velocity while it is still intact and coherent in structure. In addition, a bow wave is formed ahead of the filament for shocks with $M=1.5$, rather than the bow shock visible for $M=10$ in Fig.~\ref{Fig5}.

The velocity dispersion plots (Figs.~\ref{Fig16}(d,e,f)) show that $M=1.5$ and $M=3$ have a faster decay of velocity dispersions in all directions, in comparison to $M=10$. Indeed, the difference in the height of the initial peak indicates that the filament has been struck by a shock of differing strength, since for the milder shocks there is far less of a contrast between the velocity of the shocked and unshocked portions of the filament when the shock front first hits the cloud.

\subsubsection{$\beta_{0}$ dependence of the filament evolution}

\begin{figure*} 
\centering
      \includegraphics[width=160mm]{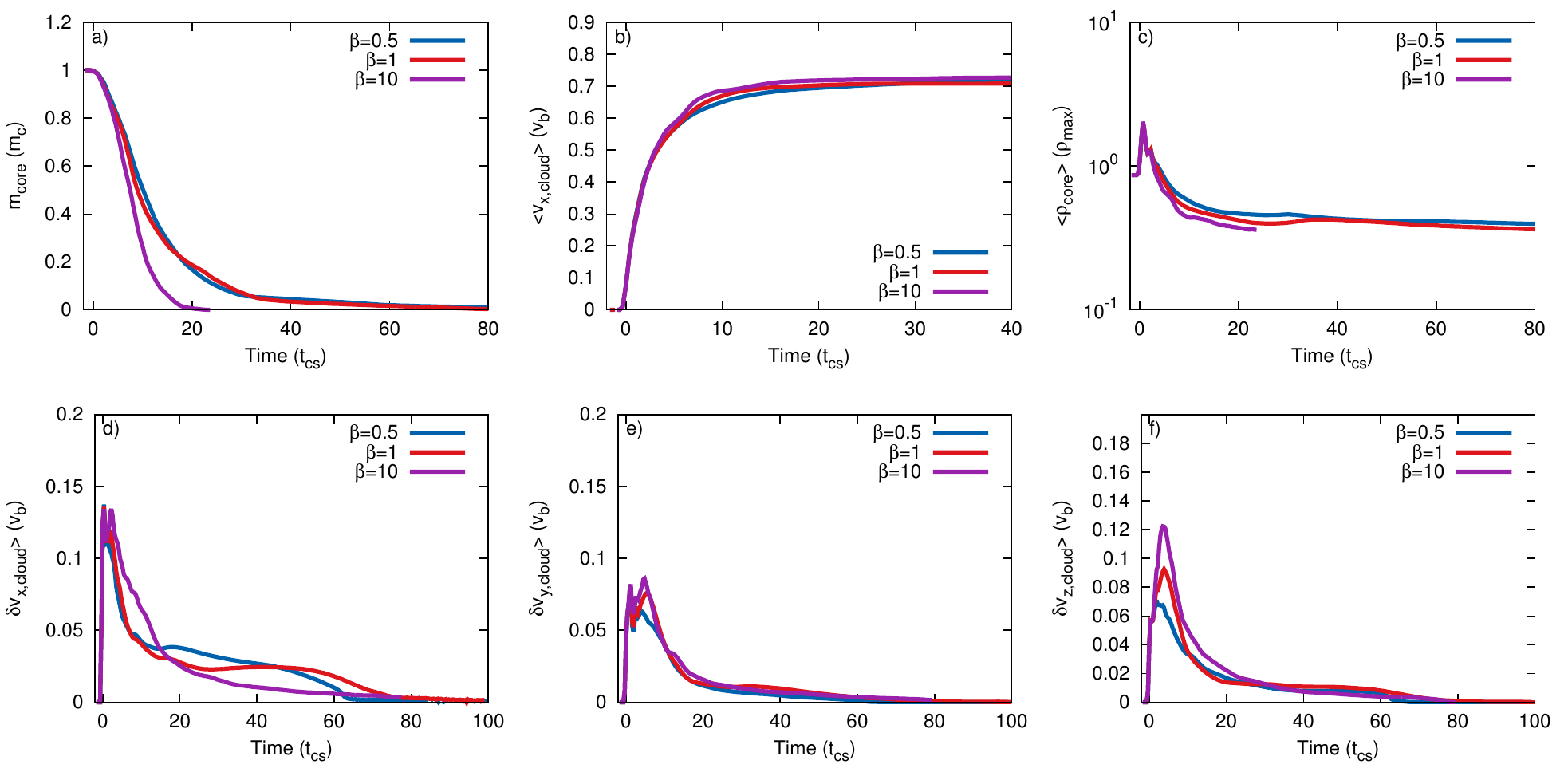}  
          \vspace{-3mm}
      \caption{Plasma beta dependence of the evolution for filaments with $l=4$ and $\theta=45^{\circ}$. The initial magnetic field is parallel to the shock normal, $M=10$, and $\chi=10$.}
  \label{Fig17}
  \end{figure*}  
  
Figure~\ref{Fig17} shows the effect of varying the plasma beta on the evolution of the filament. Figure~\ref{Fig17}(a) shows that the core mass of the model with $\beta_{0}=10$ (i.e. a weak magnetic field) is destroyed far quicker than for filaments with smaller values of $\beta_{0}$ (i.e. strong fields), since a weaker magnetic field is less able to damp the emergence of instabilities on the surface of the filament. The evolution with $\beta_{0}=0.5$ and $\beta_{0}=1$ is, however, broadly the same, and the filament morphologies for these two cases are very similar, whereas that for $\beta_{0} = 10$ shows far more fragmentation and dispersal of the cloud material. Figures~\ref{Fig17}(b-f) show that there is not a great amount of divergence between the three simulations with respect to the filament velocity, mean density, or velocity dispersion in the $y$ direction. However, the velocity dispersion in the $x$ direction does show some divergence at later times, once the structure and dynamics of the shocked filament become sensitive to the magnetic field strength, and the peak of the dispersion in the $z$ direction increases with decreasing field strength.

\subsection{Perpendicular field}

\subsubsection{Filament morphology}
\begin{figure*}
\centering
\includegraphics[width=160mm]{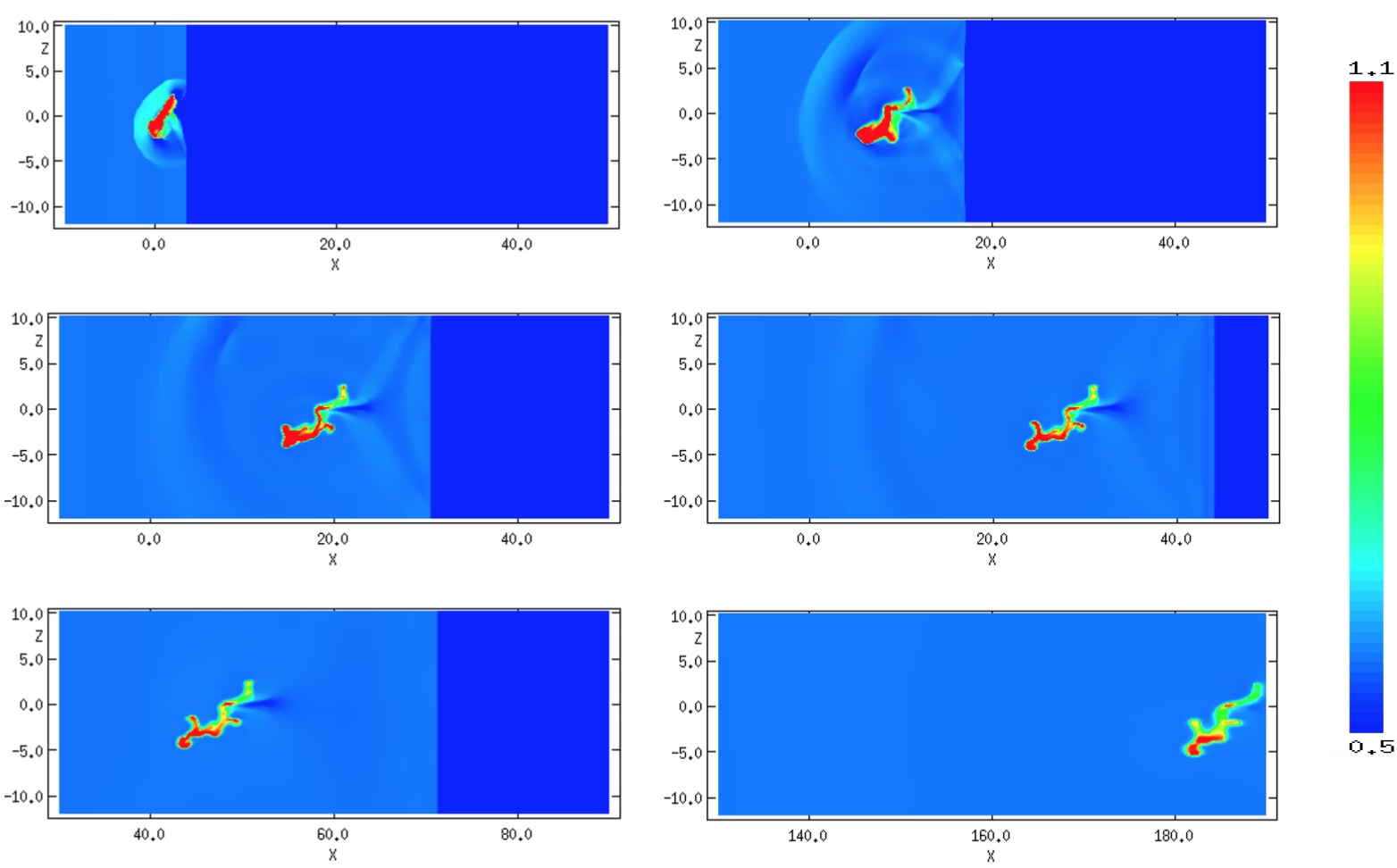}
\caption{The time evolution of the logarithmic density, scaled with respect to the ambient density, for model $m10c1b1l4o45pe$ (cf. the parallel field case in Fig.~\ref{Fig5}). The evolution proceeds left to right, top to bottom, with $t=0.95 \,t_{cs}$, $t=3.44\,t_{cs}$, $t=6.36 \,t_{cs}$, $t=8.95 \,t_{cs}$, $t=14.5 \,t_{cs}$, and $t=52.1 \,t_{cs}$. Note the shift in the $x$ axis scale for the final two panels. The initial magnetic field is perpendicular to the shock normal.}
\label{Fig18}
\end{figure*}

\begin{figure*}
\centering
\includegraphics[width=170mm]{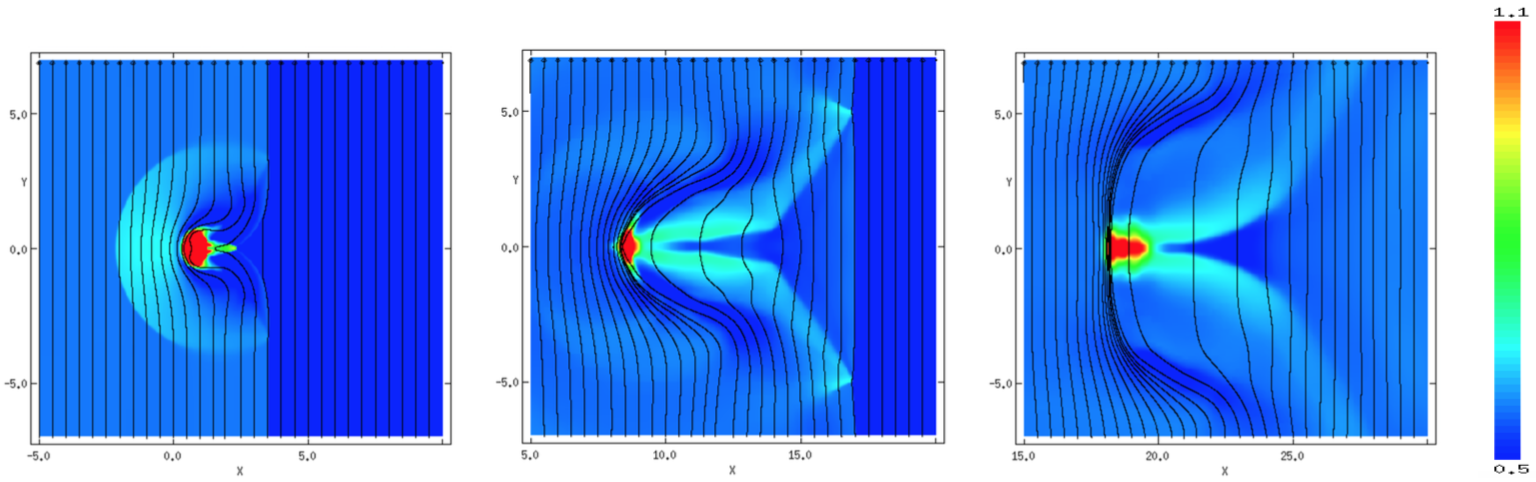}
\caption{As per Fig.18 but showing the $xy$ plane and magnetic fieldlines. The evolution proceeds left to right with $t=0.95 \,t_{cs}$, $t=3.44\,t_{cs}$, and $t=6.36 \,t_{cs}$. Note the shift in the $x$ axis scale for the final panel.}
\label{Fig19}
\end{figure*}

The time evolution of the density distribution for simulation $m10c1b1l4o45pe$ is presented in Fig.~\ref{Fig18}, with the magnetic fieldlines visible in the $xy$ plane in Fig.~\ref{Fig19}. The presence of the perpendicular (i.e. $90^{\circ}$ to the shock normal) magnetic field lines helps to protect the filament from the effects of the shock front and subsequent post-shock flow. Here, the field lines bend around the filament, allowing the flow to move along them and shielding the filament from rapid mass loss via ablation. In the filaments set at an initial angle to the shock front (the ``oblique'' filaments), the filaments are drawn out into long tendrils and are swept downstream in the flow. These filaments lose very little mass until near the end of the simulation. A small linear void is formed downstream of the filament, but this is much smaller than the void created in the parallel field scenario. As with the parallel field, oblique filaments do not form any significant linear structure along their axis because they are asymmetrical to the shock front. Compared to the parallel field case in Fig.~\ref{Fig5}, we observe that the perpendicular field ensures that the filament maintains a higher density, and produces a more rapid initial acceleration of the filament downstream. The latter is caused by the release of the tension that builds up in the field lines as they re-straighten.

\begin{figure*}
\centering
\includegraphics[width=160mm]{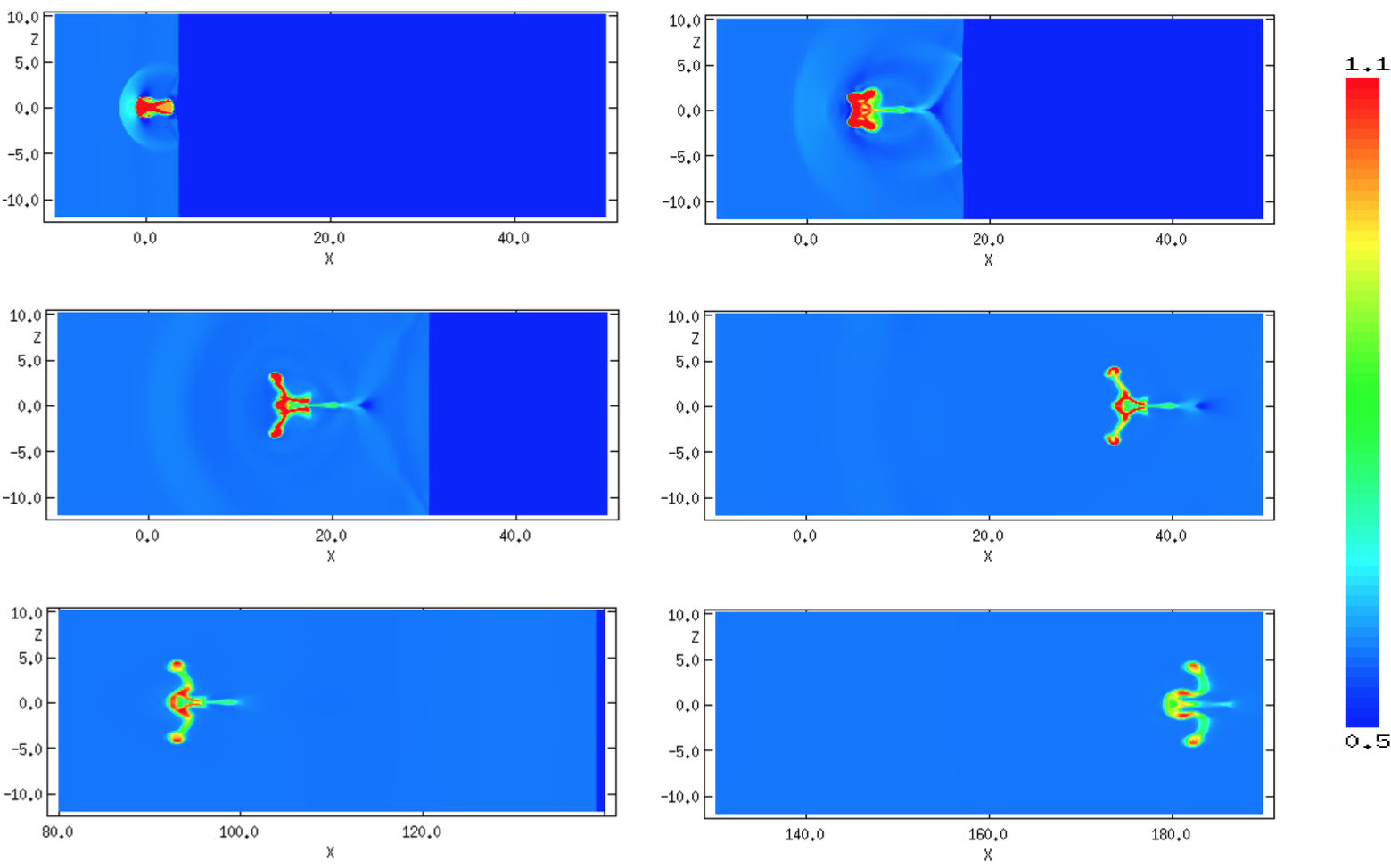}
\caption{The time evolution of the logarithmic density, scaled with respect to the ambient density, for model $m10c1b1l4o90pe$ (cf. the parallel field case in Fig.~\ref{Fig6}). The evolution proceeds left to right, top to bottom, with $t=0.95 \,t_{cs}$, $t=3.55 \,t_{cs}$, $t=6.10 \,t_{cs}$, $t=11.7\, t_{cs}$, $t=27.9\, t_{cs}$, and $t=52.2\, t_{cs}$. Note the shift in the $x$ axis scale for the bottom two panels. The initial magnetic field is perpendicular to the shock normal.}
\label{Fig20}
\end{figure*}

\begin{figure*}
\centering
\includegraphics[width=170mm]{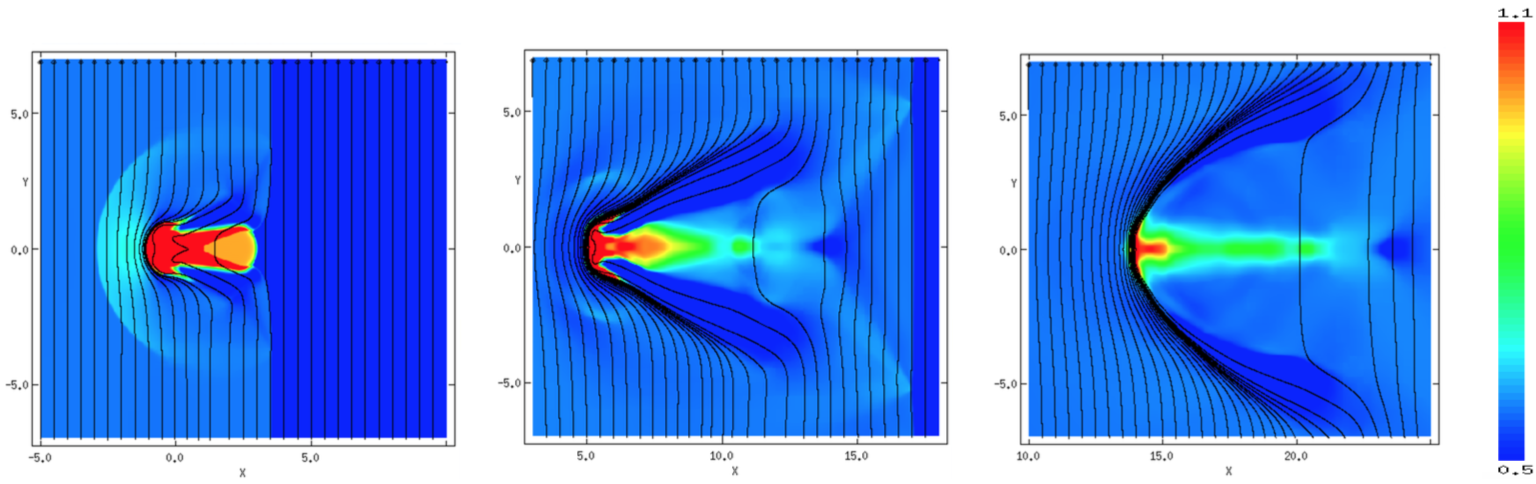}
\caption{As per Fig.20 but showing the $xy$ plane and magnetic fieldlines. The evolution proceeds left to right with $t=0.95 \,t_{cs}$, $t=3.55\,t_{cs}$, and $t=6.10 \,t_{cs}$. Note the shift in the $x$ axis scale for the final panel.}
\label{Fig21}
\end{figure*}

Figure~\ref{Fig20} shows snapshots of the density distribution for simulation $m10c1b1l4o90pe$, again with the fieldlines in the $xy$ plane shown in Fig.~\ref{Fig21}. In the parallel field case, a flux rope would be expected to form on the axis behind the filament. However, with a perpendicular magnetic field this is not observed. Instead, low density filament material forms a linear structure along the axis and, in line with the parallel field scenario's flux rope, this structure persists for some time. As in the parallel field case, the filament with $l=4$ and $\theta=85^{\circ}$ begins to form a similar structure to this filament but the symmetrical nature of the evolving filament is quickly destabilised.

\begin{figure*}
\centering
\includegraphics[width=160mm]{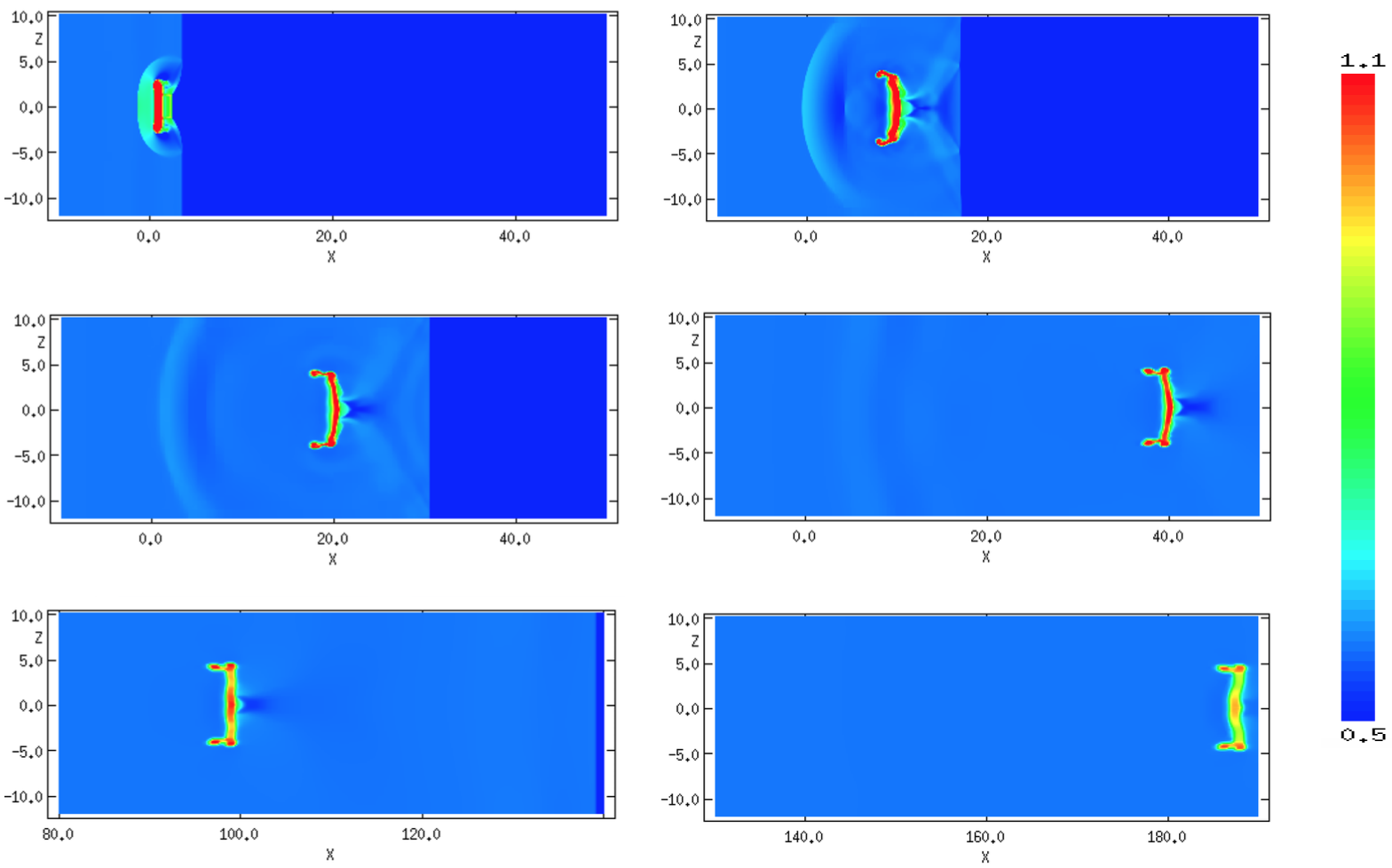}
\caption{The time evolution of the logarithmic density, scaled with respect to the ambient density, for model $m10c1b1l4o0pe$ (cf. the parallel field case in Fig.~\ref{Fig7}). The evolution proceeds left to right, top to bottom, with $t=0.95 \,t_{cs}$, $t=3.54 \,t_{cs}$, $t=6.12 \,t_{cs}$, $t=11.7 \,t_{cs}$, $t=27.9 \,t_{cs}$, and $t=52.2 \,t_{cs}$. Note the shift in the $x$ axis scale for the bottom two panels. The initial magnetic field is perpendicular to the shock normal.}
\label{Fig22}
\end{figure*}

\begin{figure*}
\centering
\includegraphics[width=170mm]{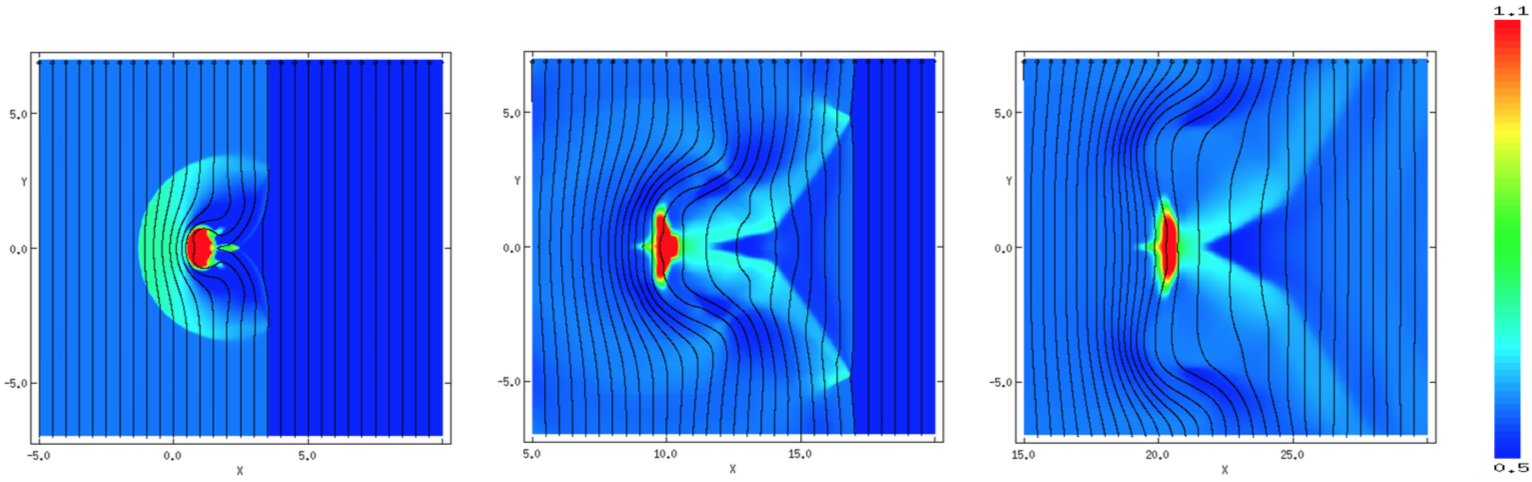}
\caption{As per Fig.22 but showing the $xy$ plane and magnetic fieldlines. The evolution proceeds left to right with $t=0.95 \,t_{cs}$, $t=3.54\,t_{cs}$, and $t=6.12 \,t_{cs}$. Note the shift in the $x$ axis scale for the final panel.}
\label{Fig23}
\end{figure*}

The density distribution for the filament in simulation $m10c1b1l4o0pe$ is depicted in Fig.~\ref{Fig22} and Fig.~\ref{Fig23}. The morphology of this filament at early times (i.e. $t=3.54 \,t_{cs}$) is very similar to that with a parallel field, except that the wings of this filament are swept backwards into the flow. From an observational point of view it may appear as if the filament has been struck by a shock travelling toward the $-x$ direction, and this may render the observational interpretation of such structures problematic. The beginnings of a very short, but broad, flux rope are present but this feature does not grow over time.

\begin{figure*}
\centering
\includegraphics[width=160mm]{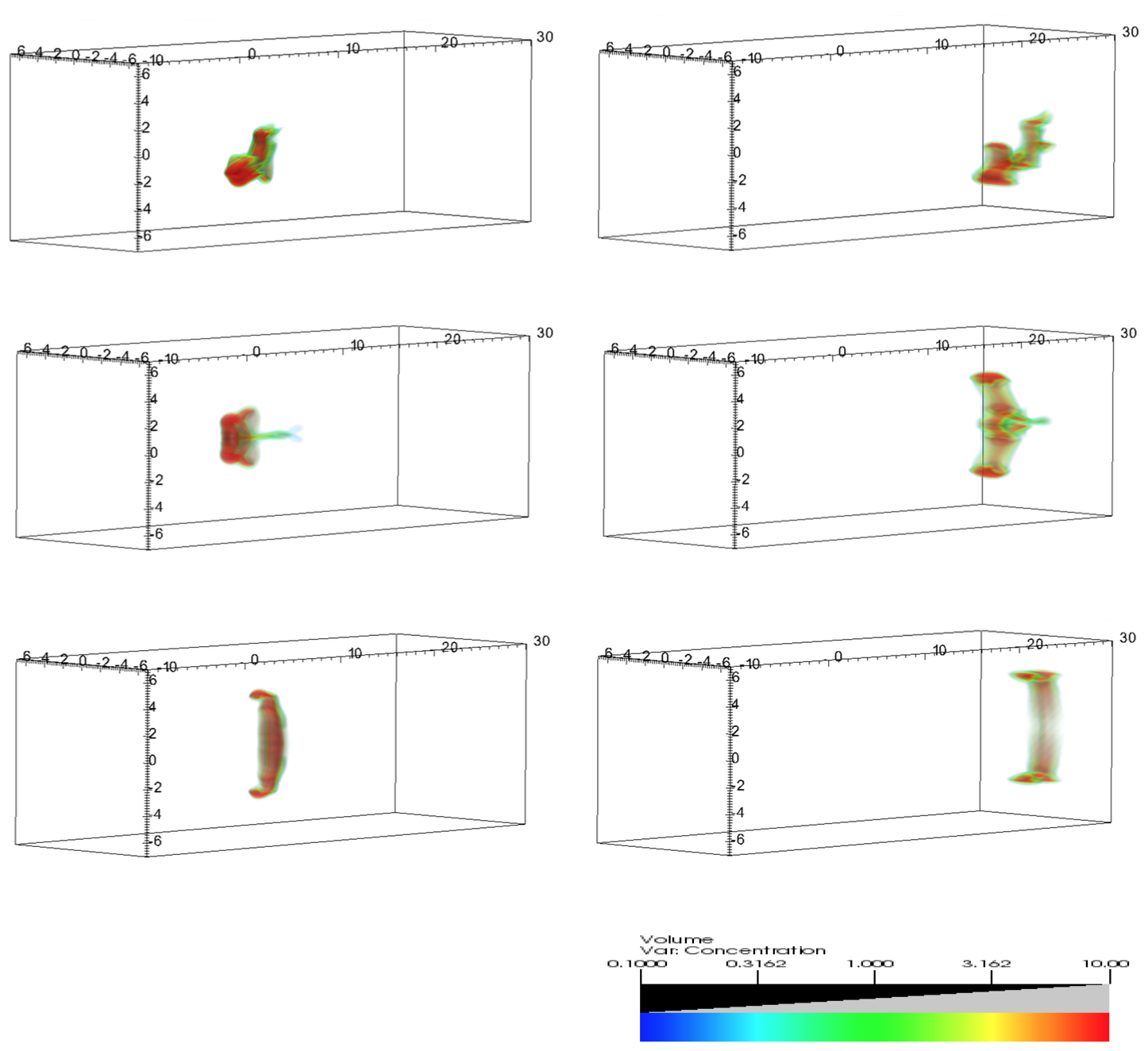} 
\caption{3D volumetric renderings of models $m10c1b1l4o45pe$ (top), $m10c1b1l4o90pe$ (middle), and $m10c1b1l4o0pe$ (bottom) at $t= 3.44 \,t_{cs}$ (left-hand column) and $t=8.95 \,t_{cs}$ (right-hand column). The initial magnetic field is perpendicular to the shock normal.}
\label{Fig24}
\end{figure*}

Figure~\ref{Fig24} shows a 3D volumetric rendering of the time evolution of the density of filament material in simulations $m10c1b1l4o45pe$, $m10c1b1l4o90pe$, and $m10c1b1l4o0pe$, clearly showing a ``sheet-like'' structure at the upstream end of the filament. Because only the filament material is shown, other features such as the bow shock are not displayed in these plots.

\subsubsection{Effect of filament length and orientation on the evolution of the core mass}
\begin{figure*} 
\centering
     \includegraphics[width=110mm]{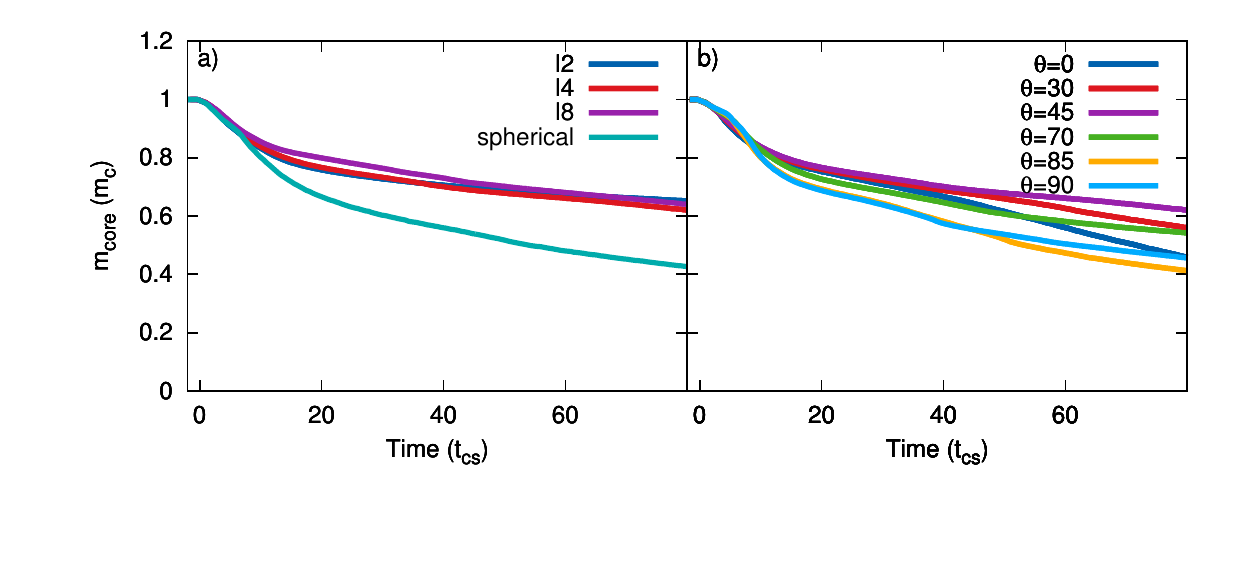} 
     \vspace{-8mm}       
   \caption{Time evolution of the core mass, $m_{core}$, for (a) a filament with variable length and an orientation of $45^{\circ}$, and (b) $l=4$ with variable orientation, in an initial perpendicular magnetic field.}
  \label{Fig25}
  \end{figure*}

Amongst all the quantities being tracked, the reduction in the core filament mass shows the most dramatic difference between simulations with parallel and perpendicular magnetic fields. Figure~\ref{Fig25} shows the evolution of the core mass for filaments in a perpendicular field. The first point of note is that these filaments are very slow to lose their mass. Indeed, in all cases the filaments still comprised a significant amount of mass (between two and three fifths of the initial mass) by $t = 80 \, t_{cs}$. This is in direct contrast to the filaments in a parallel field. Whilst filaments with $l=4$ and $\theta=85^{\circ}$ and $90^{\circ}$ lose their mass more quickly (in agreement with the parallel field cases) it is interesting that the filament with $l=4$ and $\theta=0^{\circ}$ has lost the most mass by $t=80 \,t_{cs}$: in the parallel field simulations it was one of the filaments which conserved their mass the longest.

Considering Fig.~\ref{Fig25}(a), the length of the filament does not appear to have a large influence over the evolution of the core mass, since all three filaments lose mass at approximately the same rate. The spherical cloud, in comparison, loses mass much more quickly, having lost approximately three fifths of its initial mass by the end of the simulation, as opposed to the two fifths that the other filaments have lost. Similar to the parallel magnetic field case, where the spherical cloud evolved in a similar manner to the filaments with $\theta = 0^{\circ}$, the spherical cloud in this case evolves in a similar manner to the filament with $\theta = 90^{\circ}$.

\subsubsection{Effect of filament length and orientation on the mean velocity and the velocity dispersion}
\begin{figure*} 
\centering
     \includegraphics[width=110mm]{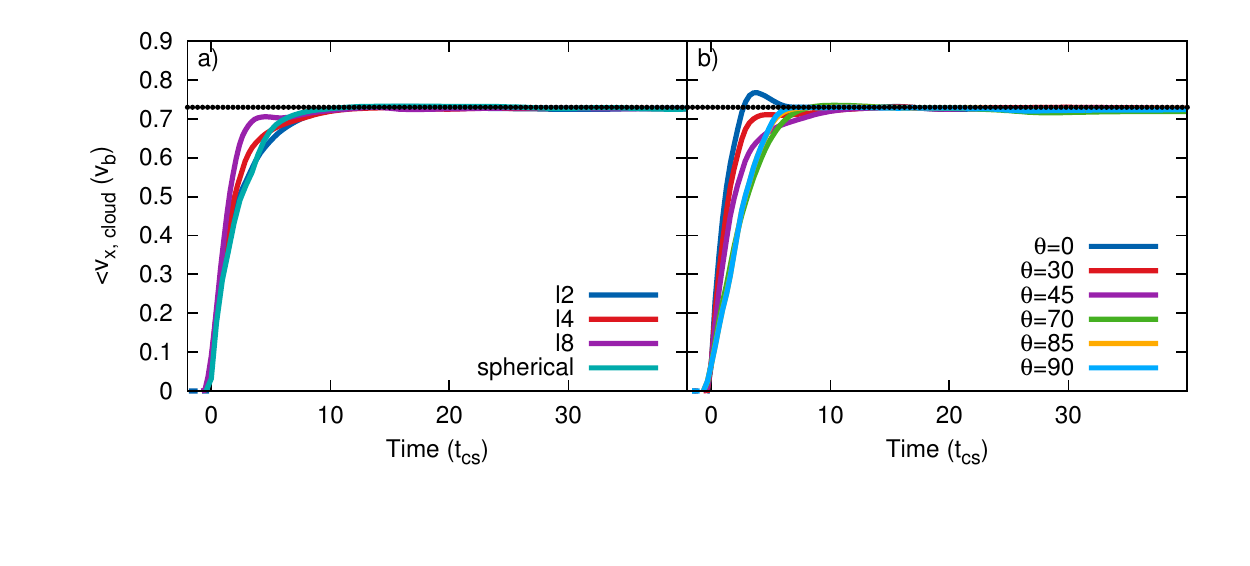}   
     \vspace{-8mm}     
  \caption{Time evolution of the filament mean velocity, $\langle v_{x} \rangle$, for (a) a filament with variable length and an orientation of $45^{\circ}$, and (b) $l=4$ with variable orientation, in an initial perpendicular magnetic field. The dotted black line indicates the velocity of the post-shock flow.}
  \label{Fig26}
  \end{figure*}

The plots showing the mean filament velocity in the $x$ direction (Fig.~\ref{Fig26}) reveal that the filaments in all cases are accelerated to the velocity of the post-shock flow more rapidly than those in a parallel magnetic field. We expect the acceleration to be faster due to i) the increased magnetic pressure which builds up on the upstream side of the filament, and ii) the `snapping back' of the field lines due to the magnetic tension which builds up as the field is dragged around the filament. In contrast to Fig.~\ref{Fig10}(b), the filament with $l=4$ and $\theta=30^{\circ}$ levels off after the initial acceleration, before accelerating again to reach the post-shock flow velocity. Additionally, the filament with $l=4$ and $\theta=0^{\circ}$ overshoots, before asymptoting to the velocity of the post-shock flow. 

The length of the filament has little effect on the mean velocity, with all three filaments initially accelerating at the same rate. However, the filament with $l=8$ and $\theta=45^{\circ}$ exhibits the ``levelling-off'' seen in plot (b), a feature not present in Fig.~\ref{Fig10}(a). The spherical cloud continues to smoothly and rapidly accelerate without levelling off and thus reaches the post-shock flow velocity earlier than the three filaments.

\begin{figure*} 
\centering    
      \includegraphics[width=110mm]{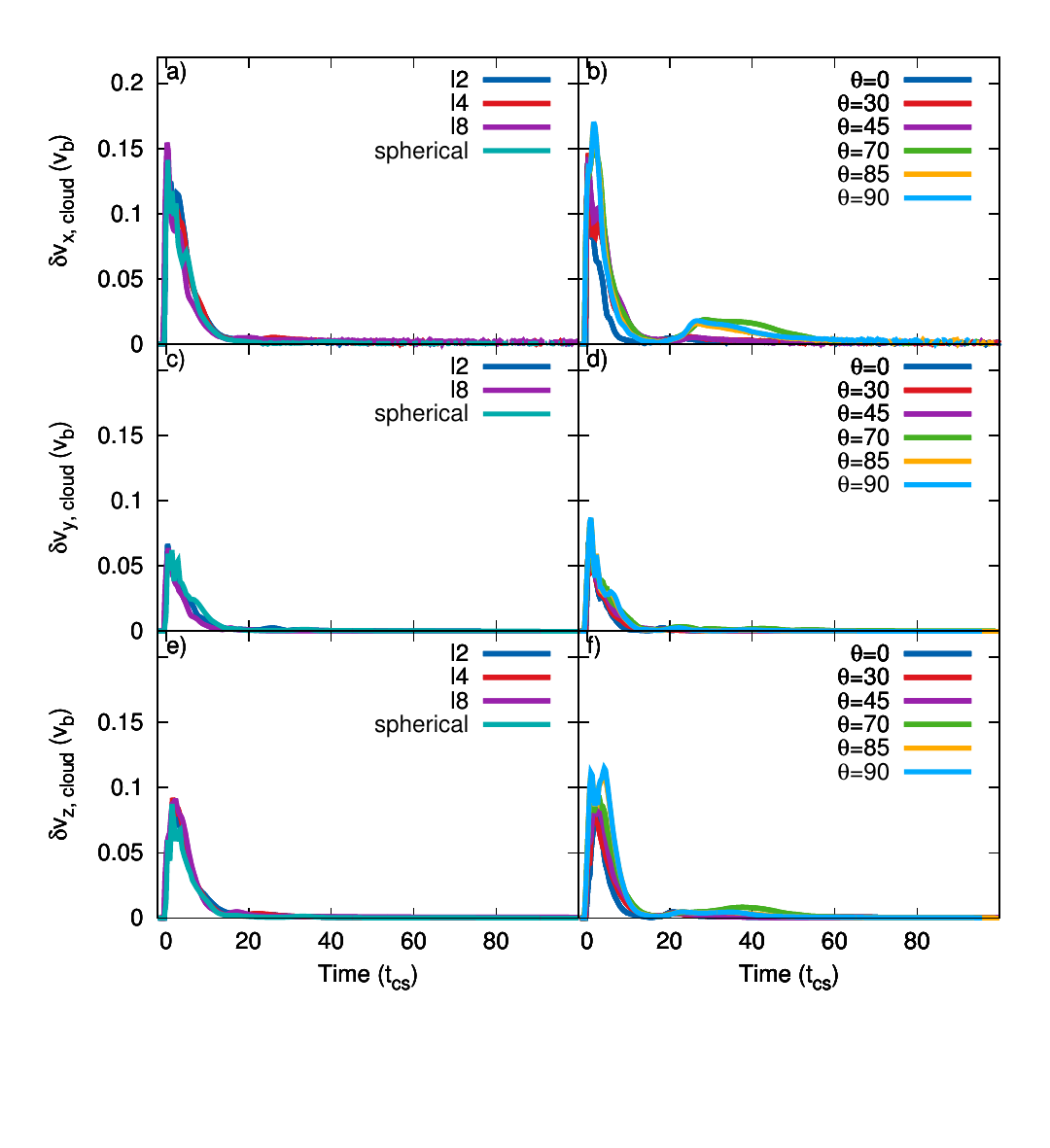}      
      \vspace{-12mm}       
   \caption{Time evolution of the filament velocity dispersion in the $x$, $y$, and $z$ directions, $\delta v_{x,y,z}$, for a filament with variable length and an orientation of $45^{\circ}$ (left-hand column), and $l=4$ with variable orientation (right-hand column) in an initial perpendicular magnetic field.}
  \label{Fig27}
   \end{figure*}

With regard to the velocity dispersion plots, the length of the cloud is shown to have even less of an influence on the evolution of this parameter than in the case of a parallel field (compare Figs.~\ref{Fig27}(a-c) to Figs.~\ref{Fig11}(a-c)). However, there is a clear split in Figs.~\ref{Fig27}(d,f) between those filaments which are more ``end on'' to the shock front, and those which are more ``broadside'' to it. As in the parallel field case, those filaments with orientations of $\theta > 45^{\circ}$ have a greater initial dispersion in the $x$ and $z$ directions, whilst filaments of varying length have very similar velocity dispersions in all directions. In all the velocity dispersion plots the peak of the dispersions is lower than those with a parallel field, indicating that the section of filament closest to the shock front has undergone less compression in the perpendicular field case. 

\subsubsection{Effect of filament length and orientation on the mean density}
\begin{figure*} 
\centering
    \includegraphics[width=110mm]{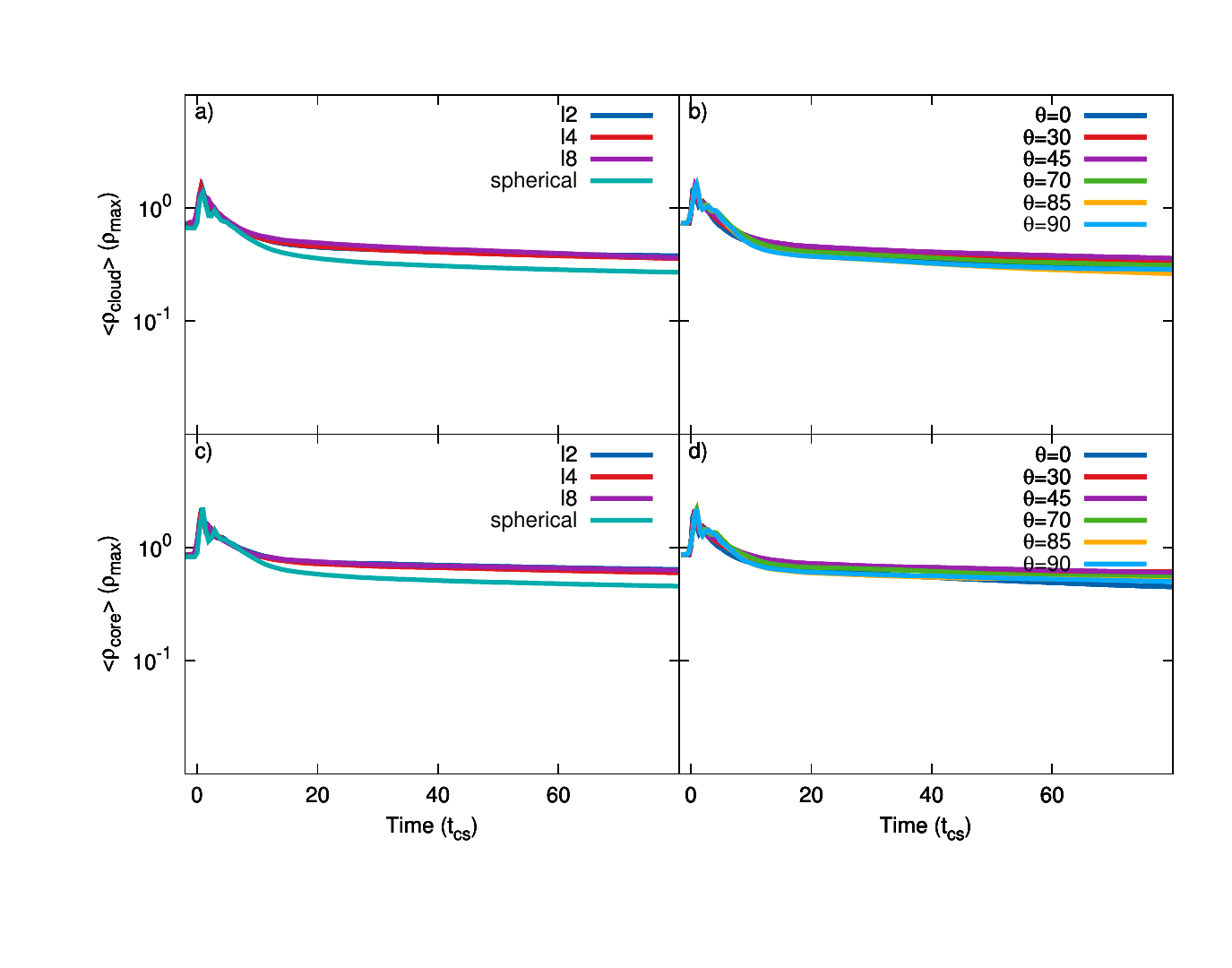}    
    \vspace{-10mm}       
  \caption{Time evolution of the mean density of the filament, $\langle \rho_{cloud}\rangle$ (top), and filament core, $\langle \rho_{core}\rangle$ (bottom), normalised to the initial maximum filament density, for filaments with (left-hand column) variable length and $\theta=45^{\circ}$, and (right-hand column) $l=4$ and a variable orientation, in an initial perpendicular magnetic field.}
   \label{Fig28}
  \end{figure*}

The mean density plots (Fig.~\ref{Fig28}) for both $\langle \rho_{cloud}\rangle$ and $\langle \rho_{core}\rangle$, in terms of the filament orientation, show very little difference between the simulations. However, as in the parallel magnetic field case, the filaments with orientations greater than $\theta=45^{\circ}$ have a slightly larger drop in mean density, overall. Plots (a) and (c) of Fig.~\ref{Fig28} show almost no change in the mean density between the simulations while the spherical cloud reduces to a much lower mean density consistent with the filaments with $\theta = 0^{\circ}$, and $90^{\circ}$, indicating that the filament length is not important for the evolution of the mean density.

\subsubsection{$\chi$ dependence of the filament evolution}
\begin{figure*}
\centering
\includegraphics[width=160mm]{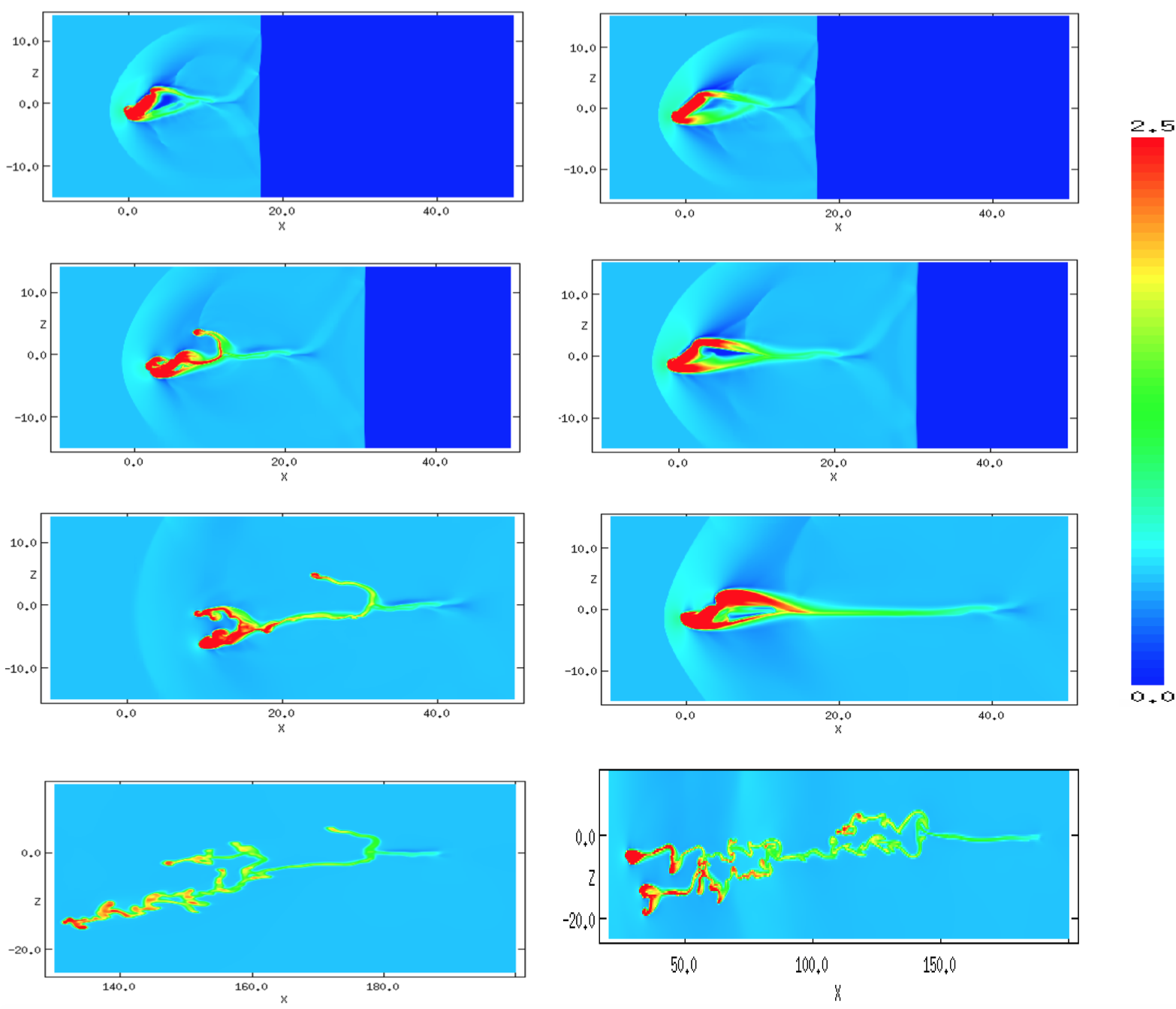}
\caption{The time evolution of the logarithmic density, scaled with respect to the ambient density, for models $m10c2b1l4o45pe$ (left-hand column) and $m10c3b1l4o45pe$ (right-hand column). The evolution proceeds top to bottom, with $t=1.08 \,t_{cs}$, $t=1.98 \,t_{cs}$, $t=3.65 \,t_{cs}$, and $t=16.6 \,t_{cs}$ for the $\chi=100$ case, and $t=0.34\, t_{cs}$, $t=0.61 \,t_{cs}$, $t=1.15 \,t_{cs}$, and $t=5.23 \,t_{cs}$ for the $\chi=1000$ case. Note the shift in the $x$ and $y$ axis scales for the final panel in each column, and the change in the logarithmic density scale compared to previous cases. The initial magnetic field is perpendicular to the shock normal.}
\label{Fig29}
\end{figure*}

\begin{figure*}
\centering
\includegraphics[width=170mm]{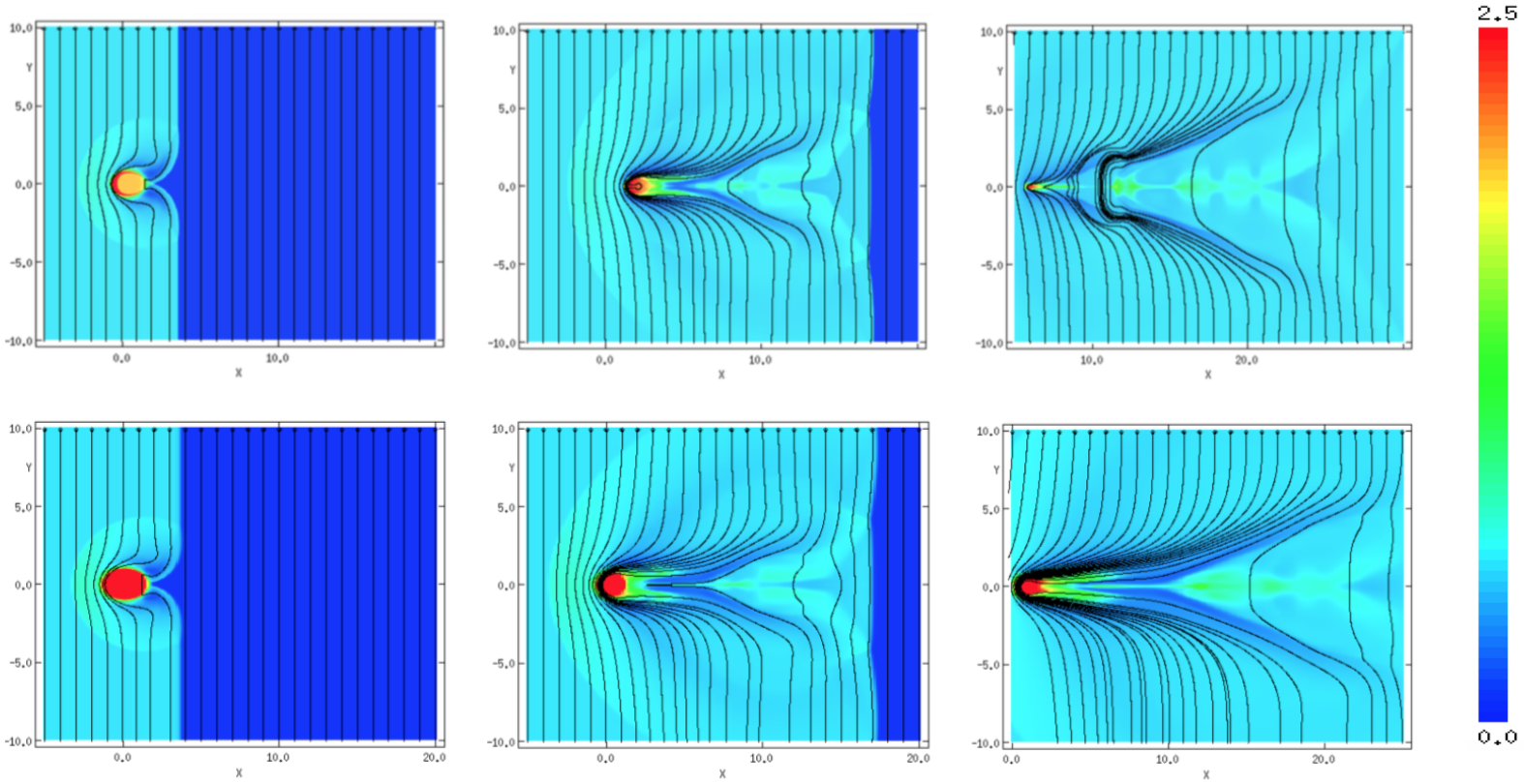}
\caption{Top row: as per Fig.29 (left-hand panels) but showing the $xy$ plane and magnetic fieldlines. The evolution proceeds left to right with $t=1.08 \,t_{cs}$, $t=1.98 \,t_{cs}$, and $t=3.65 \,t_{cs}$. Bottom row: as per Fig.29 (right-hand panels) but showing the $xy$ plane and magnetic fieldlines. The evolution proceeds left to right with $t=0.34\, t_{cs}$, $t=0.61 \,t_{cs}$, and $t=1.15 \,t_{cs}$. Note the shift in the $x$ axis scale for the final panels.}
\label{Fig31}
\end{figure*}

\begin{figure*} 
\centering     
      \includegraphics[width=160mm]{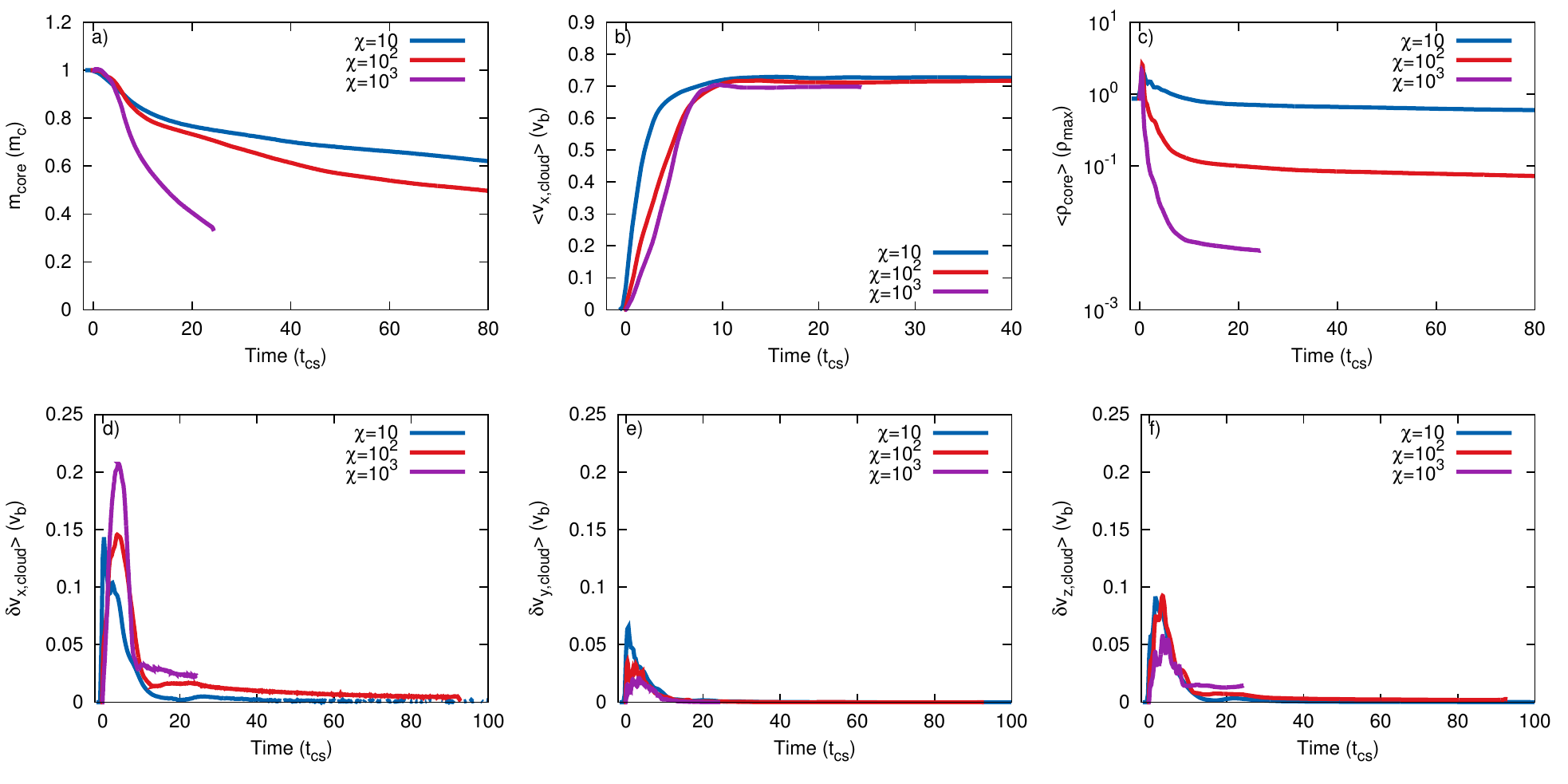} 
      \vspace{-3mm}            
  \caption{$\chi$ dependence of the evolution for filaments with $l=4$ and $\theta=45^{\circ}$. The initial magnetic field is perpendicular to the shock normal, $M=10$, and $\beta_{0}=1$. Note that although model $m10c3b1l4o45pe$ was run at a reduced resolution of $R_{16}$ it was computationally difficult to run. Therefore, the filament in this model moved off the grid before the simulation was complete.}
  \label{Fig32}
  \end{figure*}

The evolution of filaments in a perpendicular field with increasing cloud density contrasts is radically different to those in a parallel magnetic field. Figure~\ref{Fig29} shows that the filament is drawn out into long, smooth, tendril-like shapes which persist for far longer than the filaments in the parallel case (cf. Fig.~\ref{Fig13}), while the highly-turbulent features present with a parallel field are not in evidence. In addition, the magnetic fieldlines are increasingly stretched around the filament and bunched together, as seen in Fig.~\ref{Fig31}. The higher the value of $\chi$, the more drawn-out the filament is along the $x$ axis. This is evident in Fig.~\ref{Fig32}(a), where the filaments with higher values of $\chi$ retain almost two fifths of their initial mass at the end of the simulation, though that with $\chi=1000$ still has a faster mass-loss rate in agreement with the parallel field case. The mean velocity and mean density plots for both parallel and perpendicular fields are very similar. However, the velocity dispersion plots show some differences, with much less dispersion in the $x$ and $y$ directions in Figs.~\ref{Fig32}(d,e). 

\subsubsection{Mach dependence of the filament evolution}
\begin{figure*} 
\centering  
      \includegraphics[width=160mm]{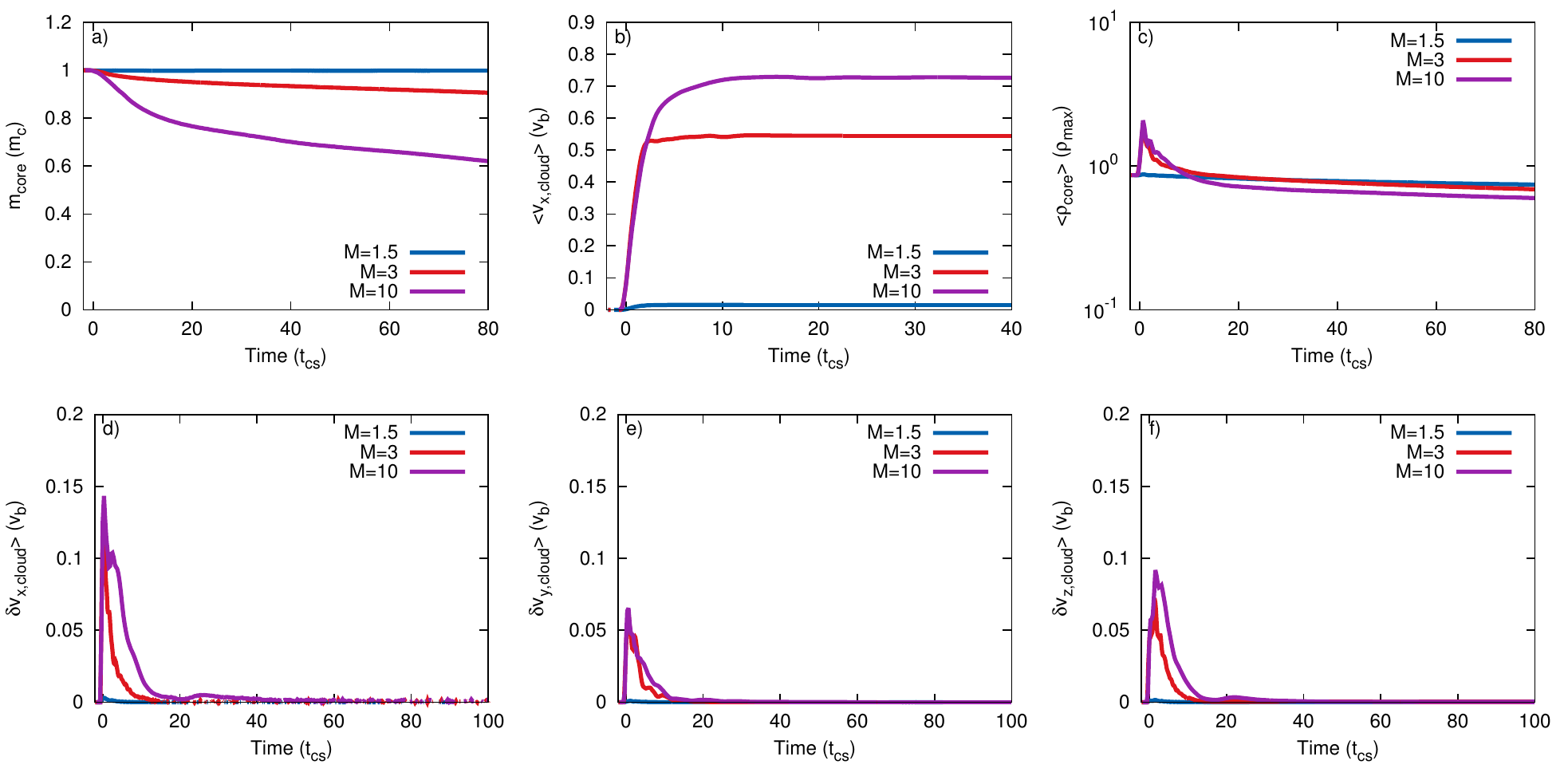}     
      \vspace{-3mm}       
  \caption{Mach number dependence of the evolution for filaments with $l=4$ and $\theta=45^{\circ}$. The initial magnetic field is perpendicular to the shock normal, $\chi=10$, and $\beta_{0}=1$.}
  \label{Fig33}
   \end{figure*}

The shock Mach number dependence of the evolution displays similar trends to that of the parallel magnetic field case. However, it can be seen from Fig.~\ref{Fig33}(a) that the filament which has been struck by a $M=1.5$ shock has lost almost no mass for the duration of the simulation (in contrast with the filament struck by an $M=10$ shock, which has lost two fifths of its mass by $t=80\,t_{cs}$). Fig.~\ref{Fig33}(b) shows that the post-shock velocity in the $M=1.5$ case is very small (and much smaller than that of the same case in a parallel field). This suggests that the combination of a mild shock and the magnetic field lines bent around the filament serve to protect the filament from compression and ablation by the flow for a considerable time. This is borne out by the morphology of the low Mach filaments, which retain the same footprint for much of the simulation (indeed, the filament with $M=1.5$ does not significantly alter its morphology at all). The velocity dispersion plots (d, e, f) show that there is far less dispersion in all directions compared with the parallel magnetic field case, though again the simulation with $M=1.5$ has almost no dispersion since its morphology has not been significantly changed by the post-shock flow during the period that the simulation was run. 

\subsubsection{$\beta_{0}$ dependence of the filament evolution}
\begin{figure*} 
\centering   
     \includegraphics[width=160mm]{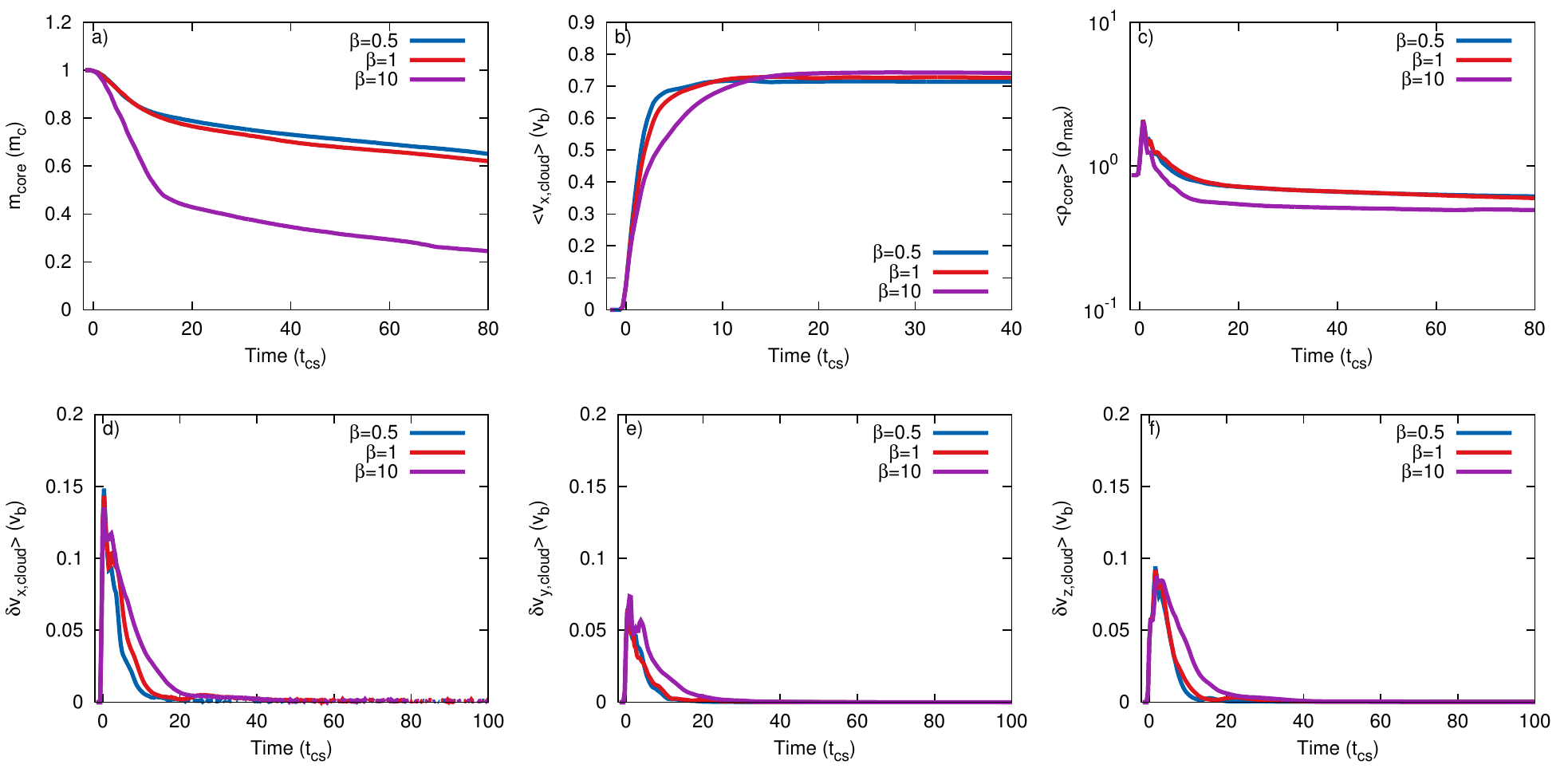}       
     \vspace{-3mm}     
  \caption{Plasma beta dependence of the evolution for filaments with $l=4$ and $\theta=45^{\circ}$. The initial magnetic field is perpendicular to the shock normal, $M=10$, and $\chi=10$.}
  \label{Fig34}
   \end{figure*}

Figure~\ref{Fig34} shows the effect of varying the plasma beta on the filament evolution. As in the parallel field case, the filament with a weak magnetic field ($\beta=10$) loses mass much more quickly than the other two, stronger, fields. The morphology of the filament in the weaker field displays similar patterns of instability to that of the parallel field, with material being stripped from the surface of the filament. In contrast, the filament in the other two strengths of field remains tightly bound for the duration of the simulation. In addition, there is again a very low amount of divergence between the simulations with regard to the velocity, velocity dispersions, and mean density, though the filament in a $\beta=10$ field takes longer to be accelerated to the velocity of the post-shock flow due to the lower upstream magnetic pressure and decreased tension in the field lines. Furthermore, its velocity dispersions decay more slowly, compared to the filaments with the stronger field strengths.

\subsection{Oblique field}

\subsubsection{Filament morphology}
The simulations run with an obliquely-orientated (i.e. at $45^{\circ}$ to the shock normal) magnetic field have very similar morphologies to those run with a perpendicularly-orientated field. For this reason, we have not included snapshots of the logarithmic density for the oblique field case. As before, filaments set at an angle to the shock front in an oblique field take on a tendril-like appearance, whilst those orientated either broadside, or end on, to the shock front produce linear features along the axis behind the filament.

\subsubsection{Effect of filament length and orientation on the core mass, mean velocity, velocity dispersion, and mean density}
\begin{figure*} 
\centering
     \includegraphics[width=110mm]{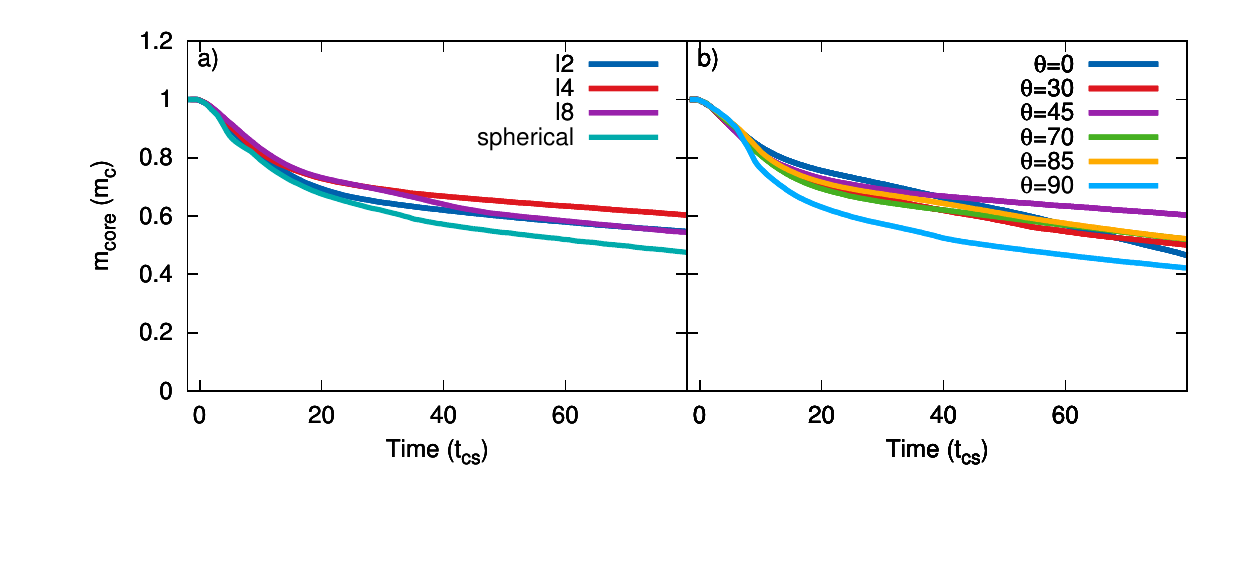} 
     \vspace{-8mm}      
   \caption{Time evolution of the core mass, $m_{core}$, for (a) a filament with variable length and an orientation of $45^{\circ}$, and (b) $l=4$ with variable orientation, in an initial magnetic field orientated at $45^{\circ}$ to the shock normal.}
  \label{Fig35}
  \end{figure*}

In terms of the evolution of the core mass, there is only a slight difference between Fig.~\ref{Fig35}(a) and Fig.~\ref{Fig25}(a). In the oblique field case, the filament with $l=4$ and $\theta=45^{\circ}$ has the most mass remaining at the end of the simulation whilst that with $l=8$ and $\theta=45^{\circ}$ loses the most mass. In the perpendicular field case, however, the rate at which each filament loses mass is reversed. Considering Fig.~\ref{Fig35}(b) and Fig.~\ref{Fig25}(b), the only difference between the two field orientations is that in the perpendicular field case the filament with $l=4$ and $\theta=85^{\circ}$ is one of two filaments which lose the most mass by the end of the simulation, but in the oblique case this filament loses mass far slower 
(at a similar rate to the filaments with $\theta=30^{\circ}$ and $\theta=70^{\circ}$).

\begin{figure*} 
\centering
    \includegraphics[width=110mm]{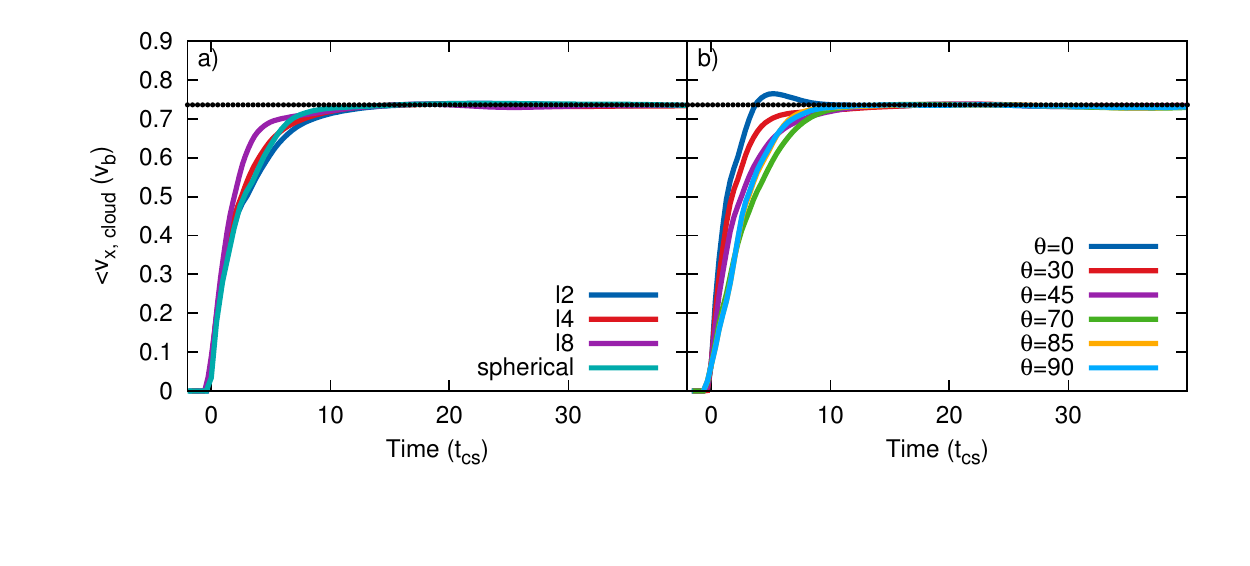}
    \vspace{-8mm}       
   \caption{Time evolution of the filament mean velocity, $\langle v_{x} \rangle$, for (a) a filament with variable length and an orientation of $45^{\circ}$, and (b) $l=4$ with variable orientation, in an initial magnetic field orientated $45^{\circ}$ to the shock normal. The dotted black line indicates the velocity of the post-shock flow.}
  \label{Fig36}
  \end{figure*}  

\begin{figure*} 
\centering   
      \includegraphics[width=110mm]{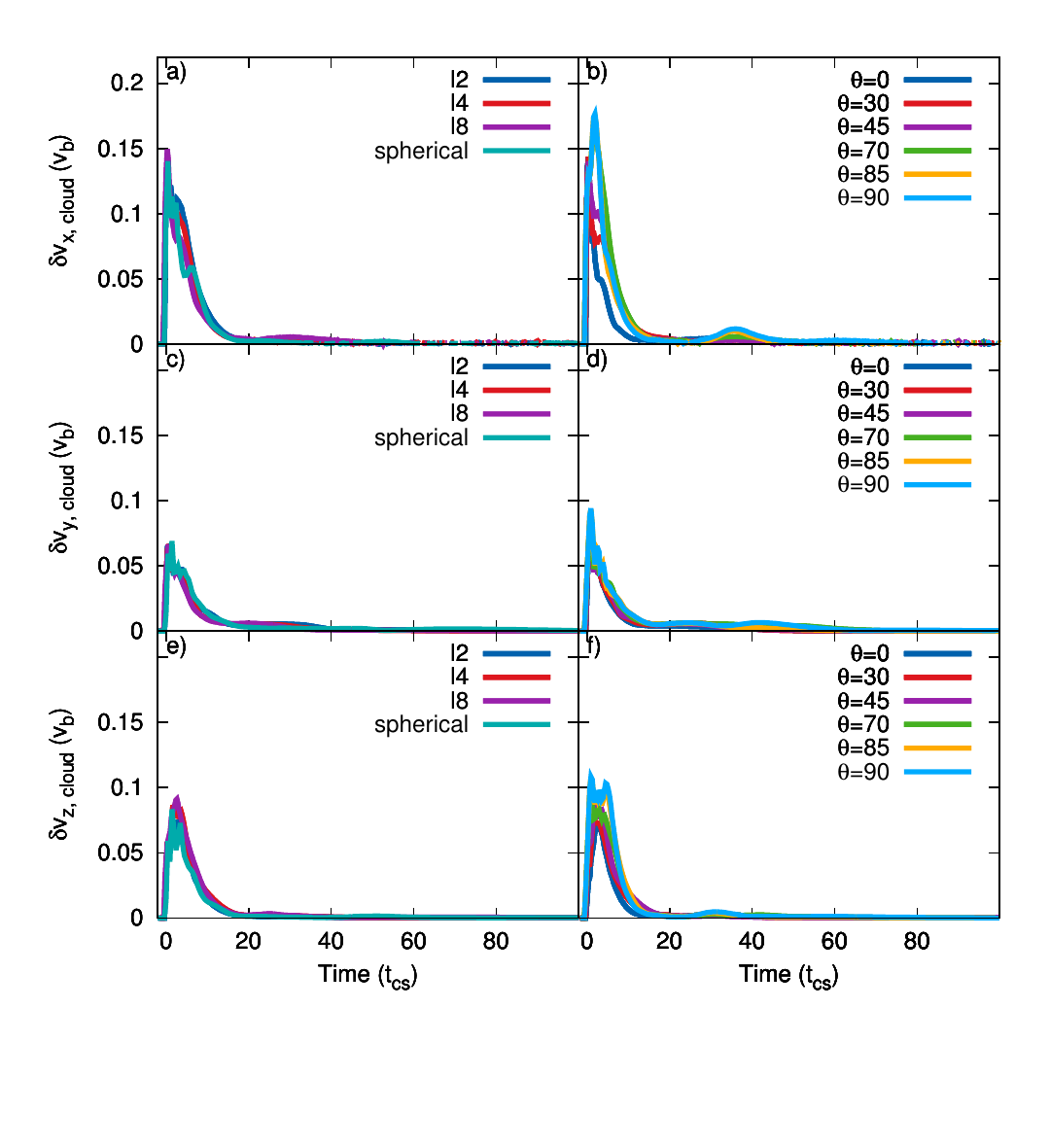}  
      \vspace{-12mm}           
  \caption{Time evolution of the filament velocity dispersion in the $x$, $y$, and $z$ directions, $\delta v_{x,y,z}$, for a filament with variable length and an orientation of $45^{\circ}$ (left-hand column), and $l=4$ with variable orientation (right-hand column) in an initial magnetic field orientated $45^{\circ}$ to the shock normal.}
  \label{Fig37}
   \end{figure*}

\begin{figure*} 
\centering
     \includegraphics[width=110mm]{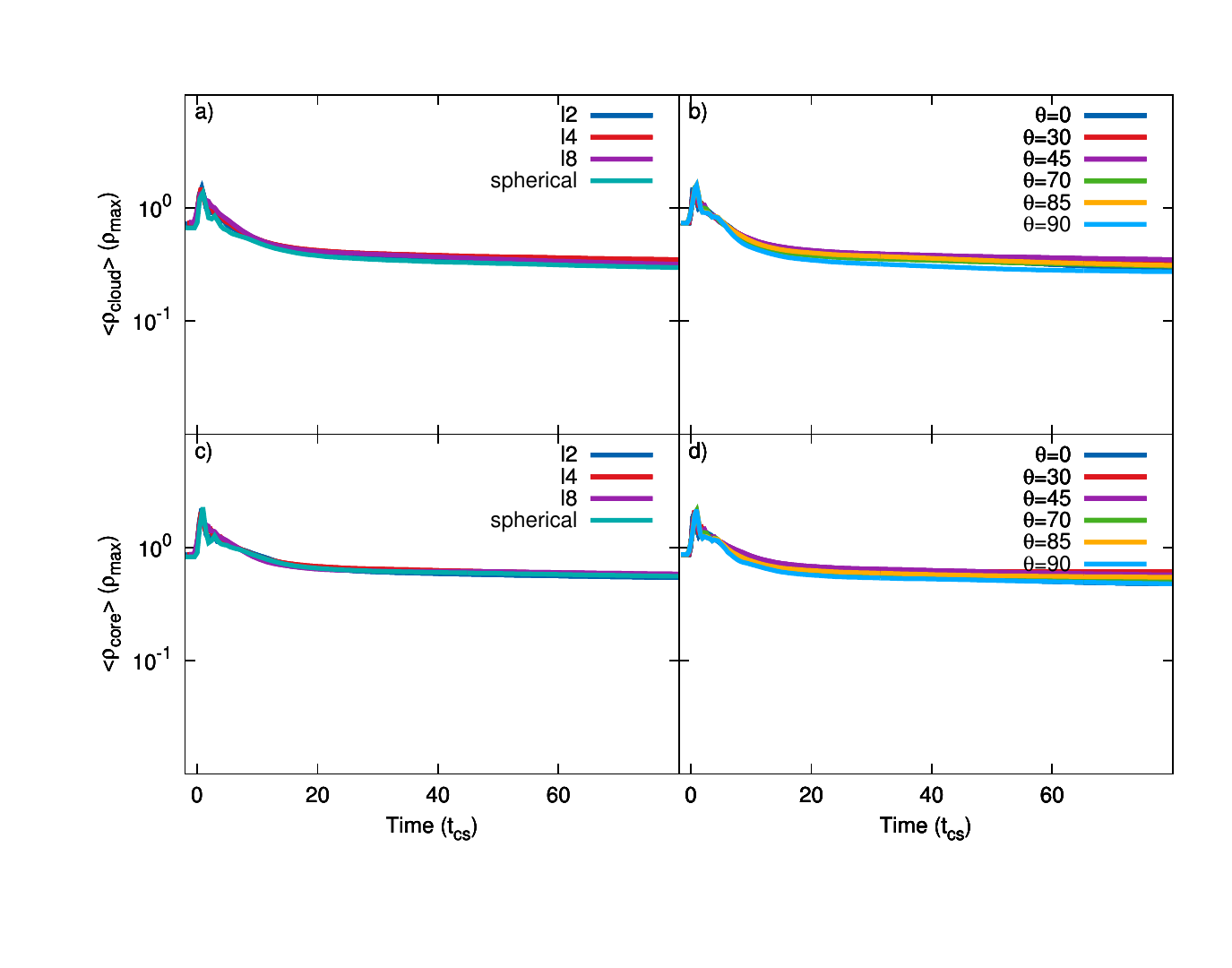}  
     \vspace{-10mm}           
   \caption{Time evolution of the mean density of the filament, $\langle \rho_{cloud}\rangle$ (top), and filament core, $\langle \rho_{core}\rangle$ (bottom), normalised to the initial maximum filament density, for filaments with (left-hand column) variable length and $\theta=45^{\circ}$, and (right-hand column) $l=4$ and a variable orientation, in an initial magnetic field orientated $45^{\circ}$ to the shock normal.}
   \label{Fig38}
  \end{figure*}

The mean velocity plots for filaments in oblique and perpendicular fields (Fig.~\ref{Fig36} and Fig.~\ref{Fig26}, respectively) are almost identical, though the filament in the oblique field with $l=8$ and $\theta=45^{\circ}$ is accelerated to the velocity of the post-shock flow much more smoothly than the same filament in the perpendicular field. The velocity dispersions for both orientations of the magnetic field are also very similar, though Fig.~\ref{Fig37}(d) does not display as large a dispersion in the $x$ direction between $t=30-40 \,t_{cs}$ as Fig.~\ref{Fig27}(d) does. In terms of the mean density (cf. Fig.~\ref{Fig38} with Fig.~\ref{Fig28}), the filaments with different orientations provide very similar plots in both the oblique and perpendicular field cases, whereas those filaments with varying lengths in the oblique field case reach a much lower mean density after the initial peak. 

\subsubsection{$\chi$, $M$, and $\beta_{0}$ dependence of the filament evolution}
\begin{figure*} 
\centering     
     \includegraphics[width=160mm]{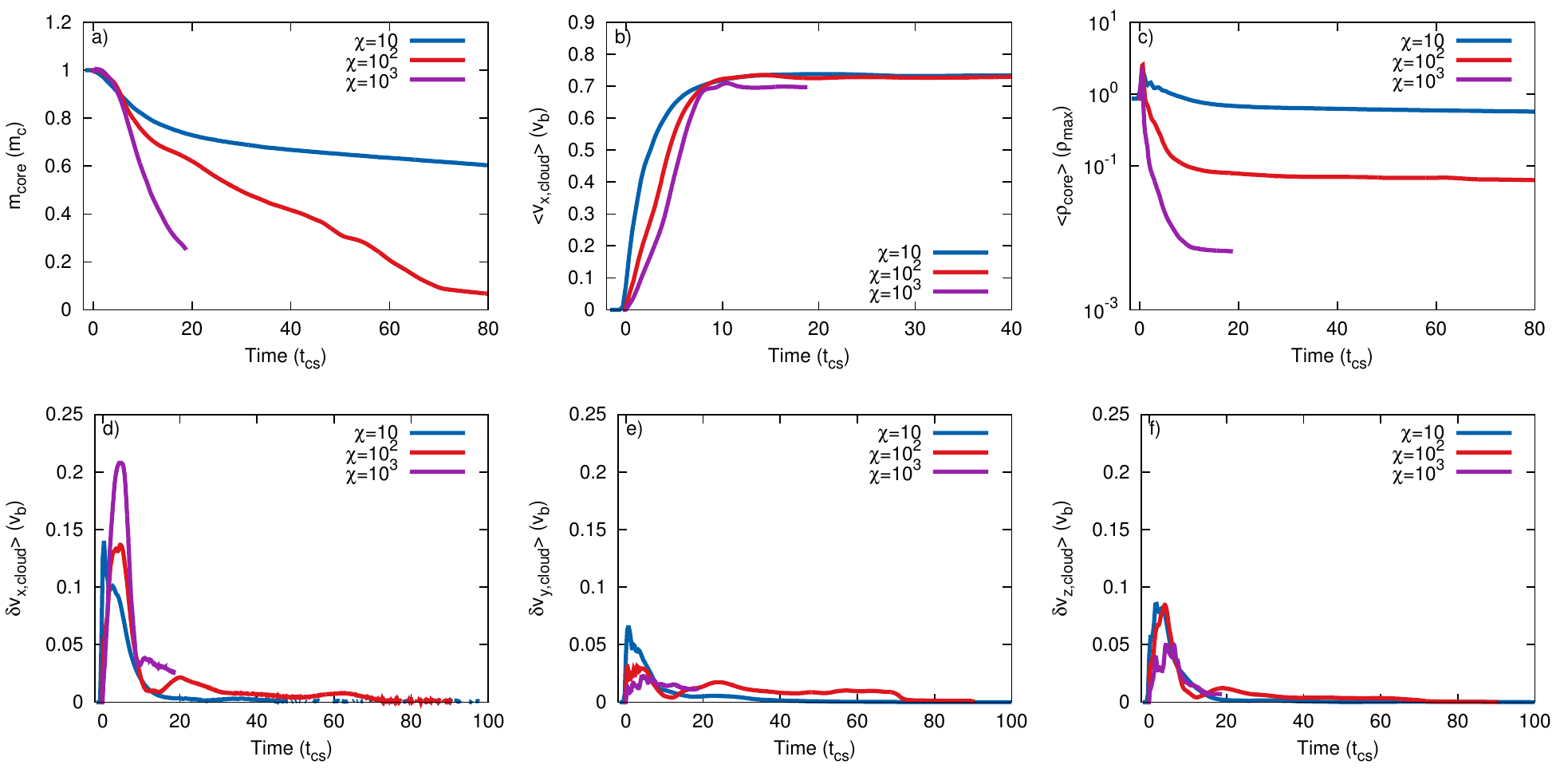} 
     \vspace{-3mm}           
  \caption{$\chi$ dependence of the evolution for filaments with $l=4$ and $\theta=45^{\circ}$. The initial magnetic field is orientated $45^{\circ}$ to the shock normal, $M=10$, and $\beta_{0}=1$. Note that although model $m10c3b1l4o45ob$ was run at a reduced resolution of $R_{16}$ it was computationally difficult to run. Therefore, the filament in this model moved off the grid before the simulation was complete.}
  \label{Fig39}
  \end{figure*}

\begin{figure*} 
\centering     
      \includegraphics[width=160mm]{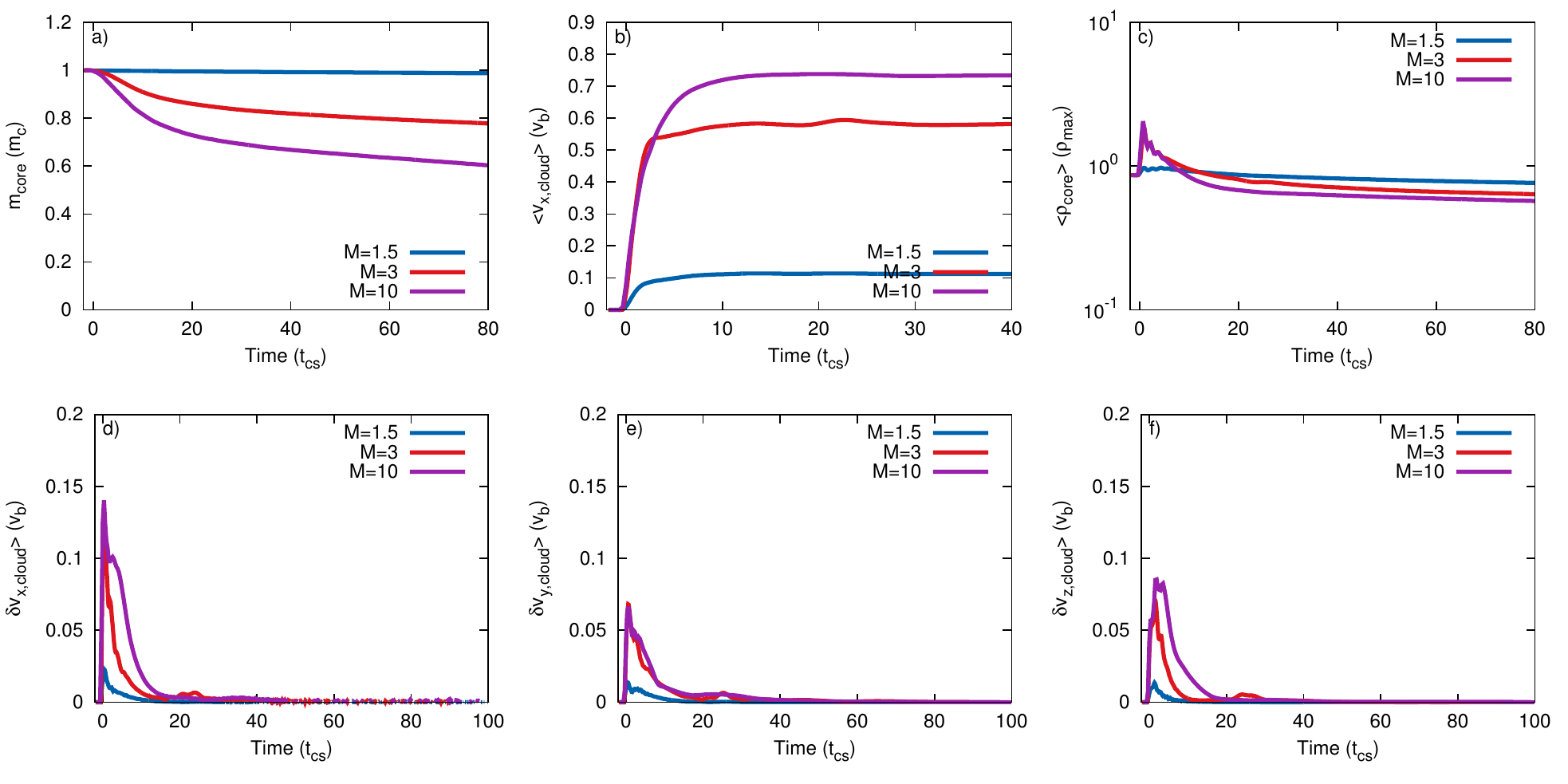}   
      \vspace{-3mm}          
   \caption{Mach number dependence of the evolution for filaments with $l=4$ and $\theta=45^{\circ}$. The initial magnetic field is orientated $45^{\circ}$ to the shock normal, $\chi=10$, and $\beta_{0}=1$.}
  \label{Fig40}
  \end{figure*}

\begin{figure*} 
\centering 
     \includegraphics[width=160mm]{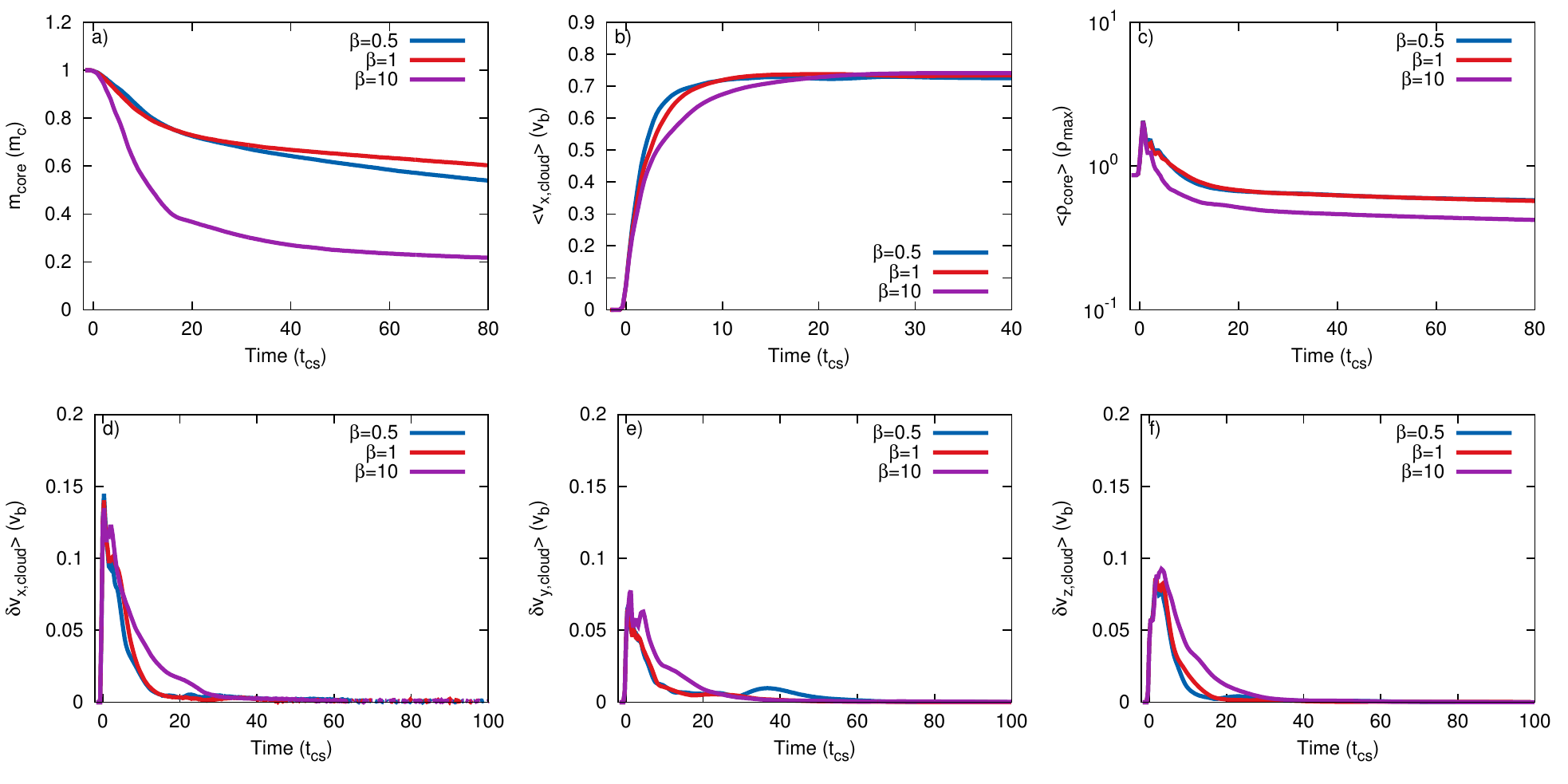}    
     \vspace{-3mm}       
   \caption{Plasma beta dependence of the evolution for filaments with $l=4$ and $\theta=45^{\circ}$. The initial magnetic field is orientated $45^{\circ}$ to the shock normal, $M=10$, and $\chi=10$.}
  \label{Fig41}
  \end{figure*}

As with the time evolution of the filaments with varying length and orientation, the dependence of the evolution on the density contrast, shock Mach number, and magnetic field strength does not significantly differ between the perpendicular and oblique field cases. In terms of the change in $\chi$, the only difference between Figs.~\ref{Fig39} and \ref{Fig32} is that the filaments with higher values of $\chi$ in the oblique field are destroyed much faster than those in a perpendicular field (though still not as rapidly as for a parallel field). Figure~\ref{Fig40} shows that the velocity of the post-shock flow is higher in the oblique field case, and thus the filament hit by a $M=1.5$ shock reaches a higher final velocity compared to the perpendicular field case. In addition, this filament has much greater velocity dispersions than the same filament in the perpendicular field case (cf. Fig.~\ref{Fig33}). The filament struck by a $M=3$ shock also loses mass at a slightly faster rate than in a perpendicular field. Considering the magnetic field strength, the main difference between the perpendicular and oblique field cases is that the filament in a field of strength $\beta_{0}=0.5$ undergoes much greater velocity dispersions in the $y$ direction at $t\simeq 40\,t_{cs}$, compared with the perpendicular field (cf. Fig.~\ref{Fig41}(e) to Fig.~\ref{Fig34}(e)).

\subsection{Timescales}
\begin{figure*} 
\centering
\includegraphics[width=110mm]{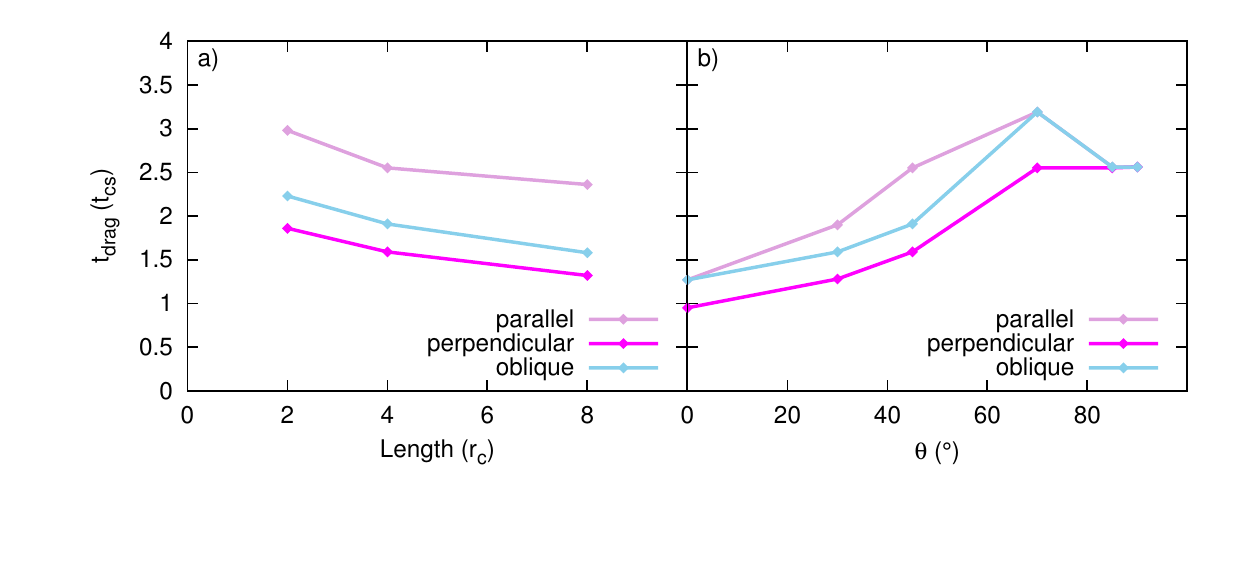} 
\vspace{-8mm}
  \caption{$t_{drag}$ (in terms of the cloud) as a function of filament length (a), where the filament has an orientation of $\theta=45^{\circ}$, and orientation (b), where the filament has a length $l=4$, for simulations with $M=10$, $\chi=10$, and $\beta=1$.}
  \label{Fig42}
  \end{figure*}

Values of $t_{drag}$, $t_{mix}$, and $t_{life}$ are noted in Tables~\ref{Table2} and \ref{Table3}. With the exception of the simulations with a cloud density contrast of 1000 in both the parallel and oblique field cases, in all other cases $t_{drag} \, < \, t_{mix}$. Figure~\ref{Fig42} shows the values of $t_{drag}$ for filaments of varying length and an orientation of $\theta=45^{\circ}$ and filaments with a length $l=4$ and varying orientations, with $M=10$, $\chi=10$, and $\beta=1$. We can see from Fig.~\ref{Fig42}(a) that $t_{drag}$ decreases at a similar rate with increasing filament length for all orientations of the magnetic field. However, the field orientation also has an influence on the value of $t_{drag}$, with filaments in a parallel field exhibiting higher values compared to those in a perpendicular field. Figure~\ref{Fig42}(b), in contrast, shows that while the field orientation has the same effect for filaments with varying $\theta$ as those with varying length, $t_{drag}$ in this case increases with increasing filament orientation, with filaments of $\theta=0^{\circ}$ exhibiting the lowest value of $t_{drag}$ (i.e. these filaments accelerate faster than the others). In addition, there is a downturn/plateauing in the value of $t_{drag}$ for filaments with orientations of $\theta \geq 70^{\circ}$. For both plots, $t_{drag}$ varies by a factor of $\sim 2.5$. $t_{drag}$ is an important indicator of the filament's acceleration within the post-shock flow; thus, in the above cases, longer filaments oriented broadside to the shock front are able to be accelerated more quickly up to the velocity of the post-shock flow.

\begin{figure*} 
\centering
\includegraphics[width=110mm]{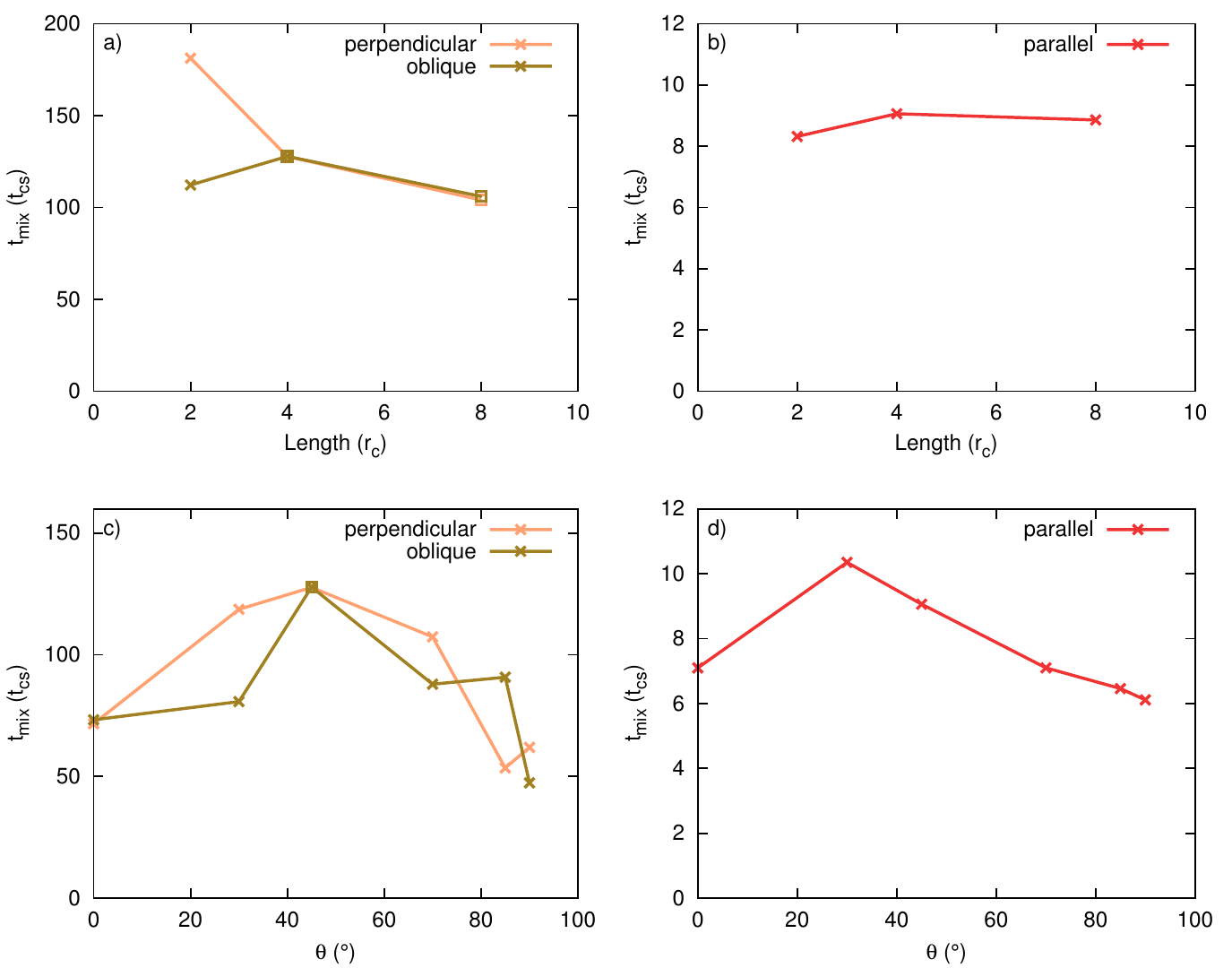} 
  \caption{$t_{mix}$, as a function of filament length (top panels), where the filament has an orientation of $\theta=45^{\circ}$, and orientation (bottom panels), where the filament has a length $l=4$, for simulations with $M=10$, $\chi=10$, and $\beta=1$, in a perpendicular/oblique field (left) and a parallel field (right). Note that data marked with a `square' represent inferred rather than actual values for $t_{mix}$ (see Table~\ref{Table3}).}
  \label{Fig43}
  \end{figure*}

Figure~\ref{Fig43} shows the change in $t_{mix}$ according to filament length and orientation, respectively. It should be noted that because MHD filaments generally exist for far longer than hydrodynamic filaments $t_{mix}$ in some of the simulations occurred after the end of the simulation. We have, therefore, plotted the simulation's final value of $t$ as $t_{mix}$ whilst emphasising that the actual $t_{mix}$ was in fact greater than this (see Tables~\ref{Table2} and \ref{Table3} for an indication of the relevant simulations). The results from \citet{Pittard16} showed that $t_{mix}$ displayed the same behaviour as $t_{drag}$ for filaments of varying length or orientation. However, our results displayed much more complex behaviour (cf. Fig.~\ref{Fig43} with Fig. 34 in \citet{Pittard16}). The results for filaments of differing length broadly showed the same trends as for $t_{drag}$, but those for filaments of varying orientation in either a perpendicular or oblique field did not. It is clear that filaments of $\theta=45^{\circ}$ in perpendicular/oblique fields are far more slow to mix in with the surrounding flow than filaments of any other orientation. $t_{mix}$ is relevant to the survival of the filament; therefore, in the above cases, filaments of length $l\leq 4$ and oriented at $\theta=45^{\circ}$ in either a perpendicular or oblique field are able to survive for significant periods of time.

\section{Discussion}
Filaments have been observed in regions such as the Taurus molecular cloud \citep{Panopoulou14}, the Lupus molecular clouds \citep{Benedettini15}, Orion A \citep{Polychroni13}, and the Pipe Nebula \citep{Peretto12}. Recent observations (e.g. from the Herschel Space Observatory) have shown filamentary structures to be highly prevalent within star-forming regions and point towards their central role in the process of star formation (e.g. \citealt{Arzoumanian11}). In addition, theoretical and numerical studies \citep{Federrath16} of such observations which followed the evolution of molecular clouds and the star formation within them, detected complex networks of filaments in all simulations and determined various filament parameters which were in excellent agreement with observations. A large proportion of prestellar cores are found to be located within dense filaments (e.g. \citealt{Schisano14}, \citealt{Konyves15}). Clusters tend to be highly concentrated at filament junctions but cores (and, thus, stars) have also been shown to form along filaments, indicating that the merger of filaments enables the formation of massive stars within clusters \citep{Schneider12}. The presence of magnetic fields and their stabilising effects on filaments have been inferred (e.g. the alignment of a filament to the ambient magnetic field \citep{Benedettini15} and the smooth morphology of some filaments \citep{Crawford05}), though there has been less discussion of this subject in the literature. Such stabilisation may have a role to play in enabling the subsequent formation of cores. In light of the importance of filamentary structures, studies of the interaction of high-speed flows with filaments, as well as the physics of filament evolution and destruction, are important for a complete understanding of the magnetohydrodynamical nature of the ISM and the process of star formation.

\subsection{SNR-filament interactions}
The interactions of spherical molecular clouds with SNR shockwaves have been well observed and studied (see \citet{Pittard15} for a comprehensive overview of the characteristics of such interactions). Whilst there are instances in the literature of the interactions of jets and winds with filaments, there have been very few studies devoted to shock-filament interactions. Therefore, a wide-ranging discussion of such observations presents difficulties. \citet{Zhou14} discuss the interaction of SNR G127.1+0.5 with an external filament. However, in this case the filament is very large and significant changes in the shock properties can be expected as it sweeps over the filament. This precludes a detailed comparison with our work.

\subsection{Entrainment of filament material}
In the current study, we found that almost all our filaments had been accelerated to the velocity of the post-shock flow by the end of the simulations. The entrainment of cold, molecular filaments has been noted in the literature (e.g. in jet-filament interactions \citep{ODea13}). Although the current work concerns the interaction of a shock with a filament there is some relevance to wind-filament/cloud interactions, since the majority of the filaments in the simulations presented in this paper survived the passage of the initial shock and were then overrun by the post-shock flow, which can be thought of as resembling a wind of the same velocity. 

\citet{Zhang15} investigated hydrodynamic isothermal wind-cloud interactions. In their simulations, they found that the ram pressure from a hot wind was not able to accelerate the cloud to observed velocities since the cloud was rapidly shredded by KH instabilities whilst it was still at a relatively low velocity. This called into question how cool gas was able to be entrained and accelerated by the surrounding flow. The authors proposed an alternative theory whereby magnetic fields could prolong the cloud's life, allowing the build-up of turbulent instabilities to occur over a much larger timescale than that implied by the hydrodynamic simulations. \citet{McCourt15} also found that tangled internal magnetic fields suppressed mixing and allowed clouds to accelerate up to the wind speed.

In a similar vein, \citet{Scannapieco15} investigated the evolution of cold spherical clouds embedded in flows of hot and fast material. They found that the velocity of the cloud was dependent on the density contrast and the velocity of the hot wind; one implication being that if $\chi \gtrsim 100$, the cloud would not be accelerated to the hot wind speed before being destroyed. In addition, the authors considered the distance travelled by the cloud and found that this was proportional to the square of the lifetime. Thus, the suppression of KH instabilities can be important in determining the distance over which the cloud moves before its destruction. In the hydrodynamic case, the distance depended almost completely on the initial cloud radius. This presented problems in that for clouds to travel distances of $\sim 100$ kpc, as observed in nearby galaxies, they would need to be the size of a galaxy in order to do so without first being destroyed. The authors suggested that magnetic fields may be one way in which the cloud's lifetime could be extended to allow them to travel such large distances.

In our study, we found that the cloud density contrast, shock Mach number, and magnetic field orientation were important for determining the lifetime of filaments. A $\chi$ of 1000 in a parallel field and a shock Mach number of 10 led to the rapid destruction of the filament by turbulent instabilities before it had reached the velocity of the post-shock flow, whereas low values of $\chi$ in a weak shock and a perpendicular or oblique field provided the best conditions for the long-term survival of the filament. Filaments struck by a weak (e.g. $M=1.5$) shock, regardless of the orientation of the magnetic field, were easily able to reach the much lower post-shock flow velocity. It should be noted, however, that our simulations did not include the effects of evaporation on the filament, which \citet{Zhang15} consider to be important for the destruction of the cloud in the presence of a magnetic field. Our simulations also reveal that the presence of a magnetic field dramatically extends the filament lifetime, allowing it to move a distance downstream many tens, hundreds, or thousands of $r_{c}$, depending on the field orientation, before the filament is finally destroyed (and in some cases the filament may not be destroyed at all).

\section{Summary and Conclusions}
This is the second in a series of papers investigating the interaction between astrophysical shocks and filaments. In this paper, we employed a magnetohydrodynamic code to investigate the evolution and destruction by an adiabatic shock of a filament embedded within a magnetised medium. In comparison to the results from the previous hydrodynamical study of filaments by \citet{Pittard16} we found that the presence of magnetic fields and an increase in the density contrast of the filament had significant effects on the evolution of the filament. We summarise our main results for each orientation of the magnetic field as follows, noting that in all comparisons the time is normalised by $t_{cs}$:

\begin{itemize} 
\item \textit{Parallel fields}:
\begin{description}
\item[(i)] Filaments which are orientated either broadside, or nearly-broadside, on to the shock front survive for far longer than those orientated end on. Unless the filament is very small, the length of the filament has no significant effect on its evolution;
\item[(ii)] Well-defined linear structures situated on the axis behind the filament are formed only when the filament is end on with respect to the shock front (i.e. orientated at $\theta=90^{\circ}$);
\item[(iii)] An increase in the cloud density contrast hastens the destruction of the cloud through the increased presence of turbulent instabilities located on the filament surface. As the density contrast increases, so does the amount of turbulence;
\item[(iv)] Low shock Mach numbers restrict the filament from fragmenting, thus significantly prolonging its life.
\end{description}
\item \textit{Perpendicular/oblique fields}:
\begin{description}
\item[(vi)] Even if the filament is end on with respect to the shock front, filaments in a perpendicularly/obliquely-orientated magnetic field do not form flux ropes;
\item[(vii)] Compared with parallel-orientated fields, perpendicular/oblique fields shield the filament to a degree from the surrounding flow, allowing the filament lifetime to be considerably extended. The filament is more greatly confined by the field and maintains a higher average density;
\item[(viii)] Filaments are more rapidly accelerated to the velocity of the post-shock flow due to the effects of the magnetic pressure and field line tension;
\item[(ix)] An increase in the filament density contrast does not initiate large turbulent instabilities, compared to the case of a parallel field;
\item[(x)] A combination of a mild (e.g. $M=1.5$) shock and a perpendicular/oblique field allows the filament to survive almost intact for a considerable length of time.
\end{description}
\end{itemize}

The work presented in this paper is difficult to apply observationally since the adiabatic simulations do not include realistic physical processes such as thermal conduction, radiative cooling, and self-gravity. In future work we will extend our investigation to include the effects of radiative cooling, and will compare synthetic observations of such simulations with actual observations in order to present a more complete picture of the evolution of filaments in the ISM. It should be noted that \citet{Banda16} explored the effects of using a quasi-isothermal equation of state to approximate the effect of radiative cooling in MHD wind-cloud simulations and found that this led to significantly longer cloud lifetimes compared to the adiabatic case; a comparison with future isothermal shock-filament interactions would, therefore, be of interest.

\section*{Acknowledgements}
This work was supported by the Science \& Technology Facilities Council [Research Grants ST/L000628/1 and ST/M503599/1]. We thank S. Falle for the use of the MG magnetohydrodynamics code used to calculate the simulations in this work and S. van Loo for adding SILO output to it. The calculations used in this paper were performed on the DiRAC Facility which is jointly funded by STFC, the Large Facilities Capital Fund of BIS, and the University of Leeds. The 3D volumetric renderings were created using the VisIt visualisation and data analysation software. The data associated with this paper are openly available from the University of Leeds data repository. \url{http://doi.org/10.5518/74}

\end{document}